\documentclass[10pt]{article}

\usepackage{graphicx}
\usepackage{amsmath,amssymb,amsfonts,amsthm}
\usepackage{mathrsfs}
\usepackage{bbm}
\usepackage{xcolor}
\usepackage{booktabs}
\usepackage[title]{appendix}
\usepackage{enumerate}

\usepackage{tikz}
\usetikzlibrary{arrows}

\usepackage{subcaption}
\usepackage[ruled,vlined]{algorithm2e}

\newcommand{\OO}{\mathcal{O}}

\newcommand{\e}{\mathbbm{E}}
\newcommand{\N}{\mathbbm{N}}
\newcommand{\A}{\mathcal{A}}
\newcommand{\NN}{\mathcal{N}}
\newcommand{\T}{\mathcal{T}}

\newtheorem{theorem}{Theorem}
\newtheorem{proposition}{Proposition}
\newtheorem{remark}{Remark}%
\newtheorem{lemma}{Lemma}

\begin{document}

\title{Tree-based Algorithms for Breakpoint Median and Near-Median Genomes
}

\author{Poly H. da Silva, Arash Jamshidpey\footnote{email: arash.jamshidpey@berkeley.edu}, and David Sankoff}
\date{}

\maketitle

\begin{abstract}
Median genomes serve as candidates for ancestral reconstruction and help study the diversity of parsimonious rearrangement solutions. However, the breakpoint median problem is NP-hard for unichromosomal genomes, and medians are non-unique, making their diversity important to study.
In this paper, we develop tree-based algorithms for studying breakpoint medians of linear unichromosomal genomes represented as unsigned permutations. The algorithms construct rooted labeled trees whose leaves encode candidate median genomes satisfying local adjacency constraints derived from the inputs. We show that the resulting reduced search spaces contain all breakpoint medians while being much smaller than the full permutation space.
We investigate these reduced spaces experimentally for three-genome instances. Our method finds exact medians and many near-optimal approximations, while early sampling of leaves often produces near-medians quickly. We also introduce constrained sampling procedures for targeting balanced near-medians, whose distances to the input genomes are more evenly distributed and hence are particularly relevant for ancestral reconstruction.
The resulting framework provides a unified method for exact enumeration, efficient sampling, and targeted exploration of median and near-median genomes. In contrast to Traveling Salesman Problem-based approaches, which typically return a single solution and do not reveal the diversity or balance of the median set, our method can enumerate all medians, sample near-medians early, and target structurally distinct candidates under the breakpoint distance.

\vspace*{0.6mm}
\noindent\textbf{keywords:} Breakpoint Distance, Medians, Genomes, Exponential Algorithms

\end{abstract}

\section{Introduction}\label{section1}
Large-scale evolutionary changes often act on genomes through
rearrangements of long chromosomal regions. When genomes share the same
set of syntenic blocks, two basic questions are how to compare their
orders and how to infer plausible ancestral orders from several observed
genomes. Introduced by~\cite{sankoff97}, the
breakpoint distance counts adjacencies between consecutive blocks that
are present in one genome but absent in another, giving a simple and
interpretable measure of gene-order dissimilarity.
Median genomes, introduced in \cite{sankoff96}, play a significant role in ancestral reconstruction and in the
small phylogeny problem. Given a distance $d$ on the space of genomes and
genomes $G_1,\ldots,G_k$, a median is a genome $G$ minimizing the total distance function $\sum_{i=1}^{k} d(\cdot ,G_i)$.
The median problem has been studied extensively under a variety of genome rearrangement models
\cite{bryant98,caprara03,tannier09,fertincombinatorics,feijao09,xu10,zanetti16}.
When the space of genomes is endowed with the breakpoint distance,
\cite{sankoff97} showed that finding a breakpoint median for unichromosomal genomes can be reduced to the Traveling Salesman Problem (TSP) on a weighted graph of genes and is therefore NP-hard. The weight of an edge is the negative of the number of input genomes in which the corresponding adjacency occurs.

Consequently, many existing approaches rely on TSP heuristics, and the genome
returned by such a method is not guaranteed to be an exact median. Moreover,
these methods typically return only a single candidate, although breakpoint
medians need not be unique. The heuristic median returned by a TSP-based
method may also be very close to one of the input genomes
\cite{haghighi12}, thereby revealing little about the diversity of
possible optimal or near-optimal ancestral reconstructions.

Genome medians in general, and breakpoint medians in particular, provide good approximations to ancestral genomes in parsimony-based phylogenetic reconstruction when evolution is assumed to proceed along parsimonious paths \cite{zheng11,jam13,miardan23}. Even when exact breakpoint medians can be computed, the median set may be large, and its members may be widely separated. In fact, we show for the first time that the number of medians of a set of k genomes grows exponentially with the number of genes. Thus, to understand the implications of breakpoint medians for parsimony-based phylogenetic reconstruction, it is important to study not only an individual optimal median, but also the structure and diversity of the full exponential set of medians, the abundance of near-medians, and the geometry of the associated search space.

\cite{jam14} proved that, with high probability, a median of $k$ genomes takes
all but a proportion of size $O_p(1)$ of its gene adjacencies from the gene
adjacencies of the input genomes. \cite{dSJS25} improved this result
and found an upper bound on the number of novel adjacencies that a median may contain, that is, adjacencies that appear in none of the input genomes.
\cite{dSJS26} studies a similar problem for multichromosomal signed genomes.
However, the constraints imposed by signs lead to significantly
different median behavior and geometry than in the unsigned case.

In this paper, we consider unsigned linear unichromosomal genomes that share the same set of $n$ syntenic blocks and represent each genome by a
permutation of $[n]=\{1,\ldots,n\}$. 
Building on the structural results of
\cite{jam14,dSJS25}, we construct a family of labeled rooted trees of height
$n-1$. Each vertex is labeled by an element of $[n]$, representing a gene or
syntenic block, and labels do not repeat along any root-to-leaf path.
As a result, the sequence of labels along each path ending at level $n-1$
defines a permutation of $[n]$, and hence a candidate genome.
For a set of random input genomes, the
output genomes of the algorithm are all approximate medians with no novel gene
adjacencies, i.e., their gene adjacencies come from those of the input genomes.
In the special case in which the genomes are at pairwise maximum breakpoint
distance, the output genomes of the algorithm are the exact medians.

Algorithms~\ref{Alg:median2} and~\ref{Alg:median2prime} are modifications of Algorithm~\ref{Alg:median} designed to include all
common adjacencies (Algorithm~\ref{Alg:median2}) or all pairwise common adjacencies
(Algorithm~\ref{Alg:median2prime}) of the input genomes. In particular, this guarantees that all
geodesic points between two independent random input genomes are found.
Algorithm~\ref{Alg:median3} relaxes Algorithm 1 by allowing each ray of the tree to include at
most $\alpha>0$ (for a given value of $\alpha$) novel gene adjacencies.
For an input set of genomes $X$, Theorem~\ref{thmalpha} bounds the number of novel adjacencies that a median may contain by $\OO_n(X)$. Consequently, setting $\alpha=\OO_n(X)$ guarantees that all medians of $X$ are represented among the outputs of Algorithm~4.
Furthermore,
Theorem~\ref{Algo4_error_bound} shows that the remaining output genomes,
which are not exact medians, indeed provide very reliable approximations to
the medians when the pairwise intersections of the input genomes are
negligible. This, in particular, implies that for independent random genomes the outputs of Algorithms~\ref{Alg:median}, \ref{Alg:median2}, and \ref{Alg:median3} are approximate
medians with high probability.
By finding upper bounds for the total number of vertices in the trees that are
the outputs of Algorithms~\ref{Alg:median}, \ref{Alg:median2}, and \ref{Alg:median3}, we show that the algorithms have
exponential complexity.

The algorithms have several important implications. For the first time,
we have access to the full set of medians and, in addition, to a large collection of near-medians, making it possible to analyze the properties of both sets. On the other hand, for the first time, we can find medians at
positions away from the corners. In particular, we can find medians located at
equidistant positions from the input genomes. When constructing the full tree
is not feasible because of its time complexity, we establish a depth-first
search procedure to efficiently sample from the outputs of Algorithm~\ref{Alg:median3} (the
leaves of the tree) and study various statistics of the space of medians and
near-medians. Our experiments indicate that these early samples capture important features of the full output space.

This paper is an extension of~\cite{dSJS25b} and is organized as follows.
Section~\ref{background} introduces the breakpoint-median framework and
establishes structural results on the median value, novel adjacencies, and
common adjacencies preserved by medians.
Section~\ref{sec:algm} develops the tree-based algorithms, including variants
that enforce common or pairwise common adjacencies, allow a bounded number of
novel adjacencies, and accommodate circular unichromosomal genomes.
Section~\ref{sec:complexity} derives bounds on the sizes of the resulting
search trees and analyzes the running times of the algorithms.
Section~\ref{sec:exp} presents experimental results for three-genome
instances, including the distribution of near-optimal solutions, the reduction
in search-space size, early depth-first sampling, and constrained sampling of
balanced candidates. Some proofs are given in the appendices.

\section{Breakpoint medians}\label{background}

We represent a unichromosomal genome by a permutation $\pi$, which is a bijection on $[n]:=\{1,\cdots,n\}$. In other words, a permutation $\pi$ can be represented by $\pi(1),\cdots,\pi(n)$, which indicates a specific order on $[n]$. When there is no risk of ambiguity, we often write $\pi_i$ instead of $\pi(i)$, and denote $\pi:=\pi_1...\pi_n$. We define the set of adjacencies of $\pi$ as $\A_\pi=\{\{\pi_i,\pi_{i+1}\}\mid 1\leq i\leq n-1\}$, where each adjacency is treated as an unordered pair. Let $S_n$ denote the set of all permutations on $[n]$. Given $x, y \in S_n$, we denote by $\mathcal{A}_{x,y} := \mathcal{A}_x \cap \mathcal{A}_y$ the set of all common adjacencies of $x$ and $y$. For a set $X\subset S_n$, we also denote by $\mathcal{A}_X:= \bigcap_{x \in X} \mathcal{A}_x$ the set of all common adjacencies of permutations in $X$. The breakpoint (bp) distance between $x$ and $y$ is defined by $$d(x, y) := n - 1 - |\mathcal{A}_{x,y}|=\frac{1}{2}|\mathcal{A}_x \Delta \mathcal{A}_y|,$$
where $\Delta$ denotes the symmetric difference between sets. The breakpoint distance is neither a geodesic distance nor an edit distance, and for this reason the notion of partial geodesics was introduced by Jamshidpey \emph{et al.}~\cite{jam14}. We can consider the breakpoint distance as a generalized edit distance that determines the parsimonious (shortest) paths of transforming one permutation to another, but with many missing points in the parsimonious path. In other words, in edit distances the length of every jump from a point in the parsimonious path to its closest point in the same path is one, while in generalized edit distances such as the breakpoint distance this length may be bigger. A partial geodesic~\cite{jam14} between $x$ and $y$ is a maximal chain $x=\pi_0,\pi_1,...,\pi_{k-1},\pi_k=y$ in $S_n$ such that $$\sum_{i=0}^{k-1} d(\pi_i,\pi_{i+1})=d(x,y).$$ We denote by $\overline{[x, y]}=\overline{[x, y]}_n$ the set of all permutations lying on partial geodesics connecting $x, y \in S_n$, and call them geodesic points of $x$ and $y$. In other words, $z\in \overline{[x, y]}$ if and only if $d(x,z)+d(z,y)=d(x,y).$\\

We now recall the definition of breakpoint medians and establish several structural results that will be used in the algorithmic construction. The first part of the section introduces the median value and derives bounds in terms of the adjacency structure of the input genomes. We then bound the number of novel adjacencies that a median may contain, that is, adjacencies not appearing in any input genome. Finally, we give sufficient conditions under which common adjacencies of the input genomes must be preserved by every median. These results form the basis for the reduced adjacency-based search spaces used by the tree algorithms developed in the following sections.

\subsection{Breakpoint medians and median-value bounds}
For $k\in \N$ and a set $X=\{x_1,...,x_k\}\subset S_n$, a breakpoint median is a permutation (genome) in $S_n$ that minimizes the total breakpoint distance function $$d_T(\cdot,X):=\sum_{i=1}^k d(\cdot, x_i).$$ The minimal value of $d_T$ is known as the median value of the set $X$, denoted by $\mu(X)=\mu_n(X)$. The set of all breakpoint medians of $X$ is denoted by $M(X)=M_n(X)$. We often omit the subscript $n$ when no ambiguity arises.

For a set of permutations $X=\{x_1,...,x_k\}$ in $S_n$ for which the pairwise breakpoint distances take the maximum value $n-1$, Jamshidpey \emph{et al.}~\cite{jam14} provide a necessary and sufficient condition  for a permutation $m$ to be a median of $X$, that is, $m$ is a median of $X$  if and only if 
$
\mathcal{A}_m \subset \bigcup_{x\in X}\mathcal{A}_{x}. 
$
Moreover, \cite{jam14} provides a necessary and sufficient condition for a permutation $\pi$ to be a geodesic point (equivalently, a median) of two permutations $x$ and $y$. A proof is provided in Appendix~\ref{appen:A}
\begin{theorem}[\cite{jam14}]\label{iff-Meidan-k=2}
    The followings are equivalent for $x,y,z\in S_n$.
    \begin{enumerate}
        \item[1)] $z\in \overline{[x,y]}$.
        \item[2)] $z\in M(\{x,y\}).$
        \item[3)] $\mathcal{A}_{x,y}\subset \mathcal{A}_{z} \subset \mathcal{A}_x \cup \mathcal{A}_y.$
    \end{enumerate}
\end{theorem}
There is no analogue of the previous result for $k \geq 3$. In particular, it is not straightforward to obtain necessary and sufficient conditions, in the sense of Theorem~\ref{iff-Meidan-k=2}, for a permutation to be a median of a general set $X\subset S_n$ when $|X| \geq 3$. More explicitly, a median need not contain all common adjacencies of the permutations in $X$; that is, there may exist a median $m$ such that $\mathcal{A}_X \nsubseteq \mathcal{A}_m$, as shown by an example of Bryant~\cite{bryant98}. However, although some medians may fail to contain all common adjacencies, there always exists at least one median with this property, namely, a median $m$ such that $\mathcal{A}_X \subseteq \mathcal{A}_m$ (cf.~\cite{bryant98}). In addition, for a general set $X$ with $|X|\geq 3$, it is not necessary that every adjacency of a median $m\in M(X)$ is an adjacency of at least one of the permutations in $X$, that is, there may exist a median $m$ such that $\A_m\nsubseteq \bigcup_{x\in X}\A_x,$ as shown by Bryant~\cite{bryant98}. 

In the remainder of this section, we derive bounds on the median value of a set $X$. Furthermore, we obtain an upper bound on the number of adjacencies of a median $m \in M(X)$ that do not appear in $\bigcup_{i=1}^k \mathcal{A}_{x_i}$, as well as an upper bound on the number of common adjacencies in $\mathcal{A}_X$ that do not appear in $\mathcal{A}_m$.

Denote by $\mathcal{P}(S)$ the set of all subsets of a set $S$. Let $X=\{x_1, ...,x_k\}\subset S_n$ and let $\mathcal{B}_X^X=\mathcal{B}_{x_1,...,x_k}^X:=\mathcal{A}_{x_1,...,x_k}$. Then, for any $j=1,\cdots,k$, let
$$
\mathcal{B}_{x_1,...,x_{j-1},x_{j+1},...x_k}^X:=\mathcal{A}_{x_1,...,x_{j-1},x_{j+1},...x_k} \setminus \mathcal{B}_{x_1,...,x_k}^X
$$
Continuing in this way, for any $i_1,\cdots, i_r\in [k]$ and $U=\{x_{i_1},...,x_{i_r}\}\subset X$, we set
$$
\mathcal{B}_U^X=\mathcal{B}_{x_{i_1},...,x_{i_r}}^X:=\mathcal{A}_U \setminus (\bigcup\limits_{U \subsetneqq V} \mathcal{B}_V^X).
$$
In other words, $\mathcal{B}_U^X$ includes all adjacencies that are common in every $x\in U$, but missing from every $y\in X\setminus U$. The proof of the following theorem is provided in Appendix~\ref{appen:A}.

\begin{theorem}\label{thm:sandwich}
Let $X=\{x_1,\ldots,x_k\}\subset S_n$. Suppose $m\in S_n$ satisfies
\begin{equation}\label{eq:sandwich}
\bigcup_{1\le i<j\le k}\mathcal{A}_{x_i x_j}\ \subseteq\ \mathcal{A}_m\ \subseteq\  \bigcup_{x\in X}\mathcal{A}_x.
\end{equation}
Then $m\in M(X)$ and, moreover,
\[
\mu(X)=(k-1)(n-1)-\sum_{r=1}^k (r-1)\sum_{1\leq i_1<\cdots<i_r\leq k}|\mathcal{B}_{x_{i_1},...,x_{i_r}}^X|.
\]
\end{theorem}

\begin{remark}\label{rem:unionB}
Using the disjoint decomposition $\{\mathcal{B}^X_U\}$ introduced above, the union of all pairwise common adjacencies can be written as
\[
\bigcup_{1\le i<j\le k}\mathcal{A}_{x_i,x_j}
=\bigcup_{r\ge 2}\ \bigcup_{\substack{U\subseteq X\\ |U|=r}} \mathcal{B}^X_U.
\]
Indeed, an adjacency belongs to the left-hand side if and only if it is present in at least two genomes, i.e.\ it belongs to $\mathcal{B}^X_U$ for some $U\subseteq X$ of size $r\ge 2$.
\end{remark}
Note that a median satisfying (\ref{eq:sandwich}) is the best possible median for a set $X$ and hence provides a lower bound for $\mu(X)$.\\

For $X\subset S_n$, let
\[
\lambda_n(X):=\sum_{r=1}^k (r-1)\sum_{1\leq i_1<\cdots<i_r\leq k}|\mathcal{B}_{x_{i_1},...,x_{i_r}}^X|,
\]
and
$
\upsilon_n(X):= \max\limits_i \sum_{j\neq i} |\mathcal{A}_{ij}|.
$
The following result gives a lower and an upper bounds, $L_n$ and $U_n$, for $\mu(X)$.
\begin{theorem}\label{thm:medianbounds2}
For any $X=\{x_1,\ldots,x_k\}\subset S_n$,
\[
\mathcal L_n:=(k-1)(n-1)-\lambda_n(X) \le\ \mu_n(X)\ \le\ \mathcal U_n:=(k-1)(n-1)-\upsilon_n(X).
\]
\end{theorem}
\begin{remark}
For $k=3$, since
$\mathcal{A}_{ij}=\mathcal{B}_{ij}^X\cup \mathcal{B}_{123}^X,$
we can write
\begin{align*}
2(n-1)-\sum_{i,j} |\mathcal{A}_{ij}| + |\mathcal{A}_{123}|
&\leq 2(n-1) -\sum_{i,j} |\mathcal{B}_{ij}^X| -2 |\mathcal{B}_{123}^X| \\
&\leq \mu(X)\leq 2(n-1)- \sum_{i,j} |\mathcal{A}_{ij}| +\min_{i,j} |\mathcal{A}_{ij}|,
\end{align*}
where the sums are over all $1\leq i<j\leq 3$.
\end{remark}

\subsection{Bounding novel adjacencies in medians}
in this section, for a set of genomes $X\subset S_n$, we obtain an upper bound for the maximum number of adjacencies of an arbitrary median that do not appear in $\bigcup_{x\in X}\A_x$. The proof of the following theorem is provided in the appendix.

\begin{theorem}[\cite{dSJS25}]\label{thmalpha}
Let $X=\{x_1,...,x_k\}\subset S_n$ be such that 
$$
d_T(x_k,X)=\min\limits_{i=1...k} d_T(x_i,X),
$$
and let $m\in M(X)$. Then
\begin{equation}\label{upbO}
|\mathcal{A}_m \setminus (\bigcup\limits_{i=1}^k \mathcal{A}_{x_i})| \leq \OO_n(X):=\sum\limits_{r=2}^{k-1} (r-1)\sum\limits_{1\leq i_1<...<i_r<k} |\mathcal{B}_{x_{i_1},...,x_{i_r}}^X|.
\end{equation}
In particular, for $k=3$, for any $m\in M(X)$
$$
|\mathcal{A}_m\setminus \bigcup\limits_{i=1}^3 \mathcal{A}_{x_i}| \leq \OO_n(\{x_1,x_2,x_3\}):=|\mathcal{B}_{x_1,x_2}^X|.
$$
\end{theorem}

\begin{remark}
    Note that the theorem makes use of the upper bound $d_T(m,X)$ $\leq d_T(x_k,X),$ for any $m\in M(X)$. In particular, for $k=3$, $d_T(x_3,X)=\min_{i=1,2,3} d_T(x_i,X)$ is equivalent to $d(x_1,x_2)=\max_{i,j} d(x_i,x_j)$, which itself is equivalent to $|\mathcal{B}_{x_1,x_2}^X|=\min_{i\neq j} |\mathcal{B}_{x_i,x_j}^X|$. In this case, $\OO_n(X)=|\mathcal{B}_{x_1,x_2}^X|=\min_{i\neq j} |\mathcal{B}_{x_i,x_j}^X|$ implies that the upper bound is the number of adjacencies common in the pair of farthest genomes, i.e., $x_1$ and $x_2$, which are missing from $x_3$.
\end{remark}

The bound in Theorem~\ref{thmalpha} indicates that medians can be searched for within a substantially smaller subset of $S_n$ determined by conditions on adjacencies.

\subsection{Common adjacencies preserved by medians}\label{Section-missing common adjacencies}
In this section, we derive sufficient conditions under which an adjacency $\{a,b\} \in \mathcal{A}_X$, for $X \subset S_n$, must appear in every median of $X$. This, in turn, yields an upper bound on the number of common adjacencies in $\mathcal{A}_X$ that may be absent from a median of $X$.\\

To further analyze the structure of medians, we associate a line graph with a permutation. More precisely, we define the map ``$\,\hat{}\,$'' by
\[
x \in S_n \longmapsto \hat{x},
\]
where $\hat{x}=([n],\mathcal A_x)$ is a graph with vertex set $[n]$ and edge set $\mathcal A_x$. In other words, two vertices $i,j$ are connected in $\hat{x}$ if $\{i,j\}\in \mathcal A_x$. Denote by $\hat{S}_n$ the set of all line graphs on $[n]$.

For a set $X\subseteq S_n$, it is also useful to define $\mathcal G(X)$ to be the union of the graphs $\hat{x}$ for $x\in X$. More precisely, let
\[
\mathcal G(X)=\bigcup_{x\in X}\hat{x},
\]
be the graph with $[n]$ as its vertex set and $\bigcup_{x\in X} \mathcal A_x$ as its edge set. In particular,
$\mathcal G(x)=\hat{x},$ for $ x\in S_n.$

For a graph $G$ and a vertex $a$ in $G$, we denote by $\mathcal N_G(a)$ the set of vertices $b$ of $G$ for which $\{a,b\}$ is an edge in $G$, i.e.
$\mathcal N_G(a)=\{b:\ b \text{ is adjacent to } a \text{ in } G\}.$
In particular, with a slight abuse of notation, for $X\subset S_n$, let
${\mathcal N}_X(a)=\mathcal N_{\mathcal G(X)}(a).$

For $X\subset S_n$, and $a,b\in [n]$, the weight of the pair $\{a,b\}$ with respect to $X$, denoted by $w_X(a,b)$, is defined to be the number of permutations $x\in X$ for which $\{a,b\}\in \mathcal A_x$. For an arbitrary graph $G$ with vertex set $[n]$, let
\[
W_X(a,G)=\bigl(w_X(a,a_1),w_X(a,a_2),\ldots,w_X(a,a_r)\bigr),
\]
where the neighbors of $a$ in $G$, namely
$\mathcal N_G(a)=\{a_1,a_2,\ldots,a_r\},$ are labeled such that
$w_X(a,a_j)\geq w_X(a,a_{j+1}),$ for $ j=1,\ldots,r-1.$

Specifically, let
\[
\overline{W}_X(a)=W_X(a,\mathcal G(X)).
\]

Also, define
\[
\varphi_X(a,G)=\sum_{b\in \mathcal N_G(a)} w_X(a,b),
\]
and, in particular,
\[
\phi_X(a)=\sum_{b\in [n]} w_X(a,b)=\varphi_X(a,\mathcal G(X)),
\]
recalling that $w_X(a,a)=0.$
Finally, for $x\in S_n$, let
\[
\bigl(\varphi_X^{+}(a,x),\varphi_X^{+}(b,x)\bigr):=\bigl(w_X(a,a'),w_X(b,b')\bigr),
\]
where $a'\in \mathcal N_{\hat{x}}(a)$ and $b'\in \mathcal N_{\hat{x}}(b)$ are located on the unique path in $\hat{x}$ connecting $a$ to $b$. We have the following proposition.

For the remainder of this section, we assume that the elements of \(X\) have pairwise distinct line-graph representations; that is, for every distinct pair \(x,y\in X\), we have \(\hat x\neq \hat y\).
\begin{proposition}\label{Suff-cond-not-median}
Let $X\subset S_n$ with $|X|=k$. For $a\neq b\in [n]$, suppose $\{a,b\}\in \mathcal A_X$ and $x\in S_n$ is such that $\{a,b\}\notin \mathcal A_x$. In addition, suppose one of the following conditions holds:
\begin{enumerate}
\item[(C1)] $\min\{|\mathcal N_{\hat x}(a)|,|\mathcal N_{\hat x}(b)|\}=1$.
\item[(C2)] $\min\{\varphi_X(a,\hat x),\varphi_X(b,\hat x)\}<k$.
\item[(C3)] $\varphi_X(a,\hat x)=\varphi_X(b,\hat x)=k$ and $W_X(a,\hat x)\neq W_X(b,\hat x)$.
\item[(C4)] $\varphi_X(a,\hat x)=\varphi_X(b,\hat x)=k$, and 
$\bigl(\varphi_X^{+}(a,x),\varphi_X^{+}(b, x)\bigr)\notin \{(i,i): \ k/2\leq i<k \}.$
\item[(C5)] $W_X(a,\hat x)=W_X(b,\hat x)=(k,0)$.
\end{enumerate}
Then $x\notin M(X)$.
\end{proposition}

Applying the above results, we derive some useful properties of medians of $X$ when $\{a,b\}\in \mathcal A_X$. The proof will be given in the appendix.

\begin{theorem}\label{phi>kk}
Let $X\subseteq S_n$, with $|X|=k$, and let $a,b\in [n]$ be such that $\{a,b\}\in \mathcal A_X$. Let $x\in M(X)$.
\begin{enumerate}
\item[(i)] Then
$\bigl(\varphi_X(a,\hat x),\varphi_X(b,\hat x)\bigr)\geq (k,k).$
\item[(ii)] If additionally,
$\varphi_X(a,\hat x)=\varphi_X(b,\hat x)=k,$
then
$W_X(a,\hat x)=W_X(b,\hat x).$
\end{enumerate}
\end{theorem}

\begin{remark}
Note that $W_X(a,\hat x)=W_X(b,\hat x)$ does not guarantee $\{a,b\}\in \mathcal A_x$ in Theorem~\ref{phi>kk}.
\end{remark}

We are ready to investigate conditions under which $\{a,b\}\in \mathcal A_x$ for $x\in M(X)$ if $\{a,b\}\in \mathcal A_X$. Let $\Gamma$ be the set of those edges (adjacencies) $\{a,b\}$ in $\mathcal G(X)$ for which
\[
\overline{W}_X(a)=\overline{W}_X(b)=(k,i,k-i),
\]
for some $k/2\leq i<k$, and let $\mathcal E(X)=\mathcal E_n(X)=|\Gamma|$.

\begin{theorem}\label{AdjX-Notin-M(X)}
Let $X\subset S_n$ such that $|X|=k$. Suppose $a,b\in [n]$ are such that $\{a,b\}\in \mathcal A_X$ and let $x\in M(X)$. Suppose $\{a,b\}\notin \mathcal A_x$. Then there exists $k/2\leq i<k$ such that
\[
\overline{W}_X(a)=\overline{W}_X(b)=(k,i,k-i).
\]
Furthermore,
\[
|\mathcal A_X\setminus \mathcal A_x|\leq |\mathcal A_X\setminus \mathcal A_{M(X)}|\leq \mathcal E_n(X).
\]
\end{theorem}
\begin{remark}
It follows from the last theorem that if $\{a,b\} \in \mathcal{A}_X$, $x \in M(X)$, and
$(|\mathcal{N}_X(a)|, |\mathcal{N}_X(b)|) \ne (3,3),$
then $\{a,b\} \in \mathcal{A}_x$. If $(|\mathcal{N}_X(a)|, |\mathcal{N}_X(b)|)=(3,3)$, but
$\overline{W}_X(a) \ne \overline{W}_X(b),$
then we still have $\{a,b\} \in \mathcal{A}_x$.
\end{remark}

\begin{remark}
Note that the converse is not true, i.e., if $x \in M(X)$ and
$\overline{W}_X(a) = \overline{W}_X(b) = (k,i,k-i),$
for some $k/2 \le i < k$, we may still have $\{a,b\} \in \mathcal{A}_x$. The example is $k=2$, in which $\mathcal{A}_X \subseteq \mathcal{A}_x$ for any median $x \in M(X)$ (see Theorem~\ref{iff-Meidan-k=2}).
\end{remark}

In the next sections, we introduce algorithms that construct specific rooted labeled trees, where each ray (a shortest path from the root to a leaf) of length $n - 1$ corresponds to a permutation determined by the sequence of labels along the path. The set of all such label sequences includes all approximate medians, thereby significantly reducing the search space. In particular, the resulting median search space consists of the set of all leaves of these trees. Although the size of this search space is still exponential, it is negligible compared to that of the full permutation space on $n$ elements.

\section{Tree-based algorithms for finding medians}\label{sec:algm}

In this section, we present a family of constructive algorithms for generating all medians of a given set of permutations $X\subset S_n$ using a tree-based construction that can be viewed in close analogy with a suffix tree. 
Rather than searching over the entire permutation group, the algorithms build permutations incrementally, one number at a time, by extending partial permutations (prefixes) while enforcing local adjacency constraints derived from $X$.   
These constraints drastically restrict the set of admissible extensions at each step and lead to a much smaller, structured search space.  
We encode these rules by rooted labeled trees whose root-to-leaf label sequences of length $n-1$ produce permutations.

To formalize this tree construction, we first recall the definition of the
neighbors of a point (i.e., a number representing a syntenic block or gene) with respect to a given set of permutations (genomes).  
Specifically, for $X=\{x_1,\ldots,x_k\}\subset S_n$ and $i\in[n]$, we define
$$\NN_X(i)=\NN_{x_1,...,x_k}(i)=\{j: \{i,j\}\in \bigcup\limits_{l=1}^k\A_{x_l}\}.$$ Note that for each $i$, $1\leq |\NN_X(i)|\leq 2k$. The equality $|\NN_X(i)|=1$ holds when $i$ satisfies both of the following conditions: $i$ is either the first or last element in each permutation $x_l$, for $1\leq l \leq k$; and $i$ is an extremity of an adjacency in $\A_X$. On the other hand, the equality $|\NN_X(i)|=2k$ holds when $i$ satisfies both of the following conditions: $i$ is neither the first nor  the last number of any permutation $x_l$, for $1\leq l \leq k$; and $i$ is not an extremity of an adjacency in $\A_{x_l,x_p}$, for any $l\neq p$. If $X$  is such that $d(x_l,x_p)=n-1$ for any $l\neq p$, then $k\leq |\NN_X(i)|\leq 2k$.

For each $i\in[n]$, we build a rooted tree whose root vertex is denoted by $\varnothing$ and whose label is fixed to be $\ell(\varnothing)=i$.  
Vertices of the tree are created inductively by extending root-to-vertex paths, as described below.

Each vertex $u$ that appears in the construction carries a label $\ell(u)\in[n]$.  
A root-to-vertex path $(u_0=\varnothing,u_1,\ldots,u_j=u)$ spells the label sequence
$(\ell(\varnothing)=i,\ell(u_1),\ldots,\ell(u_j))$.
The children of a vertex $u$ are in bijection with all $t\in [n]$ that are adjacent to $\ell(u)$ in at least one input genome and do not already appear as a label along this path; in other words,
\begin{equation}\label{eq:basic-extension}
t\in \NN_X(\ell(u))\quad\text{and}\quad 
t\notin \{\ell(\varnothing),\ell(u_1),\ldots,\ell(u_j)\}.
\end{equation}
Each such child $v$ is labeled by $\ell(v)=t$. In particular, the root has $|\NN_X(i)|$ children.

Whenever a root-to-leaf path has length $n-1$, its label sequence is a permutation $y\in S_n$.  
We refer to such a $y$ as a \emph{permutation given by a leaf}.

In the rest of this section, we first present an algorithm in Section~\ref{Alg1} for constructing trees in which every permutation $y$ given by a leaf satisfies  
$$
\A_y \subset \bigcup_{x\in X} \A_x.
$$ 
In this case, if the breakpoint distance between every pair of permutations in $X$ attains the maximum value $n - 1$, then, from Jamshidpey \emph{et al.}~\cite{jam14}, any permutation $y$ given by a leaf at level $n - 1$ is a median of $X$. Consequently, the algorithm finds all medians of $X$.  

Next, in Section~\ref{Alg2}, we construct trees where every permutation $y$ given by a leaf satisfies  
$$
\A_X \subset \A_y \subset \bigcup\limits_{x\in X} \A_x.
$$  
In this case, if the upper bound given in (\ref{upbO}) is zero—a weaker condition than requiring all pairwise distances in $X$ to be maximal—then at least one of the permutations given by a leaf of the tree is a median of $X$ (cf.~\cite{bryant98}). This allows us to determine the median value within a relatively smaller search space.  More generally, when the input genomes share few pairwise adjacencies, as is typical for random genomes, the same construction produces near-median candidates with controlled total-distance error, as proved in Theorem~\ref{thm:error-alg1}.

We further introduce in the same section a pairwise-restricted variant in which the lower bound $\A_X$ is replaced by the stronger constraint
\[
\bigcup_{i<j} \A_{x_i,x_j} \subset \A_y .
\]
Under this condition, whenever the corresponding trees has some vertices at level $n-1$, every leaf
at level $n-1$ is a median of $X$, yielding an even more sharply focused
enumeration of medians.

In Section~\ref{Alg3}, we introduce a modification of the algorithm from Section~\ref{Alg1}, providing additional flexibility to identify all medians of a general set of permutations. This is achieved by allowing permutations to contain a limited number of adjacencies not present in  
$
\bigcup_{x\in X} \A_x.
$  
The upper bound in (\ref{upbO}) ensures that all medians of $X$ are represented among the leaves of the tree constructed by this flexible algorithm.

Finally, in Section~\ref{sec:circular}, we describe a specialization of the tree-based framework to
circular unichromosomal genomes, which exploits rotational symmetry to reduce
the number of trees by a factor of $n$. The resulting construction produces the corresponding median and near-median candidates in the circular setting. This modification preserves the median
structure while yielding substantial computational speed-ups in practice.

\subsection{Medians and near-medians with no novel adjacencies}\label{Alg1}

Here, we describe the first tree-based construction, which generates permutations
$y$ satisfying
\[
\A_y \subset \bigcup_{x\in X}\A_x .
\]

Let $X=\{x_1,\ldots,x_k\}\subset S_n$.
For each $i\in[n]$, we build a rooted labeled tree $\T_X^i$ whose root is denoted by $\varnothing$ and has label $\ell(\varnothing)=i$.
Vertices are created inductively by extending root-to-vertex paths using the admissible extension rule \eqref{eq:basic-extension}.

For a vertex $u$ in $\T_X^i$, we denote by $T_u$ the ordered sequence of labels along the path from the root to $u$, and by $dig(T_u)$ the set of labels appearing in $T_u$.
Let $L_j(\T_X^i)$ be the set of vertices at level $j$, with $L_0=\{\varnothing\}$.

Whenever a root-to-leaf path has length $n-1$, its label sequence defines a permutation
$y\in S_n$ satisfying $\A_y\subset \bigcup_{x\in X}\A_x$.
We write $Y(\T_X^i)$ for the set of all such permutations produced by $\T_X^i$, and define
\[
\T_X := \{\T_X^i:1\le i\le n\},\qquad
Y(\T_X):=\bigcup_{i=1}^n Y(\T_X^i).
\]

We use the symbol $\oplus$ to denote concatenation of sequences.
If $w=(w_1,\ldots,w_\ell)$ is any finite sequence and $a$ is a symbol, we write
$w\oplus a := (w_1,\ldots,w_\ell,a)$.
This operation is used both for indexing tree vertices and for extending label sequences.
To index children in the tree, if $u=\varnothing j_1\cdots j_\ell$ is a vertex (viewed as a word) and $r\in[2k]$,
then $u\oplus r:=\varnothing j_1\cdots j_\ell r$ denotes its $r$-th child. Also, if $T_u$ is the sequence of labels along the $(\varnothing,u)$-path, then $T_{u\oplus r}=T_u\oplus \ell(u\oplus r)$. The explicit tree construction is summarized in Algorithm~\ref{Alg:median}.

\begin{algorithm}[ht]
\KwData{$X=\{x_1, ..., x_k\}\subset S_n $.}
\KwResult{Permutations $y$ such that $\A_y\subset \bigcup\limits_{l=1}^k \A_{x_l}$.}
\For{$i=1,...,n$}{
\For {$w=0,...,n$}{
$L_w\longleftarrow\emptyset$
}
$L_0\longleftarrow\{\varnothing\}$\\
$T_{\varnothing}\longleftarrow i$\\
\For{$j=1,...,n-1$}{
\For{$u\in L_{j-1}$}{
$r\longleftarrow 0$\\
\For{$t\in \NN_X(\ell(u))$}{
\If{$t \notin dig(T_u)$}{
$r \longleftarrow r+1$\\
$L_j \longleftarrow L_j\cup \{u \oplus r\}$\\
$T_{u \oplus r}\longleftarrow T_u\oplus  t$}
}
}
}
\For{$u\in L_{n-1}$}{
print $T_u$}
}
\caption{Gives permutations $y$ such that $\A_y\subset \bigcup\limits_{x\in X} \A_{x}$ \label{Alg:median} }
\end{algorithm}

Under an additional maximal-distance assumption on $X$,  every permutation produced by this construction is in fact a median. In other words, if $X=\{x_1,\ldots,x_k\}\subset S_n$ satisfies
$
d(x_\ell,x_p)=n-1,\text{ for all }\ell\ne p
$,
then $\A_{x_\ell}\cap \A_{x_p}=\emptyset$ for all $\ell\ne p$, and consequently any permutation
$y$ with $\A_y\subset \bigcup_{x\in X}\A_x$ minimizes the total breakpoint distance to $X$.
Hence, in this case, every permutation produced by the trees $\T_X^i$ is a median, and
$Y(\T_X)=M(X).$
In particular, consider the case of two permutations. 
Let $id$ denote the identity permutation and let $x\in S_n$ satisfy
$d(id,x)=n-1$.
Then $\A_{id}\cap \A_x=\emptyset$, and by the characterization of geodesic points recalled in Section~\ref{background} (see also \cite{jam14}), a permutation $y$ lies on a geodesic between $id$ and $x$ if and only if
$
\A_y \subset \A_{id}\cup \A_x .
$
Applying the above tree construction with $X=\{id,x\}$ therefore yields
$Y(\T_{id,x})=\overline{[id,x]},$
the set of all geodesic points between $id$ and $x$.
More generally, when the input genomes share few pairwise adjacencies, as is typical for random genomes, the same construction produces near-median candidates with controlled total-distance error, as proved in Theorem~\ref{thm:error-alg1}.

We illustrate the construction in Algorithm~\ref{Alg:median} using the example shown in Figure~\ref{tree}, where $x=246315$ and $X=\{id,x\}$ with $d(id,x)=n-1=5$. 
Consider the tree $\T_{id,x}^1$, whose root has label $\ell(\varnothing)=1$. 
We have 
$\NN_{id,x}(1)=\{2,3,5\}$, and therefore the root has exactly three children, with labels $2$, $3$, and $5$. 
These correspond to the three possible ways to extend the prefix $(1)$ to a length-two prefix.

Take the branch labeled $3$. 
Since $\NN_{id,x}(3)=\{1,2,4,6\}$ and $1$ already appears in the prefix, the admissible extensions from the prefix $(1,3)$ are $2$, $4$ and $6$. 
Thus the vertex with label sequence $(1,3)$ has three children labeled $2$, $4$ and $6$, producing the prefixes $(1,3,2)$, $(1,3,4)$ and $(1,3,6)$. 
The same rule is applied recursively at every vertex: from a prefix $(w_1,\ldots,w_j)$, the next symbol must belong to $\NN_{id,x}(w_j)$ and must not yet appear among $\{w_1,\ldots,w_j\}$.

Continuing in this manner, each root-to-leaf path of length $5$ produces a permutation of $\{1,\ldots,6\}$ whose adjacencies are contained in $\A_{id}\cup \A_x$. 
Since $\A_{id}\cap \A_x=\emptyset$, every such permutation lies on a partial geodesic between $id$ and $x$, and hence is a median of $\{id,x\}$. 
The set of all leaves in $\T_{id,x}^1$ therefore lists exactly the medians that start with $1$, and repeating the same construction for roots $i=2,\ldots,6$ yields all medians of the pair $\{id,x\}$. The bold edges in Figure~\ref{tree} correspond to adjacencies of $id$, while the remaining edges correspond to adjacencies of $x$.

\begin{center}
\begin{figure}[ht!]
\begin{footnotesize}\begin{center}

\begin{tikzpicture}[line cap=round,line join=round,>=triangle 45,x=0.7cm,y=0.7cm]
\clip(-3.06,-1.57) rectangle (13.69,5.65);
\draw [line width=1.8pt] (5,5)-- (0,3);
\draw (5,5)-- (5,3);
\draw (5,5)-- (11,3);
\draw [line width=1.8pt] (0,3)-- (-1,2);
\draw (0,3)-- (1,2);
\draw [line width=1.8pt] (-1,2)-- (-2,1);
\draw (-1,2)-- (-0.58,0.98);
\draw (-2,1)-- (-1.72,0.02);
\draw [line width=1.8pt] (-2,1)-- (-2.3,0.02);
\draw [line width=1.8pt] (-2.3,0.02)-- (-2.28,-1);
\draw [line width=1.8pt] (-1.72,0.02)-- (-1.72,-1);
\draw (-0.58,0.98)-- (-1,0);
\draw [line width=1.8pt] (-0.58,0.98)-- (-0.28,0.04);
\draw [line width=1.8pt] (-1,0)-- (-1,-1);
\draw [line width=1.8pt] (-0.28,0.04)-- (-0.26,-1.02);
\draw [line width=1.8pt] (1,2)-- (0.42,0.98);
\draw [line width=1.8pt] (1,2)-- (1,1);
\draw (1,2)-- (1.66,0.98);
\draw (0.42,0.98)-- (0.42,-0.04);
\draw [line width=1.8pt] (0.42,-0.04)-- (0.42,-1.04);
\draw [line width=1.8pt] (1,1)-- (1,0);
\draw (1,0)-- (1,-1);
\draw (1.66,0.98)-- (1.42,-0.02);
\draw [line width=1.8pt] (1.66,0.98)-- (2,0);
\draw [line width=1.8pt] (5,3)-- (3.55,1.87);
\draw (5,3)-- (5,2);
\draw (5,3)-- (6.43,1.93);
\draw (3.55,1.87)-- (3.55,0.87);
\draw [line width=1.8pt] (3.55,0.87)-- (3,0);
\draw (3.55,0.87)-- (4,0);
\draw [line width=1.8pt] (3,0)-- (3,-1);
\draw [line width=1.8pt] (4,0)-- (4,-1);
\draw [line width=1.8pt] (5,2)-- (4.52,0.98);
\draw (5,2)-- (5.5,0.98);
\draw (5,2)-- (4.95,0.98);
\draw (4.52,0.98)-- (4.52,-0.02);
\draw (5.5,0.98)-- (5.5,-0.02);
\draw (6.43,1.93)-- (5.97,0.86);
\draw (6.43,1.93)-- (6.83,0.93);
\draw [line width=1.8pt] (5.97,0.86)-- (6,0);
\draw [line width=1.8pt] (6.83,0.93)-- (6.48,0);
\draw (6.83,0.93)-- (7.17,0.06);
\draw (6,0)-- (6,-1);
\draw [line width=1.8pt] (11,3)-- (9.14,2.02);
\draw [line width=1.8pt] (11,3)-- (12,2);
\draw (9.14,2.02)-- (10.24,1.02);
\draw [line width=1.8pt] (9.14,2.02)-- (9.2,1.1);
\draw (9.14,2.02)-- (8.24,1.06);
\draw [line width=1.8pt] (8.24,1.06)-- (8.28,0.02);
\draw (8.28,0.02)-- (8.3,-1.02);
\draw [line width=1.8pt] (9.2,1.1)-- (9,0);
\draw (9.2,1.1)-- (9.56,0.02);
\draw (10.24,1.02)-- (10.26,0.02);
\draw (10.26,0.02)-- (10.26,-0.97);
\draw (12,2)-- (11.3,0.94);
\draw (12,2)-- (12.74,1.02);
\draw [line width=1.8pt] (11.3,0.94)-- (11,0);
\draw (11.3,0.94)-- (11.66,0);
\draw [line width=1.8pt] (11,0)-- (11,-1);
\draw [line width=1.8pt] (11.66,0)-- (11.67,-1.05);
\draw [line width=1.8pt] (12.74,1.02)-- (13,0);
\draw [line width=1.8pt] (12.74,1.02)-- (12.34,0.08);
\draw (12.34,0.08)-- (12.34,-1.02);
\draw (13,0)-- (13,-1);
\fill [color=black] (5,5) circle (1.5pt);
\draw[color=black] (4.99,5.26) node {$1$};
\fill [color=black] (0,3) circle (1.5pt);
\draw[color=black] (-0.08,3.29) node {$2$};
\fill [color=black] (5,3) circle (1.5pt);
\draw[color=black] (5.16,3.26) node {$3$};
\fill [color=black] (11,3) circle (1.5pt);
\draw[color=black] (11.15,3.26) node {$5$};
\fill [color=black] (-1,2) circle (1.5pt);
\draw[color=black] (-1.14,2.2) node {$3$};
\fill [color=black] (1,2) circle (1.5pt);
\draw[color=black] (1.13,2.26) node {$4$};
\fill [color=black] (-2,1) circle (1.5pt);
\draw[color=black] (-2.2,1.2) node {$4$};
\fill [color=black] (-0.58,0.98) circle (1.5pt);
\draw[color=black] (-0.43,1.22) node {$6$};
\fill [color=black] (-1.72,0.02) circle (1.5pt);
\draw[color=black] (-1.58,0.19) node {$6$};
\fill [color=black] (-2.3,0.02) circle (1.5pt);
\draw[color=black] (-2.46,0.19) node {$5$};
\fill [color=black] (-2.28,-1) circle (1.5pt);
\draw[color=black] (-2.54,-0.95) node {$6$};
\fill [color=black] (-1.72,-1) circle (1.5pt);
\draw[color=black] (-1.46,-0.95) node {$5$};
\fill [color=black] (-1,0) circle (1.5pt);
\draw[color=black] (-0.75,0.19) node {$4$};
\fill [color=black] (-0.28,0.04) circle (1.5pt);
\draw[color=black] (-0.11,0.19) node {$5$};
\fill [color=black] (-1,-1) circle (1.5pt);
\draw[color=black] (-0.75,-0.95) node {$5$};
\fill [color=black] (-0.26,-1.02) circle (1.5pt);
\draw[color=black] (-0.06,-0.95) node {$4$};
\fill [color=black] (0.42,0.98) circle (1.5pt);
\draw[color=black] (0.3,1.2) node {$3$};
\fill [color=black] (1,1) circle (1.5pt);
\draw[color=black] (1.2,1.2) node {$5$};
\fill [color=black] (1.66,0.98) circle (1.5pt);
\draw[color=black] (1.8,1.2) node {$6$};
\fill [color=black] (0.42,-0.04) circle (1.5pt);
\draw[color=black] (0.6,0.19) node {$6$};
\fill [color=black] (0.42,-1.04) circle (1.5pt);
\draw[color=black] (0.65,-0.95) node {$5$};
\fill [color=black] (1,0) circle (1.5pt);
\draw[color=black] (1.21,0.19) node {$6$};
\fill [color=black] (1,-1) circle (1.5pt);
\draw[color=black] (1.26,-0.95) node {$3$};
\fill [color=black] (1.42,-0.02) circle (1.5pt);
\draw[color=black] (1.6,-0.15) node {$3$};
\fill [color=black] (2,0) circle (1.5pt);
\draw[color=black] (2.21,-0.15) node {$5$};
\fill [color=black] (3.55,1.87) circle (1.5pt);
\draw[color=black] (3.5,2.02) node {$2$};
\fill [color=black] (5,2) circle (1.5pt);
\draw[color=black] (5.32,2.12) node {$4$};
\fill [color=black] (6.43,1.93) circle (1.5pt);
\draw[color=black] (6.7,2.19) node {$6$};
\fill [color=black] (3.55,0.87) circle (1.5pt);
\draw[color=black] (3.85,0.94) node {$4$};
\fill [color=black] (3,0) circle (1.5pt);
\draw[color=black] (2.88,0.23) node {$5$};
\fill [color=black] (4,0) circle (1.5pt);
\draw[color=black] (4.28,0.26) node {$6$};
\fill [color=black] (3,-1) circle (1.5pt);
\draw[color=black] (3.33,-0.95) node {$6$};
\fill [color=black] (4,-1) circle (1.5pt);
\draw[color=black] (4.26,-0.95) node {$5$};
\fill [color=black] (4.52,0.98) circle (1.5pt);%%
\draw[color=black] (4.35,1.12) node {$5$};%%
\fill [color=black] (4.95,0.98) circle (1.5pt);%%%
\draw[color=black] (4.95,0.7) node {$2$};%%%
\fill [color=black] (5.5,0.98) circle (1.5pt);%%
\draw[color=black] (5.7,1.05) node {$6$};%%
\fill [color=black] (4.52,-0.02) circle (1.5pt);
\draw[color=black] (4.62,-0.22) node {$6$};
\fill [color=black] (5.5,-0.02) circle (1.5pt);
\draw[color=black] (5.51,-0.3) node {$5$};
\fill [color=black] (5.97,0.86) circle (1.5pt);
\draw[color=black] (6.35,0.94) node {$5$};
\fill [color=black] (6.83,0.93) circle (1.5pt);
\draw[color=black] (7.11,1.18) node {$4$};
\fill [color=black] (6,0) circle (1.5pt);
\draw[color=black] (6.27,0.13) node {$4$};
\fill [color=black] (6.48,0) circle (1.5pt);
\draw[color=black] (6.81,-0.04) node {$5$};
\fill [color=black] (7.17,0.06) circle (1.5pt);
\draw[color=black] (7.5,-0.02) node {$2$};
\fill [color=black] (6,-1) circle (1.5pt);
\draw[color=black] (6.35,-0.95) node {$2$};
\fill [color=black] (9.14,2.02) circle (1.5pt);
\draw[color=black] (9.23,2.32) node {$4$};
\fill [color=black] (12,2) circle (1.5pt);
\draw[color=black] (12.27,2.26) node {$6$};
\fill [color=black] (10.24,1.02) circle (1.5pt);
\draw[color=black] (10.58,1.11) node {$6$};
\fill [color=black] (9.2,1.1) circle (1.5pt);
\draw[color=black] (9.64,1.14) node {$3$};
\fill [color=black] (8.24,1.06) circle (1.5pt);
\draw[color=black] (8.04,1.16) node {$2$};
\fill [color=black] (8.28,0.02) circle (1.5pt);
\draw[color=black] (8.56,0.1) node {$3$};
\fill [color=black] (8.3,-1.02) circle (1.5pt);
\draw[color=black] (8.66,-0.95) node {$6$};
\fill [color=black] (9,0) circle (1.5pt);
\draw[color=black] (9.25,-0.11) node {$2$};
\fill [color=black] (9.56,0.02) circle (1.5pt);
\draw[color=black] (9.87,-0.11) node {$6$};
\fill [color=black] (10.26,0.02) circle (1.5pt);
\draw[color=black] (10.61,0.15) node {$3$};
\fill [color=black] (10.26,-0.97) circle (1.5pt);
\draw[color=black] (10.56,-0.95) node {$2$};
\fill [color=black] (11.3,0.94) circle (1.5pt);
\draw[color=black] (11.62,1.09) node {$4$};
\fill [color=black] (12.74,1.02) circle (1.5pt);
\draw[color=black] (13.02,1.27) node {$3$};
\fill [color=black] (11,0) circle (1.5pt);
\draw[color=black] (11.23,0.26) node {$3$};
\fill [color=black] (11.66,0) circle (1.5pt);
\draw[color=black] (11.92,0.26) node {$2$};
\fill [color=black] (11,-1) circle (1.5pt);
\draw[color=black] (11.3,-0.95) node {$2$};
\fill [color=black] (11.67,-1.05) circle (1.5pt);
\draw[color=black] (11.94,-0.95) node {$3$};
\fill [color=black] (13,0) circle (1.5pt);
\draw[color=black] (13.28,0.26) node {$2$};
\fill [color=black] (12.34,0.08) circle (1.5pt);
\draw[color=black] (12.68,0.13) node {$4$};
\fill [color=black] (12.34,-1.02) circle (1.5pt);
\draw[color=black] (12.61,-0.95) node {$2$};
\fill [color=black] (13,-1) circle (1.5pt);
\draw[color=black] (13.35,-0.95) node {$4$};
\end{tikzpicture}
\end{center}
\end{footnotesize}
\caption[The representation of the tree $\T_{id,x}^1$, for $x=246315$, with its labels]{The representation of the tree $\T_{id,x}^1$, for $x=246315$, with its labels. In this example we have $\NN_{id,x}(1)= \{2,3,5\}$, $\NN_{id,x}(2)= \{1,3,4\}$, $\NN_{id,x}(3)= \{1,2,4,6\}$, $\NN_{id,x}(4)= \{2,3,5,6\}$, $\NN_{id,x}(5)= \{1,4,6\}$ and $\NN_{id,x}(6)=\{3,4,5\}$. Also, $d(id,x)=n-1=5$, and so each path from the root to a leaf at level $5$ constitutes a permutation in $\overline{[id,x]}$. The list of permutations in $\overline{[id,x]}$ given by this tree is: $id=123456$, $123465$, $123645$, $123654$, $124365$, $124563$, $132456$, $132465$, $136542$, $154236$, $154632$, $156432$, $156423$, $156342$ and $156324$. These are all the permutations in $\overline{[id,x]}$ that start at $1$. The bold edges represent the adjacencies of $id$ and the other edges represent the adjacencies of $x$.\label{tree}}
\end{figure}
\end{center}

\begin{remark}[Finding permutations with maximum distance from a set $X$]
Given  $X=\{x_1, ..., x_k\}\subset S_n$, denote by $\overline{\NN}_X(i)=\overline{\NN}_{x_1,...,x_k}(i):=[n]\setminus \NN_X(i)$ the complement set of $\NN_X(i)$, for $1\leq i \leq n$. Note that, if we replace $\NN_X(\cdot)$ by $\overline{\NN}_X(\cdot)$ in Algorithm~\ref{Alg:median}, we obtain all permutations with maximum distance from $X$, i.e., we find all permutations $y$ such that $d(y,x_j)=n-1$, for $1\leq j \leq k$.
\end{remark}

In general, when the number of common adjacencies between each pair of input genomes is small, as is typically the case for $k$ random genomes, the candidates produced by Algorithm~1 provide good approximations to breakpoint medians. The following theorem provides an upper bound on the error of the approximate medians obtained from Algorithm~1.

\begin{theorem}\label{thm:error-alg1}
Let $X \subset S_n$, with $|X|=k$. Then for any $i \in [n]$, $\mathcal{T}_X^i$ is a connected tree, and $X \subseteq Y(\mathcal{T}_X)$. Moreover, for every $y \in Y(\mathcal{T}_X)$,
\[
0 \le d_T(y,X) - \mu_n(X) \le \lambda_n(X).
\]
\end{theorem}

\begin{proof}
By construction, it is trivial that $\mathcal{T}_X^i$ is a connected tree for any $i \in [n]$, and $X \subseteq Y(\mathcal{T}_X)$. Now for every $y \in Y(\mathcal{T}_X)$
\[
(k-1)(n-1) - \lambda_n(X) \le \mu_n(X) \le d_T(y,X) \le (k-1)(n-1),
\]
where the first two inequalities come from Theorem~\ref{thm:medianbounds2}.

For the last inequality in the statement of the theorem, note that the maximum total distance of an element of $Y(\mathcal{T}_X)$ to $X$ occurs if there exists $\tilde{y} \in Y(\mathcal{T}_X)$ for which
\[
\mathcal{A}_{\tilde{y}} = \bigcup_{x \in X} \mathcal{B}^{X \cup \{\tilde{y}\}}_{\tilde{y}x},
\]
where each adjacency of $\tilde{y}$ is from exactly one input $x \in X$. Then for any $y \in Y(\mathcal{T}_X)$ we have
\[
\begin{aligned}
d_T(y,X) &\le d_T(\tilde{y},X)
= \sum_{x\in X}(n-1-|\mathcal{A}_{\tilde{y}x}|) \\
&= k(n-1)-\sum_{x\in X}\left|\mathcal{B}_{\tilde{y}x}^{X\cup\{\tilde{y}\}}\right|
= (k-1)(n-1).
\end{aligned}
\]
\end{proof}

\subsection{Medians including common adjacencies}\label{Alg2}

We next introduce a restricted version of the tree-based construction of
Section~\ref{Alg1} for the case where permutations in $X$ may share adjacencies.
Motivated by the fact that any geodesic point of a pair of permutations must contain all their common adjacencies,
we modify the tree construction so that these adjacencies are enforced along every
root-to-leaf path, and hence appear in every generated permutation.
More generally, the restricted construction produces permutations $y$ satisfying
\begin{equation}\label{eq:Alg2-sandwich}
\A_X \subseteq \A_y \subseteq \bigcup_{x\in X}\A_x .
\end{equation}

The common adjacencies form disjoint paths in the line graph representation of permutations, which we call
segments. Intuitively, once a path enters such a segment, it must traverse
the entire segment without branching, otherwise the resulting permutation would
fail to contain all required adjacencies.

More precisely, let $\pi\in S_n$ and $I\subset \A_\pi$.  A segment $s$ of $I$ is a maximal set of consecutive adjacencies of $I$, i.e., it is a set $$s=\{\{\pi(r),\pi(r+1)\},\{\pi(r+1),\pi(r+2)\},\cdots,\{\pi(r+k-1),\pi(r+k)\}\}\subset I$$ such that $\{\pi(r-1),\pi(r)\}, \{\pi(r+k),\pi(r+k+1)\}\notin I$, for $r>1$ and $r+k<n$. We often denote $s$ by $\|\pi(r),\cdots, \pi(r+k)\|$, and write $s\hat{\in} I $. We say that $Int(s):=\{\pi(r+1),\cdots, \pi(r+k-1)\}$ are the internal points of $s$, and $End(s):=\{\pi(r), \pi(r+k)\}$ are the end points of $s$.
Generalizing the idea, the internal and end points of $I\subset \mathcal{A}_\pi$ are defined by $$ Int(I):=\bigcup_{s\hat{\in} I} Int(s), \ \ \  
End(I):=\bigcup_{s\hat\in I} End(s).
$$
Note that the above definitions do not depend on a specific choice of $\pi$, that is, the definitions remain intact if we replace $\pi$ by any $\pi'$ for which $I\subset \mathcal A_{\pi'}$.

These segment definitions lead to simple but crucial constraints on the tree construction: segments impose forced continuation rules along any valid construction path.
Consider a segment $s\hat{\in}\A_X$ and an endpoint $j\in End(s)$.
Write $\overline{j}$ for the other endpoint of $s$ and let $j^*$ be the unique symbol such that $\{j,j^*\}\in s$.
Then, in order to satisfy $\A_X\subseteq \A_y$, once a partial construction reaches an endpoint $j$ of a segment whose other endpoint $\overline{j}$ has not yet been used, the next symbol is forced to be $j^*$, and the construction must continue without branching until it reaches $\overline{j}$.

Moreover, no internal point of $\A_X$ can be used as the root label:
if $i\in Int(\A_X)$ then two distinct adjacencies in $\A_X$ are incident to $i$, and a permutation produced by a tree rooted at $i$ cannot realize both simultaneously at the beginning of the path.

Accordingly, for each $i\in [n]\setminus Int(\A_X)$ we construct a rooted labeled
tree $\overline{\T}^{\,i}_{X}$ with root label $\ell(\varnothing)=i$.
Vertices are created as in Algorithm~\ref{Alg:median}, except that the branching rule
is modified as follows:
\begin{itemize}
\item If $\ell(u)\in End(\A_X)$ and $\overline{\ell(u)}\notin dig(T_u)$, then $u$ has exactly one child $v$ and the label of $v$ is $\ell(v)=\ell(u)^*$, where $\{\ell(u),\ell(u)^*\}$ is the adjacency in
the corresponding segment of $\mathcal{A}_X$.
\item Otherwise, $u$ branches as in \eqref{eq:basic-extension}, labels of the children correspond to all $t\in \NN_X(\ell(u))\setminus dig(T_u)$.
\end{itemize}

After $n-1$ extension steps, each leaf at level $n-1$ corresponds to a permutation $y$ satisfying \eqref{eq:Alg2-sandwich}.  Let
\[
\overline{\T}_X:=\{\overline{\T}_X^i: i\in[n]\setminus Int(\A_X)\},
\qquad
Y(\overline{\T}_X):=\bigcup_{i\in[n]\setminus Int(\A_X)} Y(\overline{\T}_X^i).
\]

For a set $X$ where the upper bound in (\ref{upbO}) is equal to zero (recall that this condition is weaker than requiring all permutations in $X$ to be at maximum pairwise distance), from~\cite{bryant98}, there exists at least one $y\in Y(\overline{\T}_X)$ that is a median of $X$. More precisely, in this case, any $y'\in Y(\overline{\T}_X)$ such that 
$$
\sum\limits_{l=1}^k d(x_l,y')=\underset{y\in Y(\overline{\T}_X)}{\min} \left( \sum\limits_{l=1}^k d(x_l,y)\right),
$$
is a median of $X$. Thus, in addition to finding some medians of $X$, this algorithm also finds the median value of $X$ efficiently. The tree construction process is described in Algorithm~\ref{Alg:median2}.
\begin{algorithm}[ht]
\KwData{$X=\{x_1, ..., x_k\}\subset S_n $.}
\KwResult{Permutations $y$ such that $\A_{x_1,...,x_k}\subset\A_y\subset \bigcup\limits_{l=1}^k \A_{x_l}$.}
\For{$i\in \{1,...,n\}\setminus Int(\A_X)$}{
\For {$w=0,...,n$}{
$L_w\longleftarrow\emptyset$\\
}
$L_0\longleftarrow\{\varnothing\}$\\
$T_{\varnothing}\longleftarrow i$\\
\For{$j=1,...,n-1$}{
\For{$u\in L_{j-1}$}{
$r\longleftarrow 0$\\
\If {$\ell(u)\in End(\A_X)$ and $\overline{\ell(u)}\notin dig(T_u)$}{
$r \longleftarrow r+1$\\
$L_j \longleftarrow L_j\cup \{u \oplus r\}$\\
$T_{u \oplus r}\longleftarrow T_u\oplus  \ell(u)^*$\\
}
\Else{
\For{$t\in \NN_X(\ell(u))$}{
\If{$t \notin dig(T_u)$}{
$r \longleftarrow r+1$\\
$L_j \longleftarrow L_j\cup \{u \oplus r\}$\\
$T_{u \oplus r}\longleftarrow T_u\oplus  t$\\}
}
}
}
}
\For{$u\in L_{n-1}$}{
print $T_u$}
}
\caption{Generates all permutations $y$ such that $\A_{x_1,...,x_k}\subset\A_y\subset \bigcup\limits_{l=1}^k \A_{x_l}$ \label{Alg:median2}}
\end{algorithm}

This construction also has a useful graph-theoretic interpretation. More precisely, Algorithm~2 may be viewed as Algorithm~1 applied to a contracted version of the union graph \(\mathcal G(X)\). Namely, each maximal common segment \(s\) of \(\mathcal A_X\) is contracted to a single effective edge joining the two vertices in \(\operatorname{End}(s)\), while the vertices in \(\operatorname{Int}(s)\) are suppressed. We denote the resulting contracted graph by \(\widetilde{\mathcal G}(X)\). The contracted edge represents the traversal of the whole common segment. Thus, once the construction enters a contracted edge of \(\widetilde{\mathcal G}(X)\) through one of its adjacent vertices (endpoints), the continuation is deterministic by crossing the corresponding contracted edge. Branching is then resumed only after the contracted edge has been crossed, or at vertices not contained in any common segment. 

If $d(x_l,x_p)=n-1$ for all $l\neq p$, then $\A_X=\emptyset$ and the forced-continuation condition never triggers; Algorithm~\ref{Alg:median2} reduces to Algorithm~\ref{Alg:median}.
More generally, for any $X$, each tree $\overline{\T}_X^i$ is a subtree of $\T_X^i$.
The restriction is exactly that (i) the root cannot lie in $Int(\A_X)$, and (ii) along each segment $s\hat{\in}\A_X$, once a path hits an endpoint it continues deterministically along $s$ until it reaches the other endpoint.

The restriction \eqref{eq:Alg2-sandwich} can be further strengthened by
requiring that all adjacencies shared by any pair of input permutations
must be present in the constructed permutation.
Define
\[
\mathcal A_X^* := \bigcup_{1\le i<j\le k} \mathcal A_{x_i,x_j}.
\]
We then consider permutations $m$ satisfying the stronger condition
\begin{equation}\label{eq:Alg2-pairwise-sandwich}
\mathcal A_X^* \subseteq \mathcal A_m \subseteq \bigcup_{x\in X}\mathcal A_x.
\end{equation}

This condition is directly motivated by Theorem~\ref{thm:sandwich}, which shows
that any $m\in S_n$ satisfying \eqref{eq:Alg2-pairwise-sandwich} is
necessarily a median of $X$ with the best possible median value, i.e., the total distance of $m$ to $X$ attains the lower bound $\mathcal{L}_n$ in Theorem~\ref{thm:medianbounds2}. Consequently, whenever a tree-based construction enforcing
\eqref{eq:Alg2-pairwise-sandwich} is nonempty, every leaf at level $n-1$
corresponds to a median, and no further optimization over the output set is
required.

Algorithmically, this pairwise-restricted variant is obtained by replacing
$\mathcal A_X$ with $\mathcal A_X^*$ throughout the construction described
above.
All segment definitions, forced-continuation rules, and admissible root labels
are determined with respect to $\mathcal A_X^*$ rather than $\mathcal A_X$.
In particular, internal points of $\mathcal A_X^*$ are excluded as possible
roots, and once a construction path reaches an endpoint of a segment
$s\hat\in\mathcal A_X^*$, it is forced to traverse the entire segment before
branching is allowed.

When the resulting trees are nonempty at level $n-1$, this construction enumerates a subset of
$Y(\overline{\mathcal T}_X)$ consisting exclusively of exact medians.
Compared to Algorithm~\ref{Alg:median2}, the pairwise-restricted variant may
produce fewer candidates or even be empty, but in return it provides a direct
and certifying procedure for median enumeration that bypasses any distance
evaluation.
In practice, this variant is particularly effective when the input permutations
share many pairwise adjacencies, yielding a small number of highly constrained
trees whose leaves are all optimal solutions.

We refer to this pairwise-restricted variant as Algorithm~\ref{Alg:median2prime}.
In analogy with Algorithm~\ref{Alg:median2}, the corresponding rooted trees are
denoted by $\tilde{\mathcal T}^i_X$, where  all segment constraints and forced continuations are defined using
the pairwise adjacency set $\mathcal A_X^*$ instead of $\mathcal A_X$. Let $Y(\tilde{\mathcal T}^i_X)$ be the set of permutations generated by $\tilde{\mathcal T}^i_X$ at level $n-1$. Note that $Y(\tilde{\mathcal T}^i_X)$ may be empty. We write
\[
\tilde{\mathcal T}_X
:=\{\tilde{\mathcal T}^i_X : i\in[n]\setminus Int(\mathcal A_X^*)\},
\qquad
Y(\tilde{\mathcal T}_X)
:=\bigcup_i Y(\tilde{\mathcal T}^i_X)
\]
for the resulting collection of trees and the set of permutations they generate.

\begin{algorithm}[ht]
\KwData{$X=\{x_1,\ldots,x_k\}\subset S_n$.}
\KwResult{Permutations $m$ satisfying 
$\mathcal A_X^*\subseteq \mathcal A_m \subseteq \bigcup_{x\in X}\mathcal A_x$.}

Compute $\mathcal A_X^*=\bigcup_{1\le i<j\le k}\mathcal A_{x_i,x_j}$\;

\For{$i\in [n]\setminus Int(\mathcal A_X^*)$}
{
  construct a rooted tree $\tilde{\mathcal T}^i_X$ as in
  Algorithm~\ref{Alg:median2}, with all segment and forced-continuation rules
  defined using $\mathcal A_X^*$\;
}

Output all permutations labeling leaves at level $n-1$ of the trees
$\tilde{\mathcal T}^i_X$\;

\caption{: Pairwise-restricted tree construction}
\label{Alg:median2prime}
\end{algorithm}

\begin{remark}[Geometric strength of adjacency constraints]
The distinction between the constraints
$\mathcal A_X\subseteq \mathcal A_y$ and
$\mathcal A_X^*\subseteq \mathcal A_y$
has a natural geometric interpretation.
The set $\mathcal A_X$ consists of adjacencies shared by \emph{all} input
permutations and therefore captures only globally stable structures.
In contrast, $\mathcal A_X^*$ aggregates adjacencies that are stable across
\emph{at least one pair} of inputs, encoding a richer collection of locally
consistent features.
From the perspective of the tree construction, replacing $\mathcal A_X$ by
$\mathcal A_X^*$ increases the number and length of forced segments, thereby
reducing branching and shrinking the space of admissible permutations.
Geometrically, this corresponds to restricting attention to a smaller region of
the geodesic hull of $X$ where multiple pairwise geodesic constraints intersect.
While the stronger constraint may render the construction empty in some cases,
Theorem~\ref{thm:sandwich} shows that whenever feasible permutations exist, they
are necessarily exact medians.
Thus $\mathcal A_X^*$ provides a sharper but more selective geometric filter
than $\mathcal A_X$, trading completeness for optimality.
\end{remark}

The following theorems determine how close the outputs of Algorithm 2 and Algorithm 3 are to a median.

\begin{theorem}\label{thm:error-alg2}
Let $X \subseteq S_n$ with $|X|=k$. For any $i \in [n]$, $\overline{\mathcal{T}}_X^{\,i}$ is a connected tree, and
$X \subseteq Y(\overline{\mathcal{T}}_X).$
Moreover, the following statements hold.

\begin{enumerate}
\item[(i)] For $k=2$, we have 
$Y(\overline{\mathcal{T}}_X) \subseteq M(X),$
and for any $y \in Y(\overline{\mathcal{T}}_X)$,
\[
\mu(X) = d_T(y,X) = n-1 - |\mathcal{A}_X|.
\]

\item[(ii)] For $k \ge 3$, and for any $y \in Y(\overline{\mathcal{T}}_X)$,
\[
0 \le d_T(y,X) - \mu_n(X) \le \lambda_n(X) - (k-1)|\mathcal{A}_X|.
\]
\end{enumerate}
\end{theorem}
\begin{proof}
Part (i) is trivial from Theorem~\ref{iff-Meidan-k=2}. For part (ii), note that for every $y\in Y(\overline{\mathcal{T}}_X)$,
\[
(k-1)(n-1) - \lambda_n(X) \le \mu_n(X) \le d_T(y,X) \le (k-1)(n-1-|\mathcal{A}_X|),
\]
where the first inequality comes from Theorem~\ref{thm:medianbounds2}. For the last inequality, the maximum possible total distance of elements of $Y(\overline{\mathcal{T}}_X)$ to $X$ occurs if there exists $\tilde{y} \in Y(\overline{\mathcal{T}}_X)$ such that
\[
\mathcal{A}_{\tilde{y}} = \mathcal{A}_X \cup \left( \bigcup_{x \in X} \mathcal{B}^{X \cup \{\tilde{y}\}}_{\tilde{y}x} \right).
\]

Then for any $y \in Y(\overline{\mathcal{T}}_X)$,
\[
d_T(y,X) \le d_T(\tilde{y},X) = \sum_{x \in X} \left( n-1 - |\mathcal{A}_X| - \left|\mathcal{B}^{X \cup \{\tilde{y}\}}_{\tilde{y}x}\right| \right)
\]
\[
= (k-1)(n-1-|\mathcal{A}_X|).
\]
\end{proof}
\begin{remark}
The error bound in Theorem~\ref{thm:error-alg2} is independent of the size of
$\mathcal{A}_X$. Moreover, this bound is small when the pairwise-common
adjacency sets
$\mathcal{A}_{x_i,x_j}\setminus\mathcal{A}_X$ are small for $i,j\in [k]$. Hence, under
this condition, Algorithm~\ref{Alg:median2} remains a reliable method for obtaining good
median approximations whether $|\mathcal{A}_X|$ is small or large. The
former regime arises, for example, when $X$ consists of independent
permutations sampled uniformly from $S_n$. The latter arises when the
input genomes share a linear number of adjacencies, so that
$|\mathcal{A}_X|\approx cn$ for some $0<c\leq 1$. In this regime, the
input genomes belong to a more closely related family, and the common
segments, represented by contracted edges, substantially reduce the
search space and, consequently, the running time.
\end{remark}
The following theorem follows Thorem~\ref{thm:sandwich}.
\begin{theorem}
Let $X \subseteq S_n$ with $|X|=k$. For any $i \in [n]$, $\tilde{\mathcal{T}}_X^{\,i}$ is a connected tree, and $Y(\tilde{\mathcal{T}}_X)\subseteq M(X)$. If $Y(\tilde{\mathcal{T}}_X)$ is non-empty, then for any $y \in Y(\tilde{\mathcal{T}}_X)$,
\[
d_T(y,X) = (k-1)(n-1) - \lambda_n(X).
\]
\end{theorem}

\begin{remark}
Algorithm~3 is not guaranteed to have any solutions. More precisely, $Y(\tilde{\mathcal{T}}_X)$ may be empty for some $X \subseteq S_n$.
\end{remark}
\subsection{Budgeted flexible branching for finding all medians} \label{Alg3} 

Algorithm~\ref{Alg:median} searches for medians inside the union constraint
$\A_y\subseteq \bigcup_{x\in X}\A_x$.  For a general input set $X\subset S_n$, however,
a true median may contain some ``outside'' adjacencies.  Theorem~\ref{thmalpha}
shows that the number of such outside adjacencies is at most $\OO_n(X)$ for every
median $m\in M(X)$. To apply this result to $k$ independent random permutations, namely $\mathcal X=\{\xi_1,...,\xi_k\}\subset S_n$, recall that a sequence of random variables $(Z_n)_{n\in \mathbb Z_+}$ converges in probability to a random variable $Z$, as $n$ goes to infinity, if for any $\varepsilon>0$, $\mathbb P(|Z_n-Z|>\varepsilon)\rightarrow 0$. We know that $\OO_n(\mathcal X)$ is very small with high probability. More explicitly, from~\cite{dSJS25}, we know that 
 $$\frac{\OO_n(\mathcal X)}{a_n}\rightarrow 0, \ n\rightarrow \infty,$$ 
 in probability, for any sequence $(a_n)_{n\in \N}$ diverging to $\infty$, such that $a_n/n \rightarrow 0$, as $n\rightarrow \infty$. 

 Thus, in the random regime, allowing $\alpha=\OO_n(\mathcal X)$ outside adjacencies enlarges the search space only slightly, while still capturing all medians with high probability.
 
Motivated by this, we enlarge the search space by allowing up to $\alpha$ outside
adjacencies, where $\alpha$ will later be chosen as $\alpha=\OO_n(X)$, for $X\subset S_n$.
For $\alpha\ge 0$, we say that a permutation $\pi\in S_n$ is
$\alpha$-freedom with respect to $X$ if
\[
\bigl|\A_\pi\setminus \textstyle\bigcup_{x\in X}\A_x\bigr|\le \alpha .
\]
In particular, taking $\alpha=\OO_n(X)$ guarantees that every median of $X$ is
$\alpha$-freedom.

We implement this flexibility by augmenting the tree construction with a
\emph{budget} parameter.  For each $i\in[n]$ we construct a rooted labeled tree
$\T^i_{X,\alpha}$ with root $\varnothing$ and $\ell(\varnothing)=i$, together with
a budget value $\tau(u)=\tau_u\in\{0,1,\ldots,\alpha\}$ attached to each vertex $u$.
Intuitively, $\tau(u)$ counts how many additional outside adjacencies we may still
introduce after reaching $u$.  We initialize $\tau(\varnothing)=\alpha$.

As before, a vertex $u$ represents the partial permutation $T_u$ spelled by the
labels along the root-to-$u$ path, and $dig(T_u)$ denotes the set of used symbols along $T_u$.
The children of $u$ are created by extending $T_u$ by one new symbol, with the
following rule. If $\tau(u)=0$, then no further outside adjacency is allowed.  The vertex $u$ branches exactly as in \eqref{eq:basic-extension}, i.e., the labels of the children are in bijection with all $t\in \NN_X(\ell(u))\setminus dig(T_u)$, and each child $v$ has
$\ell(v)=t$ and $\tau(v)=0$.
If $\tau(u)>0$, then $u$ branches to \emph{all} unused symbols:
its children are in bijection with the set $[n]\setminus dig(T_u)$.  For each $t\in [n]\setminus dig(T_u)$ we create a child $v$ with $\ell(v)=t$. The budget is updated according to whether the newly created adjacency $\{\ell(u),t\}$ lies in the union adjacency set
\begin{equation}\label{budget-def}
\tau(v)=
\begin{cases}
\tau(u), & t\in \NN_X(\ell(u)),\\
\tau(u)-1, & t\notin \NN_X(\ell(u)).
\end{cases}
\end{equation}

After $n-1$ extension steps, each leaf $u$ at level $n-1$ spells a permutation
$y=T_u$.  By construction, $y$ is $\alpha$-freedom with respect to $X$.
We write $Y(\T^i_{X,\alpha})$ for the set of leaf permutations produced by
$\T^i_{X,\alpha}$ and define
\[
\T_{X,\alpha}:=\{\T^i_{X,\alpha}:1\le i\le n\},\qquad
Y(\T_{X,\alpha}):=\bigcup_{i=1}^n Y(\T^i_{X,\alpha}).
\]
When $\alpha=\OO_n(X)$, Theorem~\ref{thmalpha} implies that
\[
M(X)\subseteq Y(\T_{X,\OO}).
\]
Moreover, for any leaf $u$, the value $\tau(u)$ records the remaining budget, so
$\alpha-\tau(u)$ is exactly the number of outside adjacencies used along $T_u$.
Finally, $\T_X^i$ from Algorithm~\ref{Alg:median} is a subtree of $\T^i_{X,\alpha}$
(obtained by forbidding the spending of the budget), as illustrated in
Figure~\ref{treealpha}.  The explicit construction is summarized in
Algorithm~\ref{Alg:median3}.

\begin{center}
\begin{figure}[ht!]
\begin{center}
\definecolor{ffqqqq}{rgb}{1,0,0}
\definecolor{qqqqcc}{rgb}{0,0,0.8}
\begin{scriptsize}
\begin{tikzpicture}[line cap=round,line join=round,>=triangle 45,x=0.6cm,y=0.5cm]
\clip(-5.02,-4.52) rectangle (14.22,4.04);
\draw [ color=qqqqcc] (5,3.5)-- (-2,2);
\draw [ color=qqqqcc] (5,3.5)-- (3,2);
\draw [ color=qqqqcc] (5,3.5)-- (9,2);
\draw   (5,3.5)-- (13,2);
\draw [ color=ffqqqq] (-2,2)-- (-0.5,0);
\draw   (-2,2)-- (-2,0);
\draw [ color=qqqqcc] (-2,2)-- (-4,0);
\draw [ color=qqqqcc] (3,2)-- (1,0);
\draw [ color=ffqqqq] (3,2)-- (3,0);
\draw   (3,2)-- (5,0);
\draw   (9,2)-- (7,0);
\draw [ color=qqqqcc] (9,2)-- (9,0);
\draw [ color=qqqqcc] (9,2)-- (11,0);
\draw   (13,2)-- (12.5,0);
\draw   (13,2)-- (13.5,0);
\draw   (-4,0)-- (-3.5,-2);
\draw   (-2,0)-- (-2.5,-2);
\draw   (-2,0)-- (-1.5,-2);
\draw [ color=qqqqcc] (-4,0)-- (-4.5,-2);
\draw [ color=ffqqqq] (-0.5,0)-- (-0.5,-2);
\draw   (1,0)-- (0.5,-2);
\draw [ color=qqqqcc] (1,0)-- (1.5,-2);
\draw   (3,0)-- (2.5,-2);
\draw [ color=ffqqqq] (3,0)-- (3.5,-2);
\draw   (5,0)-- (4.5,-2);
\draw   (7,0)-- (6.5,-2);
\draw   (7,0)-- (7.5,-2);
\draw [ color=qqqqcc] (9,0)-- (8.5,-2);
\draw   (9,0)-- (9.5,-2);
\draw [ color=qqqqcc] (11,0)-- (10.5,-2);
\draw   (11,0)-- (11.5,-2);
\draw   (12.5,0)-- (12.5,-2);
\draw   (13.5,0)-- (13.5,-2);
\draw [ color=qqqqcc] (-4.5,-2)-- (-4.5,-4);
\draw   (-3.5,-2)-- (-3.5,-4);
\draw [ color=ffqqqq] (-0.5,-2)-- (-0.5,-4);
\draw   (0.5,-2)-- (0.5,-4);
\draw [ color=qqqqcc] (1.5,-2)-- (1.5,-4);
\draw   (2.5,-2)-- (2.5,-4);
\draw [ color=ffqqqq] (3.5,-2)-- (3.5,-4);
\draw   (5,0)-- (5.5,-2);
\draw   (7.5,-2)-- (7.5,-4);
\draw [ color=qqqqcc] (8.5,-2)-- (8.5,-4);
\draw [ color=qqqqcc] (10.5,-2)-- (10.5,-4);
\draw   (11.5,-2)-- (11.5,-4);
\draw   (12.5,-2)-- (12.5,-4);
\draw   (13.5,-2)-- (13.5,-4);
\fill [color=black] (5,3.5)  circle (1.5pt);
\draw[color=black] (4.96,3.9) node {$1$};
\fill [color=black] (-2,2)  circle (1.5pt);
\draw[color=black] (-2.02,2.38) node {$2$};
\fill [color=black] (3,2)  circle (1.5pt);
\draw[color=black] (2.96,2.32) node {$3$};
\fill [color=black] (9,2)  circle (1.5pt);
\draw[color=black] (9.2,2.22) node {$4$};
\fill [color=black] (13,2)  circle (1.5pt);
\draw[color=black] (13.16,2.32) node {$5$};
\fill [color=black] (-0.5,0)  circle (1.5pt);
\draw[color=black] (-0.26,0.16) node {$5$};
\fill [color=black] (-2,0)  circle (1.5pt);
\draw[color=black] (-1.74,0.14) node {$4$};
\fill [color=black] (-4,0)  circle (1.5pt);
\draw[color=black] (-4.24,0.22) node {$3$};
\fill [color=black] (1,0)  circle (1.5pt);
\draw[color=black] (0.98,0.38) node {$2$};
\fill [color=black] (3,0)  circle (1.5pt);
\draw[color=black] (3.26,0.18) node {$4$};
\fill [color=black] (5,0)  circle (1.5pt);
\draw[color=black] (5.22,0.12) node {$5$};
\fill [color=black] (7,0)  circle (1.5pt);
\draw[color=black] (6.78,0.16) node {$2$};
\fill [color=black] (9,0)  circle (1.5pt);
\draw[color=black] (9.26,0.08) node {$3$};
\fill [color=black] (11,0)  circle (1.5pt);
\draw[color=black] (11.26,0.12) node {$5$};
\fill [color=black] (12.5,0)  circle (1.5pt);
\draw[color=black] (12.7,0.08) node {$4$};
\fill [color=black] (13.5,0)  circle (1.5pt);
\draw[color=black] (13.74,0.1) node {$2$};
\fill [color=black] (-3.5,-2)  circle (1.5pt);
\draw[color=black] (-3.3,-1.88) node {$5$};
\fill [color=black] (-2.5,-2)  circle (1.5pt);
\draw[color=black] (-2.28,-1.88) node {$3$};
\fill [color=black] (-1.5,-2)  circle (1.5pt);
\draw[color=black] (-1.3,-1.86) node {$5$};
\fill [color=black] (-4.5,-2)  circle (1.5pt);
\draw[color=black] (-4.32,-1.86) node {$4$};
\fill [color=black] (-0.5,-2)  circle (1.5pt);
\draw[color=black] (-0.28,-1.82) node {$4$};
\fill [color=black] (0.5,-2)  circle (1.5pt);
\draw[color=black] (0.74,-1.86) node {$4$};
\fill [color=black] (1.5,-2)  circle (1.5pt);
\draw[color=black] (1.72,-1.86) node {$5$};
\fill [color=black] (2.5,-2)  circle (1.5pt);
\draw[color=black] (2.72,-1.86) node {$2$};
\fill [color=black] (3.5,-2)  circle (1.5pt);
\draw[color=black] (3.84,-1.86) node {$5$};
\fill [color=black] (4.5,-2)  circle (1.5pt);
\draw[color=black] (4.88,-1.86) node {$4$};
\fill [color=black] (6.5,-2)  circle (1.5pt);
\draw[color=black] (6.84,-1.86) node {$3$};
\fill [color=black] (7.5,-2)  circle (1.5pt);
\draw[color=black] (7.86,-1.88) node {$5$};
\fill [color=black] (8.5,-2)  circle (1.5pt);
\draw[color=black] (8.86,-1.9) node {$2$};
\fill [color=black] (9.5,-2)  circle (1.5pt);
\draw[color=black] (9.84,-1.9) node {$5$};
\fill [color=black] (10.5,-2)  circle (1.5pt);
\draw[color=black] (10.8,-1.68) node {$2$};
\fill [color=black] (11.5,-2)  circle (1.5pt);
\draw[color=black] (11.8,-1.68) node {$3$};
\fill [color=black] (12.5,-2)  circle (1.5pt);
\draw[color=black] (12.82,-1.84) node {$3$};
\fill [color=black] (13.5,-2)  circle (1.5pt);
\draw[color=black] (13.9,-1.88) node {$3$};
\fill [color=black] (-4.5,-4)  circle (1.5pt);
\draw[color=black] (-4.12,-3.88) node {$5$};
\fill [color=black] (-3.5,-4)  circle (1.5pt);
\draw[color=black] (-3.16,-3.86) node {$4$};
\fill [color=black] (-0.5,-4)  circle (1.5pt);
\draw[color=black] (-0.16,-3.86) node {$3$};
\fill [color=black] (0.5,-4)  circle (1.5pt);
\draw[color=black] (0.8,-3.88) node {$5$};
\fill [color=black] (1.5,-4)  circle (1.5pt);
\draw[color=black] (1.8,-3.88) node {$4$};
\fill [color=black] (2.5,-4)  circle (1.5pt);
\draw[color=black] (2.8,-3.86) node {$5$};
\fill [color=black] (3.5,-4)  circle (1.5pt);
\draw[color=black] (3.8,-3.88) node {$2$};
\fill [color=black] (5.5,-2)  circle (1.5pt);
\draw[color=black] (5.8,-1.9) node {$2$};
\fill [color=black] (7.5,-4)  circle (1.5pt);
\draw[color=black] (7.8,-3.88) node {$3$};
\fill [color=black] (8.5,-4)  circle (1.5pt);
\draw[color=black] (8.8,-3.92) node {$5$};
\fill [color=black] (10.5,-4)  circle (1.5pt);
\draw[color=black] (10.8,-3.92) node {$3$};
\fill [color=black] (11.5,-4)  circle (1.5pt);
\draw[color=black] (11.92,-3.86) node {$2$};
\fill [color=black] (12.5,-4)  circle (1.5pt);
\draw[color=black] (12.8,-3.86) node {$2$};
\fill [color=black] (13.5,-4)  circle (1.5pt);
\draw[color=black] (13.8,-3.86) node {$4$};
\end{tikzpicture}
\end{scriptsize}
\caption[Representation of $\T_{X,\OO }^1$, for $X=\{id=12345, 52341, 23145\}$, where $\OO_5(X)=1$]{Representation of $\T_{X,\OO }^1$, for $X=\{id=12345, 52341, 23145\}$, where $\OO_5(X)=1$. The subtree induced by the blue edges is $\overline{ \T }_X^1$ and the subtree induced by the blue and red edges is $\T_X^1$. The median value of $X$ is $\mu (X)= 4$ and  $14523$ is the unique median given by the tree  $\T_{ X,\OO }^1$ which is different from the input permutations. In this example, all medians given by the tree  $\T_{X,\OO}^1$ are actually in the subtree $\overline{\T}_X^1$. Also, $13254$ is an example of a permutation in the set $\{y\in S_n; \A_X\subset \A_y\subset \bigcup_{x\in X} \A_x \}$ that is not a median for $X$.\label{treealpha}}
\end{center}
\end{figure}
\end{center}

\begin{algorithm}[ht!]
\KwData{$X=\{x_1, ..., x_k\}\subset S_n $.}
\KwResult{$\alpha$-freedom permutations with respect to $X$}
\For{$i=1,...,n$}{
\For {$w=0,...,n$}{
$L_w\longleftarrow\emptyset$\\
}
$L_0\longleftarrow\{\varnothing\}$\\
$T_{\varnothing}\longleftarrow i$\\
$\tau_{\varnothing}\longleftarrow \alpha$\\
\For{$j=1,...,n-1$}{
\For{$u\in L_{j-1}$}{
$r\longleftarrow 0$\\
\If{$\tau_u>0$}{
\For{$t=1,...,n$}{
\If{$t \notin dig(T_u)$}{
$r \longleftarrow r+1$\\
$L_j \longleftarrow L_j\cup \{u \oplus r\}$\\
$T_{u \oplus r}\longleftarrow T_u\oplus  t$\\
\If{$t\notin \NN_X(\ell(u))$}{
$\tau_{u \oplus r}\longleftarrow \tau_u-1$}
\Else{$\tau_{u \oplus r}\longleftarrow \tau_u$}}
}
}
\Else{
\For{$t\in \NN_X(\ell(u))$}{
\If{$t \notin dig(T_u)$}{
$r \longleftarrow r+1$\\
$L_j \longleftarrow L_j\cup \{u \oplus r\}$\\
$T_{u \oplus r}\longleftarrow T_u\oplus  t$\\
$\tau_{u \oplus r}\longleftarrow \tau_u$}
}
}
}
}
\For{$u\in L_{n-1}$}{
print $T_u$}
}
\caption{$\alpha$-freedom permutations with respect to $X$\label{Alg:median3}}
\end{algorithm}

Figure~\ref{treealpha} shows the tree $\T_{X,\OO}^1$ constructed for $X=\{id=12345,52341,23145\}$, where $\OO_5(X)=1$, with initial budget
$\tau_{\varnothing}=1$. Since $\tau_{\varnothing}>0$, the root branches to all unused symbols $2,3,4,5$. Extensions $\{1,t\}$ that are not adjacencies in any permutation of $X$ consume one unit of budget, while extensions in $\NN_X(1)$ preserve it; in the figure, the child labeled $5$ has $\tau=0$, whereas the other children still have
$\tau=1$.
Along any branch, vertices with positive budget branch freely, while once
$\tau_u=0$ the construction is restricted to neighbors in $\NN_X(\ell(u))$.
The subtree induced by the blue edges is $\overline{\T}_X^1$ from
Algorithm~\ref{Alg:median2}, and the larger subtree induced by the blue and red
edges is $\T_X^1$ from Algorithm~\ref{Alg:median}.
Each root-to-leaf path of length $4$ spells a permutation of $[5]$ together with
the number of outside adjacencies used.
In this example, $14523$ is the unique median appearing among the leaves of $\T_{X,\OO}^1$ which is not one of the input permutations.

The next theorem provides an upper bound for the error of the approximate medians that are outcomes of Algorithm~4.
\begin{theorem}\label{Algo4_error_bound}
Let $X \subseteq S_n$, with $|X|=k$. For any $\alpha \in \mathbb{Z}_+$ and $i \in [n]$, $\mathcal{T}_{X,\alpha}^{\,i}$ is a connected rooted tree, and $X \subseteq Y(\mathcal{T}_{X,\alpha})$. Moreover, for every $y \in Y(\mathcal{T}_{X,\alpha})$, we have
\[
0 \le d_T(y,X) - \mu_n(X) \le \lambda_n(X) + \alpha.
\]
\end{theorem}
\begin{proof}
Similarly to the previous algorithms, we can see, for every $y \in Y(\mathcal{T}_{X,\alpha})$
\[
(k-1)(n-1) - \lambda_n(X) \le \mu_n(X) \le d_T(y,X) \le (k-1)(n-1) + \alpha.
\]

For the last inequality, note that the maximum possible total distance of elements of $Y(\mathcal{T}_{X,\alpha})$ occurs if there exists $\tilde{y} \in Y(\mathcal{T}_{X,\alpha})$ such that
\[
\mathcal{A}_{\tilde{y}} = \mathcal{B}_{\tilde{y}}^{X \cup \{\tilde{y}\}} \cup \left( \bigcup_{x \in X} \mathcal{B}_{\tilde{y}x}^{X \cup \{\tilde{y}\}} \right),
\]
where $\left|\mathcal{B}_{\tilde{y}}^{X \cup \{\tilde{y}\}}\right| = \alpha$. Hence,
\[
\begin{aligned}
d_T(y,X) &\le d_T(\tilde{y},X)
= \sum_{x \in X} \left( n-1 - \left|\mathcal{B}_{\tilde{y}x}^{X \cup \{\tilde{y}\}}\right| \right) \\
&= k(n-1) - (n-1-\alpha)
= (k-1)(n-1) + \alpha.
\end{aligned}
\]
\end{proof}

The following theorem gives the relationships among the outputs of Algorithms~{1-4}. The proof follows directly from the constructions.

\begin{theorem}
Let $X \subseteq S_n$. Then for any $\alpha \in \mathbb{Z}_+$, we have
\[
\tilde{\mathcal{T}}_X \subseteq \overline{\mathcal{T}}_X \subseteq \mathcal{T}_X \subseteq \mathcal{T}_{X,\alpha},
\]
\[
Y(\tilde{\mathcal{T}}_X) \subseteq Y(\overline{\mathcal{T}}_X) \subseteq Y(\mathcal{T}_X) \subseteq Y(\mathcal{T}_{X,\alpha}).
\]
\end{theorem}
\subsection{Circular unichromosomal genomes}\label{sec:circular}

In this subsection we briefly discuss how the tree-based constructions can be
adapted to circular unichromosomal genomes, and how this modification leads to
substantial computational savings.

A circular genome on $n$ symbols can be viewed as a permutation up to cyclic
rotation. A convenient way to generate a uniformly random circular permutation
is to fix the symbol $1$ and place the remaining symbols $2,\ldots,n$ uniformly
at random around it. Equivalently, one may generate a uniformly random
permutation $x\in S_n$ and then identify all cyclic rotations of $x$ as the same
circular genome.

In the linear unichromosomal case, the algorithms construct one tree for each
possible root symbol, resulting in $n$ distinct trees. For circular genomes,
however, all cyclic rotations are equivalent. It therefore suffices to fix a
single reference symbol (e.g.\ $1$) as the root and construct only one tree.
This eliminates the factor-of-$n$ redundancy present in the linear setting and
yields an approximately $n$-fold reduction in running time.

The tree construction proceeds exactly as in the linear case, with one important
modification: the final extension step must close the cycle. Concretely, a leaf
at level $n-1$ is accepted only if its last symbol is adjacent to the root
symbol in $\mathcal{G}(X)$, so that the resulting permutation represents a valid circular genome.
With this additional constraint, each accepted leaf corresponds to a circular
unichromosomal genome satisfying the same adjacency restrictions as in the
linear model.

As a result, the circular version of the algorithm explores a
significantly smaller search space while preserving the structure of medians and geodesic
points. In practice, this allows us to reach substantially larger genome sizes. 

\section{Complexity Analysis}\label{sec:complexity}

We analyze the running time of Algorithms~\ref{Alg:median},\ref{Alg:median2} and \ref{Alg:median3}.
Let
\[
\mathcal{G}(X)=\bigl([n],\ \bigcup_{i=1}^k \A_{x_i}\bigr)
\]
denote the union adjacency graph of the input set
$X=\{x_1,\dots,x_k\}\subset S_n$, as defined in Section~\ref{Section-missing common adjacencies},
and let
$\Delta_X=\Delta=\max_{v\in[n]}|\mathcal{N}_X(v)|\le 2k$. All algorithms explore root-to-leaf paths that correspond to simple paths in subgraphs of \(\mathcal G(X)\). The complexity bounds follow from elementary properties of
such paths in bounded--degree graphs.
We use $O^*(\cdot)$ notation to suppress polynomial factors in $n$ and $k$.

\subsection{Time complexity of Algorithm~\ref{Alg:median}}

Algorithm~\ref{Alg:median} explores all simple paths of length $n-1$ in $\mathcal G(X)$,
starting from every possible root. A vertex at level $t$ in the search tree, i.e. a vertex whose graph distance from the root is $t$, corresponds to a partial path of length $t$,
and its children correspond to all of its unused neighbors in $\mathcal{G}(X)$.

We define the branching factor at level $t\in \{0,\cdots,n-2\}$, denoted by $b(t)$, as the maximum number of children of a vertex at level $t$. Note that, $b(0)\le \Delta_X$, and at level $t\ge 1$, at most $n-t-1$ vertices remain unused, and therefore the branching
factor satisfies
\begin{equation}
b(t)\le \min(\Delta_X-1,\,n-t-1).
\label{eq:branching}
\end{equation}

Recall that $Y(\mathcal{T}_X)$ denotes the total number of permutations (leaves) produced by
Algorithm~\ref{Alg:median}.
Using~\eqref{eq:branching}, we obtain
\begin{equation}\label{eq:L1-product}
|Y(\mathcal{T}_X)|
\;\le\;
n \Delta_X \prod_{t=1}^{n-2}\min(\Delta_X-1,\,n-1-t).
\end{equation}

\begin{lemma}[Leaf bound for Algorithm~\ref{Alg:median}]
\label{lem:alg1-leaf}
For $X\subset S_n$, we have
\[
|Y(\mathcal{T}_X)|
\;\le\;
n\,\Delta_X!\,(\Delta_X-1)^{\,n-1-\Delta_X}.
\]
\end{lemma}

\begin{proof}
We have $\Delta_X-1 \ge n-1-t$ iff $t\ge n-\Delta_X$. This implies
\[
\prod_{t=1}^{n-2}\min(\Delta_X-1,n-1-t)
=
(\Delta_X-1)!\,(\Delta_X-1)^{\,n-1-\Delta_X}.
\]
The proof follows (\ref{eq:L1-product}).
\end{proof}

\begin{theorem}[Time complexity of Algorithm~\ref{Alg:median}]\label{maintheorem_complexity_1}
\label{thm:alg1-time}
Algorithm~\ref{Alg:median} runs in time
\[
T_1(X)
=
O\!\left(n^2\,\Delta_X\,\Delta_X!\,(\Delta_X-1)^{\,n-1-\Delta_X}
\right).
\]
In particular,
\[
T_1(X)=O^*(\Delta_X^n)\le O^*((2k)^n).
\]
\end{theorem}

\begin{proof}
Algorithm~\ref{Alg:median} constructs a search tree whose vertices correspond to partial permutations (simple paths) in the adjacency graph
$\mathcal G(X)$. Each element of $Y(\mathcal T_X)$ corresponds to a leaf at level $n-1$ and encodes a full permutation.

At level \(0\), there are \(n\) possible roots. From a root, the number of
children is at most \(\Delta_X\). At any later level \(t\ge 1\), the
previous vertex of the partial path has already been used, and therefore
the number of admissible unused neighbors is at most
\[
\min(\Delta_X-1,n-t-1).
\]
Thus, the number of vertices at each level is bounded above by the same
branching product used in Lemma~1. In particular, the number of vertices at
any level is at most
\[
n\Delta_X!\,(\Delta_X-1)^{n-1-\Delta_X}.
\]
Since the depth of the search tree is at most \(n-1\), the total number of
visited vertices is bounded by
\[
O\!\left(
n^2\Delta_X!\,(\Delta_X-1)^{n-1-\Delta_X}
\right).
\]

At each visited vertex, Algorithm~1 enumerates the unused neighbors of the
current symbol in \(\mathcal G(X)\), which costs \(O(\Delta_X)\) time.
Therefore,
\[
T_1(X)
=
O\!\left(
n^2\Delta_X\,\Delta_X!\,(\Delta_X-1)^{n-1-\Delta_X}
\right).
\]

Finally,
\[
\Delta_X\,\Delta_X!\,(\Delta_X-1)^{n-1-\Delta_X}
\le
\Delta_X\,\Delta_X^{\Delta_X}\Delta_X^{n-1-\Delta_X}
=
\Delta_X^n.
\]

Hence
$T_1(X)=O(n^2\Delta_X^n)=O^*(\Delta_X^n)$.
Since \(\Delta_X\le 2k\), this implies
$T_1(X)=O^*((2k)^n)$.
\end{proof}

\subsection{Time complexity of Algorithm~\ref{Alg:median2}}

Algorithm~\ref{Alg:median2} refines the basic tree construction by enforcing forced
continuation along common segments of \(\mathcal A_X\). As discussed in
Section~\ref{Alg2}, vertices belonging to the interior of a common segment cannot
serve as branching points: once a path enters such a segment, it must traverse
the entire segment deterministically. Algorithmically, this has the effect of
contracting each maximal common segment into a single block (an effective edge).

In other words, as explained in Section~\ref{Alg2}, Algorithm~\ref{Alg:median2} may be viewed as
Algorithm~\ref{Alg:median} running on a smaller contracted graph
\(\widetilde{\mathcal G}(X)\), constructed in Section~\ref{Alg2}, with the additional rule that reaching an endpoint of an effective contracted edge whose other endpoint has not yet appeared in the partial path forces traversal of that edge before any ordinary branching is considered.
Thus the vertices in
\(Int(\mathcal A_X)\) are suppressed, the vertices in
\(End(\mathcal A_X)\) remain as endpoints of effective edges,
and the vertices in
$[n]\setminus
\bigl(Int(\mathcal A_X)\cup
End(\mathcal A_X)\bigr)$
remain as ordinary vertices.

Let
$n_{\mathrm{eff}}:=n-|Int(\mathcal A_X)|$
denote the number of effective vertices of $\tilde{\mathcal G}(X)$ after contracting the maximal
common segments of \(\mathcal A_X\), and let
\[
\widetilde \Delta_X
:=
\max_{v\notin Int(\mathcal A_X)}
|\mathcal N_{\widetilde{\mathcal G}(X)}(v)|
\]
be the maximum degree of the contracted graph. Applying the leaf bound of
Lemma~\ref{lem:alg1-leaf} to \(\widetilde{\mathcal G}(X)\) would therefore give the same bound
as for Algorithm~\ref{Alg:median}, but with \(n\) replaced by \(n_{\mathrm{eff}}\) and
\(\Delta_X\) replaced by \(\widetilde\Delta_X\). We now show that the latter
replacement is not needed: the bound can be written directly in terms of the
original degree parameter \(\Delta_X\). Recall that \(Y(\overline{\mathcal T}_X)\) denotes the set of permutations
generated by Algorithm~\ref{Alg:median2}.

\begin{lemma}[Leaf bound for Algorithm~\ref{Alg:median2}]\label{lem:alg2-leaf}
For \(X\subset S_n\),
\[
|Y(\overline{\mathcal T}_X)|
\le
n_{\mathrm{eff}}\Delta_X!(\Delta_X-1)^{n_{\mathrm{eff}}-1-\Delta_X}.
\]
\end{lemma}

\begin{proof}
By the contracted-graph interpretation above, Algorithm~\ref{Alg:median2} is Algorithm~\ref{Alg:median}
applied to \(\widetilde{\mathcal G}(X)\), with the additional rule that reaching an endpoint of an effective contracted edge whose other endpoint has not yet appeared in the partial path forces traversal of that edge before any ordinary branching is considered. Hence Lemma~\ref{lem:alg1-leaf} gives
\[
|Y(\overline{\mathcal T}_X)|
\le
n_{\mathrm{eff}}\widetilde\Delta_X!
(\widetilde\Delta_X-1)^{
n_{\mathrm{eff}}-1-\widetilde\Delta_X}.\]
If \(\widetilde\Delta_X=\Delta_X\), this is exactly the desired bound.

It remains to consider the case \(\widetilde\Delta_X<\Delta_X\). Since
contraction preserves the degree of every vertex outside
\(Int(\mathcal A_X)\), this can happen only if the maximum
degree in \(\mathcal G(X)\) is attained at vertices of
\(Int(\mathcal A_X)\). Every such vertex has degree \(2\).
Thus \(\Delta_X=2\) and \(\widetilde\Delta_X<2\). Since
\(\widetilde{\mathcal G}(X)\) is connected, this implies
\(\widetilde\Delta_X=1\), so \(\widetilde{\mathcal G}(X)\) is a single edge.
Equivalently, \(\mathcal G(X)\) is a line graph. This is the trivial case:
either \(|X|=1\), or \(X=\{x,y\}\) with \(A_x=A_y\). In both cases,
Algorithm~\ref{Alg:median2} produces only the two orientations of the same line graph, and
therefore
\[
|Y(\overline{\mathcal T}_X)|=2 < 
n_{\mathrm{eff}}\Delta_X!
(\Delta_X-1)^{n_{\mathrm{eff}}-1-\Delta_X}
=
2\cdot 2!\cdot 1^{-1}
=
4.
\]
Hence, the desired inequality also holds in this case.
\end{proof}

\begin{theorem}[Time complexity of Algorithm~\ref{Alg:median2}]\label{maintheorem_complexity_2}
Algorithm~2 runs in time
\[
T_2(X)=
O\left(
n_{\mathrm{eff}}\, n\Delta_X\Delta_X!(\Delta_X-1)^{n_{\mathrm{eff}}-1-\Delta_X}
\right).
\]
In particular,
\[
T_2(X)=O^*(\Delta_X^{n_{\mathrm{eff}}})
\le
O^*((2k)^{n_{\mathrm{eff}}}).
\]
\end{theorem}
\begin{proof}
By the contracted-graph interpretation, Algorithm~\ref{Alg:median2} operates in the same way as Algorithm~\ref{Alg:median} on the contracted graph
\(\widetilde{\mathcal G}(X)\), together with the rule that if the current path reaches an endpoint of an effective contracted edge and the other endpoint has not yet been used, then the path must next traverse that effective edge before any other branching choice is allowed. As in the proof of Theorem~\ref{maintheorem_complexity_1}, the number of
vertices at each level of the search tree is bounded above by the
corresponding leaf bound. Hence, by Lemma~\ref{lem:alg2-leaf}, each level of
\(\overline{\mathcal T}_X\) contains at most
\[
n_{\mathrm{eff}}\Delta_X!(\Delta_X-1)^{n_{\mathrm{eff}}-1-\Delta_X}
\]
vertices. Since the effective depth is at most \(n_{\mathrm{eff}}-1\), the
total number of visited effective vertices, in the contracted search tree, is bounded by
\[
O\left(
n_{\mathrm{eff}}^2
\Delta_X!(\Delta_X-1)^{n_{\mathrm{eff}}-1-\Delta_X}
\right),
\]
and thus the number of vertices in the expanded search tree across all $n-1$ levels is bounded by 
\[
O\left(
n_{\mathrm{eff}}\, n
\Delta_X!(\Delta_X-1)^{n_{\mathrm{eff}}-1-\Delta_X}
\right).
\]
The remainder of the proof follows the same argument as in Theorem~\ref{maintheorem_complexity_1}.
\end{proof}
\begin{remark}[Effect of common segments]
Although the worst-case bound for Algorithm~\ref{Alg:median2} matches that of Algorithm~\ref{Alg:median},
the forced-continuation rule can substantially shrink the search tree in
practice: long segments in \(\mathcal A_X\) eliminate branching over many
consecutive levels and reduce the number of admissible roots, since
\(Int(\mathcal A_X)\) cannot be chosen as a root label.
\end{remark}

\subsection{Time complexity of Algorithm~\ref{Alg:median3}}\label{subsec:complexity-alg3}

Algorithm~\ref{Alg:median3} extends Algorithm~\ref{Alg:median} by allowing
budgeted use of adjacencies outside the union
$\bigcup_{x\in X}\mathcal A_x$.
Each vertex $u$ in the search tree carries a remaining budget
$\tau(u)\in\{0,1,\ldots,\alpha\}$, where $\tau(\varnothing)=\alpha$ at the root.
At a vertex $u$ at level $t$ (i.e.\ after $t$ extension steps), there are exactly
$n-t-1$ unused symbols available for the next label.

If $\tau(u)=0$, Algorithm~\ref{Alg:median3} branches exactly as in
Algorithm~\ref{Alg:median}, so the number of children is at most
$\min(\Delta_X-1,\ n-t-1)$, for $t\ge 1$.
If $\tau(u)>0$, the vertex branches to \emph{all} unused symbols, so it has exactly
$n-t-1$ children.  Along any root-to-leaf path, the budget can be decreased at
most $\alpha$ times; equivalently, there can be at most $\alpha$ levels at which
the algorithm takes a step $t\notin \NN_X(\ell(u))$ and hence uses one unit of
outside-adjacency budget.

\begin{lemma}[Leaf bound for Algorithm~\ref{Alg:median3}]\label{lem:alg3-leaf}
For \(\alpha\ge 0\), we have
\[
|Y(\mathcal T_{X,\alpha})|
\le
n
\sum_{r=0}^{\alpha}
\binom{n-1}{r}
\frac{(n-1)!}{(n-1-r)!}
\Delta_X^{\,n-1-r}.
\]
Hence, for fixed \(\alpha\),
\[
|Y(\mathcal T_{X,\alpha})|
=
O\left(n^{2\alpha+1}\Delta_X^{\,n-1-\alpha}\right).
\]
\end{lemma}

\begin{proof}
Fix a root \(i\in[n]\). A root-to-leaf path has \(n-1\) extension steps,
indexed by \(t=0,\ldots,n-2\). For a given path, let
\(B\subseteq\{0,\ldots,n-2\}\) be the set of levels at which the chosen next
symbol is not in \(\mathcal N_X(\ell(u))\), so that one unit of
outside-adjacency budget is used. Then \(|B|\le \alpha\).

Now fix such a set \(B\), with \(|B|=r\). At a level \(t\in B\), the algorithm
may choose any unused symbol, so the number of choices is at most
\(n-t-1\). At a level \(t\notin B\), the chosen symbol must belong to
\(\mathcal N_X(\ell(u))\), and therefore the number of choices is at most
\(\Delta_X\), independently of the previous choices made along the path.
Thus, for this fixed set \(B\), the number of possible root-to-leaf paths using the outside adjacencies in steps $t\in B$ is
at most
\[
\left(
\prod_{t\in B}(n-t-1)
\right)
\Delta_X^{\,n-1-r}.
\]
Summing over all subsets \(B\) with \(|B|\le\alpha\), and then multiplying by
the \(n\) possible roots, gives
\[
|Y(\mathcal T_{X,\alpha})|
\le
n
\sum_{r=0}^{\alpha}
\sum_{\substack{B\subseteq\{0,\ldots,n-2\}\\ |B|=r}}
\left(
\prod_{t\in B}(n-t-1)
\right)
\Delta_X^{\,n-1-r}.
\]

For each fixed \(r\), the product over \(B\) is bounded by choosing the
\(r\) largest values among \(1,\ldots,n-1\). Hence
\[
\prod_{t\in B}(n-t-1)
\le
\frac{(n-1)!}{(n-1-r)!}.
\]
There are \(\binom{n-1}{r}\) choices of \(B\), which gives
\[
|Y(\mathcal T_{X,\alpha})|
\le
n
\sum_{r=0}^{\alpha}
\binom{n-1}{r}
\frac{(n-1)!}{(n-1-r)!}
\Delta_X^{\,n-1-r}.
\]

Finally, for fixed \(\alpha\),
\[
\binom{n-1}{r}\le n^r,
\qquad
\frac{(n-1)!}{(n-1-r)!}\le n^r,
\]
and \(r\le\alpha\). Since \(\Delta_X\le n\), each term is bounded by
\(O(n^{2\alpha}\Delta_X^{\,n-1-\alpha})\). Multiplying by the \(n\) root
choices and summing over the fixed number \(\alpha+1\) of possible values of
\(r\) gives
\[
|Y(\mathcal T_{X,\alpha})|
=
O\left(n^{2\alpha+1}\Delta_X^{\,n-1-\alpha}\right).
\]
\end{proof}
\begin{theorem}[Time complexity of Algorithm~\ref{Alg:median3}]
For fixed \(\alpha\ge 0\), Algorithm~\ref{Alg:median3} runs in time
\[
T_4(X)
=
O\left(
n^{2\alpha+2}\Delta_X^{\,n-\alpha}
\right).
\]
In particular,
\[
T_4(X)=O^*(\Delta_X^{\,n-\alpha}).
\]
\end{theorem}

\begin{proof}
By Lemma~\ref{lem:alg3-leaf}, the number of leaves generated by Algorithm~\ref{Alg:median3} is
\[
O\left(
n^{2\alpha+1}\Delta_X^{\,n-1-\alpha}
\right)
\]
for fixed \(\alpha\). The same level-by-level counting argument used in the
proof of Theorem~\ref{maintheorem_complexity_1} shows that each level of
\(\mathcal T_{X,\alpha}\) satisfies the same upper bound. Since the depth is
at most \(n-1\), the total number of visited vertices is therefore
\[
O\left(
n^{2\alpha+2}\Delta_X^{\,n-1-\alpha}
\right).
\]
Multiplying by the local enumeration cost \(O(\Delta_X)\), as in the proof of
Theorem~\ref{maintheorem_complexity_1}, gives
\[
T_4(X)
=
O\left(
n^{2\alpha+2}\Delta_X^{\,n-\alpha}
\right).
\]
The \(O^*\)-bound follows by suppressing polynomial factors in \(n\).
\end{proof}

\section{Experimental results}\label{sec:exp}

The goal of this section is to empirically study the structure and size of the
search spaces generated by our tree-based algorithms, as well as the geometry of medians and near-medians under the breakpoint distance.
We analyze the structure of the algorithmically generated search spaces and their
relationship to optimal and near-optimal solutions, together with practical
strategies for exploring large and structured solution spaces.

The experimental analysis is structured as follows.
First, we study the distribution of total breakpoint distances over the reduced
search space produced by Algorithm~\ref{Alg:median3}, quantifying how tightly
candidate permutations cluster around the median value. Next, we compare the size of the reduced search space generated by Algorithm~\ref{Alg:median3}, that is, the set of candidate permutations \(Y(\mathcal T_{X,\OO})\), with the size \(n!\) of the full permutation space \(S_n\). We then turn to sampling-based strategies. In Subsection~\ref{Section:sampling1}, we introduce a depth-first leaf-traversal procedure for the search trees generated by Algorithm~\ref{Alg:median3}. This procedure visits leaves one by one, recording a candidate permutation whenever a branch reaches level \(n-1\), and therefore provides a natural way to collect early samples without completing the full traversal of the tree. We use this approach to show that near-medians can be obtained quickly without enumerating the full reduced search space. Finally, we use the same depth-first traversal procedure to construct medians and near-medians satisfying additional geometric constraints. In particular, we target candidates that are far from the input genomes and are approximately equidistant from them; for three random input genomes, such balanced candidates have distances close to \(2n/3\) from each input.

Unless otherwise stated, all experiments in the following subsections consider
sets $X=\{x_1,x_2,x_3\}\subset S_n$ of three permutations, where $x_1=id$
and $x_2,x_3$ are sampled uniformly at random subject to the constraint
$\OO_n(X)\leq 3$.
Each reported statistic is averaged over 100 independent runs for each value of
$n$.

\subsection{Near-optimal medians and total distance distribution}\label{subsec:Ki}

We begin by analyzing how the candidate medians generated by
Algorithm~\ref{Alg:median3} are distributed according to their total
breakpoint distance from the input set.

\begin{figure}[ht!]
\begin{center}
\includegraphics[scale=0.55]{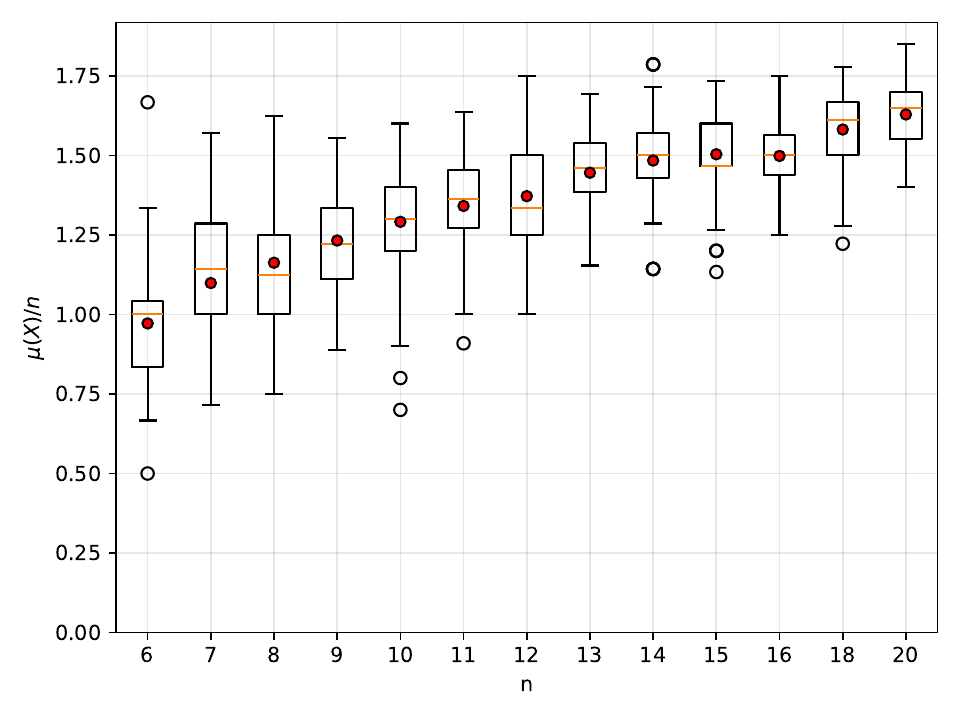}
\end{center}
\caption{Distribution of the normalized median value $\mu(X)/n$ over 100 independent runs for each genome size $n$. Boxplots show variability, and the red dot denotes the sample mean.}\label{plot1}
\end{figure}

For a given run, let $\mu(X)=\min_{y\in S_n}\sum_{x\in X} d(y,x)$ denote
the median value.
Figure~\ref{plot1} shows boxplots of the normalized median value $\mu(X)/n$ over 100 independent runs for each genome size $n$, with the red dot indicating the sample mean.
The distribution shifts upward with $n$, and both its central tendency and mean increase.
While variability remains for moderate $n$, the relative dispersion decreases, indicating increasing concentration. Overall, the mean normalized median value increases with $n$ and appears to approach $2$ as $n\to\infty$, in accordance with the last theorem of~\cite{jam14}.

Although not all permutations in $Y(\T_{X,\OO})$ are exact medians, many
of them have total distance very close to the median value, indicating that they serve as good approximations. To formalize this, we define for each integer $j\ge 0$
\[
K_j(\T_{X,\OO})
:=\left\{
y\in Y(\T_{X,\OO}) :
\sum_{x\in X} d(y,x)-\mu(X)=j
\right\}.
\]
Thus $K_0(\T_{X,\OO})=M(X)$ is the set of exact medians, while
$K_j(\T_{X,\OO})$ for small $j>0$ consists of approximate medians
whose total distance exceeds the optimum by only $j$ breakpoint units.

\begin{figure}[ht!]
\begin{center}
\includegraphics[width=1\linewidth]{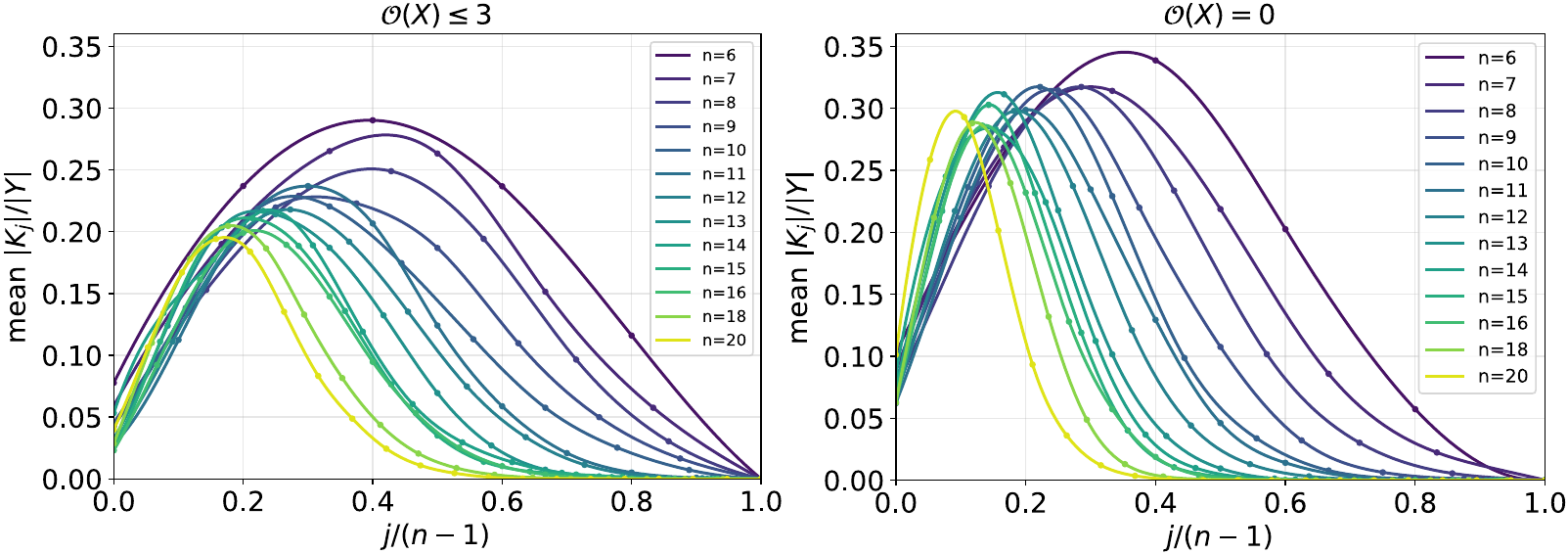}
\caption{Mean proportions
$|K_j(\mathcal T_{X,\OO})|/|Y(\mathcal T_{X,\OO})|$ as a function of
the normalized excess distance $j/(n-1)$ for two families of input
instances. Left: instances satisfying $\OO_n(X)\leq 3$. Right: instances
satisfying $\OO_n(X)=0$. The curves show similar near-optimal
concentration, with the stricter $\OO_n(X)=0$ case producing sharper
peaks.}\label{kfig}
\end{center}
\end{figure}

To study the concentration of near-optimal solutions, we compute for
each $j$ the mean proportion
$|K_j(\T_{X,\OO})|/|Y(\T_{X,\OO})|$
over all runs.
Figure~\ref{kfig}~(left) shows that a significant portion of permutations in $Y(\T_{X,\OO})$ have total distances concentrated near the minimum, indicating that while most are not exact medians, many are close approximations.

In fact, across all tested values of $n$, the union $K_0 \cup K_1 \cup K_2$ consistently contains over 27\% of $Y(\T_{X,\OO})$, confirming the abundance of near-optimal solutions in the reduced space. For example, at $n=8$, this set accounts for 42\% of candidates; for $n=14$, it still covers 37\% despite the increase in size. Table~\ref{tab:approx_medians} summarizes these proportions numerically for selected values of $n$.

\begin{table}[ht!]
\centering
\caption{Proportion of genomes in $K_0 \cup K_1 \cup K_2$ over $Y(\T_{X,\OO})$, averaged over 100~runs.}
\begin{tabular}{ccccc}
\toprule
$n$ & $|K_0|$ & $|K_1|$ & $|K_2|$ & Total proportion (\%) \\
\midrule
6  & 7.77\% & 23.94\% & 29.31\% & 61.02\% \\
8  & 4.25\% & 15.29\% & 23.14\% & 42.68\% \\
10 & 4.37\% & 13.83\% & 21.85\% & 40.05\% \\
12 & 3.03\% & 11.11\% & 19.24\% & 33.38\% \\
14 & 5.33\% & 13.42\% & 18.47\% & 37.22\% \\
15 & 2.62\% & 11.09\% & 19.08\% & 32.79\%\\
16 & 2.31\% &9.22\% &16.09\% &27.62\% \\
18 & 2.86\% &11.01\% &17.80\% &31.67\% \\
20 & 3.99\% &10.64\% &16.87\% & 31.50\% \\
\bottomrule
\end{tabular}
\label{tab:approx_medians}
\end{table}

In addition, to illustrate how the near-optimal
distribution changes for a more restrictive input class, we 
computed the same quantities for instances satisfying $\OO_n(X)=0$.
These two experiments should be viewed as comparisons between two
families of input instances, rather than as a paired comparison on the same sets
$X$.
Figure~\ref{kfig} shows that the qualitative behavior is similar in both cases: a substantial
fraction of the reduced search space lies close to the median value.
However, the curves for $\OO_n(X)=0$ are more sharply peaked and more
concentrated near small excess distance, while the curves for
$\OO_n(X)\le 3$ are broader. This is consistent with the fact that
allowing up to three adjacencies outside the input adjacency set produces
a larger and more flexible reduced search space.

Beyond proportions of near-optimal solutions, we also examine the overall
distribution of total breakpoint distances attained within the reduced search
space $Y(\T_{X,\OO})$.
For each run, we consider the distribution of total distances
$\sum_{x\in X} d(y,x)$ over all $y\in Y(\T_{X,\OO})$, and record its minimum,
maximum, mean, and variance.
To facilitate comparison across genome sizes, all distances are normalized by
$n$.

\begin{table}[ht!]
\centering
\caption{Summary statistics of normalized total distance, averaged over 100 runs.}
\label{tab:dist_stats}
\begin{tabular}{cccccc}
\toprule
$n$ &
$\min \delta_X/n$ &
$\max \delta_X/n$ &
$\e(\delta_X/n)$ &
$\mathrm{Var}(\delta_X/n)$ &
$\max j$ \\
\midrule
6  & 0.97 & 1.66 & 1.34 & 0.0314 & 4.13 \\
8  & 1.16 & 1.79 & 1.52 & 0.0197 & 5.03 \\
10 & 1.29 & 1.83 & 1.60 & 0.0128 & 5.47 \\
12 & 1.37 & 1.87 & 1.66 & 0.0095 & 6.03 \\
14 & 1.48 & 1.89 & 1.71 & 0.0068 & 5.71 \\
15 & 1.50 & 1.91 & 1.74 & 0.0057 & 6.04 \\
16 & 1.49 & 1.91 & 1.73 & 0.0057 & 6.62 \\
18 & 1.58 & 1.93 & 1.78 & 0.0043 & 6.21 \\
20 & 1.63 & 1.94 & 1.81 & 0.0032 & 6.17 \\
\bottomrule
\end{tabular}
\end{table}
More precisely, given $X$, let $\mathscr{Y}_X$ be a permutation chosen uniformly at random from $Y(\mathcal{T}_{X,\OO})$. For a given $X$, define as the total distance of $\mathscr{Y}_X$ to $X$
\[
\delta_X=d_T(\mathscr{Y}_X,X).\]
Then for a given $X$, the expected value of $\delta_X$, namely the average of the total distance of the leaves for one run of $X$, is given by
\[
\mathbb E(\delta_X)=\frac{1}{|Y(\mathcal{T}_{X,\OO})|}\sum_{y\in Y(\mathcal{T}_{X,\OO})} d_T(y, X).
\]
The variance $\mathrm{Var}(\delta_X)=\mathbb E(\delta_X^2)-\mathbb E(\delta_X)^2$ can be computed similarly. We also denote,
\[
\min \delta_X:=\min_{y\in Y(\mathcal{T}_{X,\OO})} d_T(y,X) \ \ \ \text{and} \ \ \ \max \delta_X:=\max_{y\in Y(\mathcal{T}_{X,\OO})} d_T(y,X).
\]

Table~\ref{tab:dist_stats} reports these statistics averaged over $100$
independent runs for each value of $n$.
The minimum normalized distance, which coincides with the normalized median
value $\mu(X)/n$, increases steadily with $n$.
The mean normalized distance over $Y(\T_{X,\OO})$ also increases with $n$ and
appears to approach a limiting value close to $2$, consistent with the behavior
observed for exact medians. Moreover, the mean of the maximum normalized distance grows only mildly and remains relatively close to the mean, suggesting that even the largest distances observed in the reduced search space are not far from the bulk of the distribution.
The variance of the normalized total distance decreases as $n$ grows, suggesting an increasing concentration of total distances within the reduced search space. Although $Y(\T_{X,\OO})$ contains many candidates, their normalized total distances appear to cluster in a relatively narrow band above the optimal value. At the same time, the maximum excess distance $\max\{j : K_j(\T_{X,\OO})\neq\emptyset\}$ remains moderate and grows only slowly over the tested values of $n$, indicating that the observed support of the distance distribution expands mildly with genome size.

These observations suggest that the candidate medians generated by Algorithm~\ref{Alg:median3} form a structured and relatively concentrated search space, as Theorem~\ref{Algo4_error_bound} clearly indicates. While exact medians constitute only a fraction of $Y(\T_{X,\OO})$, many candidate genomes have total distances close to the optimum, and the maximum observed excess distance $\max\{j:K_j(\T_{X,\OO})\neq\emptyset\}$ remains moderate over the tested range of $n$, increasing only mildly despite the growth in genome size. This provides quantitative support for Theorem~\ref{Algo4_error_bound} and for the effectiveness of sampling-based strategies, which we investigate in the following subsections.

\subsection{Search-space reduction}
To quantify the algorithm’s efficiency, we compare the size of the reduced search
space generated by Algorithm~\ref{Alg:median3} to the full permutation group $S_n$. Table~\ref{tab:space_reduction} demonstrates that the number of permutations explored by Algorithm~\ref{Alg:median3} represents only a tiny fraction of $n!$, yet suffices to find all exact and many near-optimal medians. For instance, when $n = 18$, the reduced search space contains less than $0.000008\%$ of the full $18! \approx 6.4 \times 10^{15}$ permutations.

\begin{table}[ht!]
\centering
\caption{Reduction in search space by Algorithm~\ref{Alg:median3}. Candidate and median counts are averages over 100 runs.}
\begin{tabular}{@{\!}c@{\!~~~}c@{\!~~~}c@{\!~~~}c@{\!~~~}c@{\!}}
\toprule
$n$ & Total candidates & Total medians & $n!$ & $\frac{\text{candidates}}{n!}$ (\%) \\
\midrule
6  & 220.14 & 20.18 & 720 & 30.58\% \\
8  & 2720.92 & 74.08 & 40320 & 6.75\% \\
10 & 29238.34 & 455.3 & 3628800 & 0.81\% \\
12 & 352857.2 & 2249.5 & $4.79 \times 10^8$ & 0.07\% \\
14 & 2870803.16 & 30328.38 & $8.72 \times 10^{10}$ & 0.003\% \\
15 & 13867110.74 & 30730.84 & $1.31 \times 10^{12}$ &0.0011\%\\
16 & 39189139.2 & 50400.46 & $2.09 \times 10^{13}$ &0.00019\%\\
18 & 492933428.9 & 438337.68 & $6.4 \times 10^{15}$ &0.000008\%\\
20 & 4527522588 & 4415864.14 & $2.43\times 10^{18}$ &$1.86\times 10^{-7}$\%\\
\bottomrule
\end{tabular}
\label{tab:space_reduction}
\end{table}

The growth of the reduced search space is still exponential in \(n\), although with a much smaller scale than the full permutation space. A least-squares fit of the logarithm of the mean number \(C_n\) of candidates in Table~\ref{tab:space_reduction} gives
\[
\log C_n \approx -1.77+1.205n,
\]
corresponding to an empirical multiplicative growth factor \(e^{1.205}\approx 3.34\) for each increase of \(n\) by one. This is consistent with the upper bound from Lemma~\ref{lem:alg3-leaf}, which gives a candidate-space size of order $O\!\left(n^{2\alpha+1}\Delta_X^{\,n-1-\alpha}\right).$
For fixed \(\alpha\), this bound is exponential in \(n\) up to polynomial factors. Thus, the experiments show an important effect: Algorithm~\ref{Alg:median3} reduces the search space by many orders of magnitude relative to \(S_n\), but the resulting reduced tree still grows exponentially, explaining the rapid increase in runtime for larger \(n\).

For all tested instances with $n\le 12$, the mean runtime per run is below one
second, with very small variability across runs.
The runtime increases gradually with $n$, reflecting the growth of the reduced
search space.
For larger instances, this growth becomes more pronounced; for example, at
$n=16$ the mean runtime is approximately $5.6$ minutes.
Memory usage remains similarly well controlled across all tested values of $n$:
the mean resident set size stays close to $35$--$40$~MB, with negligible
variability across runs. 
The reported memory usage corresponds to a streaming enumeration of candidate
permutations, in which the algorithm counts candidates and accumulates
statistics on the fly without storing the full set
$Y(\mathcal{T}_{X,\OO})$ in memory.
As a result, memory consumption remains modest (on the order of tens of
megabytes) even when the total number of candidates is very large (the mean runtime and memory usage were measured on a 2.3 GHz quad-core Intel Core i7 machine with 32 GB RAM).

Although it is known that, given a set of genomes $X$, there may exist medians that do not contain all adjacencies in $\mathcal{A}_X$, we verified that for the input sets tested, all medians returned by Algorithm~\ref{Alg:median3} contained the full set of common adjacencies $\mathcal{A}_X$ shared by the input genomes. 
While Algorithm~\ref{Alg:median3} was run with $\OO_n(X) \leq 3$, allowing up to three adjacencies outside the union of the input adjacencies, we observed that such instances were extremely rare—and when they occurred, each involved only a single external adjacency. For example, at $n = 6$, only 0.04 medians per run (roughly 0.35\% of all medians) included one adjacency not present in the union of the inputs. At $n = 12$, the mean was 0.48 per run (under 0.015\%). For all other values of $n \leq 15$, no external adjacencies were observed. These results indicate that nearly all medians are already covered in the zero-freedom  case ($\alpha=0$), that is, when every adjacency is drawn from the input genomes. In practice, therefore, we can use Algorithm~\ref{Alg:median}, which corresponds to the zero-freedom version of Algorithm~\ref{Alg:median3}, to recover most of the medians while achieving substantial speed-ups.

Finally, we emphasize again that the strength of this approach is not limited to the
exact medians found inside the search space. Although only a fraction of the
permutations generated by Algorithm~\ref{Alg:median3} may be exact medians,
Theorem~\ref{Algo4_error_bound} shows that the total distance of every genome
generated by Algorithm~\ref{Alg:median3} is within a controlled additive error
of the median value. Thus, when the allowed number of outside adjacencies is
small, as is the case, with high probability, for \(k\) random genomes, the entire space
\(Y(\mathcal T_{X,\alpha})\) consists of near-median candidates. In particular,
as \(n\) grows, this error becomes small on the normalized scale, so the
generated genomes provide good approximations to breakpoint medians.

\subsection{Sampling of medians and near-medians from the leaves}\label{Section:sampling1}

Full enumeration of the reduced search space $Y(\mathcal T_{X,\OO})$
becomes increasingly demanding as $n$ grows, even though the space is
dramatically smaller than $S_n$.
The strong concentration phenomena observed in the previous subsections
suggest that it may not be necessary to enumerate the entire space in
order to obtain high-quality solutions.
We therefore investigate sampling-based strategies that exploit the
structure of the tree to rapidly produce medians and near-medians.

The tree $\mathcal T_{X,\OO}$ is naturally organized by levels. For each
choice of the first element, the root corresponds to a one-letter partial
candidate, and vertices at level $j$ correspond to partial candidates of
length $j+1$. Thus leaves at level $n-1$ represent complete candidate
permutations. Algorithm~\ref{Alg:median3} is written in a level-by-level
form: at step $j$, it constructs the set $L_j$ of all partial candidates
of length $j+1$ from the previous level $L_{j-1}$. This presentation is
convenient for describing the full search space and for proving that the
leaves contain all $\alpha$-freedom permutations.
For the sampling experiments, however, we implement the same search tree
using a depth-first traversal. At each partial candidate $u$, the
admissible one-step extensions of $u$ are placed in a fixed order,
according to the order in which they are generated by the algorithm. The
depth-first traversal first visits the first admissible extension of
$u$, then the first admissible extension of that new vertex, and so on,
continuing down one branch as far as possible. If the branch reaches
level $n-1$, it represents a complete candidate permutation, and this
candidate is recorded as a sampled leaf. If the branch cannot be extended
before reaching level $n-1$, it is discarded.
After a branch has either produced a complete candidate or reached a
dead end, the algorithm backtracks to the most recent partial candidate
that still has an unvisited admissible extension. It then visits the next
extension of that vertex in the fixed order, and again follows that new
branch as far as possible. Thus, if the ordered children of a vertex are
$v_1,\ldots,v_m$, the traversal explores the entire subtree rooted at
$v_1$ before moving to $v_2$, then explores the entire subtree rooted at
$v_2$ before moving to $v_3$, and so on. The sampled candidates are
therefore the first complete leaves encountered in this depth-first
order.
This changes only the order in which complete candidates are encountered,
not the set of complete candidates that can be generated. The advantage
is that the algorithm does not need to store all vertices at a given
level. It keeps only the current branch and the information needed to
continue the search. This makes depth-first traversal especially
suitable for early sampling: the first complete leaves can be obtained
without constructing the entire tree, or even an entire level of the
tree.

Formally, for each $i\in[n]$, we define a depth-first search (DFS) process
$\gamma_{i,0},\gamma_{i,1},\ldots,$
where $\gamma_{i,0}=\varnothing$ is the root vertex and
$\ell(\gamma_{i,0})=i$. Whenever the process is at an
already-created vertex $u$, it either creates a new vertex $v$ or moves
backward toward the root $\varnothing$. In the former case, we call $v$ a
child of $u$ and $u$ the parent of $v$. Let $\mathcal{C}_t(u)$ denote the set
of children of $u$ created up to and including time $t$, and let
$\iota(v)=u$ denote the parent of $v\neq\varnothing$. We also set
$\iota(\varnothing)=\varnothing$.

As before, for each vertex $u$, let $\ell(u)\in[n]$ denote its label, and let
$T_u$ be the sequence of labels, in order, along the path in
$\mathcal T_{X,\OO}^i$ connecting $\varnothing$ to $u$. Then
$\operatorname{dig}(T_u)$ denotes the set of labels of the vertices in $T_u$.
Define
\[
\zeta_t(u):=
\begin{cases}
\mathcal{N}_X\bigl(\ell(u)\bigr)\setminus
\left(
\operatorname{dig}(T_u)\cup
\left\{\ell(v):v\in\mathcal{C}_t(u)\right\}
\right),
& \text{if }\tau(u)=0,\\[6pt]
[n]\setminus
\left(
\operatorname{dig}(T_u)\cup
\left\{\ell(v):v\in\mathcal{C}_t(u)\right\}
\right),
& \text{if }\tau(u)>0,
\end{cases}
\]
where the budget function $\tau$ is defined as in (\ref{budget-def}). Thus, $\zeta_t(u)$ is the set of admissible labels that have not yet been
used by time $t$. The algorithm proceeds as follows.
Start from
$\gamma_{i,0}=\varnothing$.
At each time $t$, suppose that $\gamma_{i,t}=u$. If
$\zeta_t(u)\neq\varnothing$, create a new vertex
$v:=u\oplus\left|\mathcal{C}_t(u)\right|$,
and set
$\ell(v):=\min\zeta_t(u)$, $\gamma_{i,t+1}:=v$.
If $\zeta_t(u)=\varnothing$, move to the parent of $u$ by setting
$\gamma_{i,t+1}:=\iota(u)$.
Print $T_{\gamma_{i,t+1}}$ whenever $\gamma_{i,t+1}$ is at level $n-1$.
Stop the process at the first time $t$ such that
$\gamma_{i,t+1}=\gamma_{i,t}=\varnothing$.
By construction, the process starts and ends at $\varnothing$, and each edge
is traversed exactly twice.

As a preliminary illustration of the cost of early sampling,
Table~\ref{tab:runtime_sampling} shows the time required in one
representative run to collect the first $1000$ leaves for selected
values of $n$.
These examples indicate that the early portion of the search tree can
often be explored very quickly, even for large genome sizes.
The runtime may vary across instances depending on $\alpha$ and the
resulting branching structure of $\mathcal T_{X,\OO}$.

\begin{table}[ht!]
\centering
\caption{Illustrative single-run time required to collect the first $1000$
leaves of $\mathcal T_{X,\OO}$.}
\label{tab:runtime_sampling}
\begin{tabular}{c@{~~~~~~~}c@{~~~~~~~}c@{~~~~~~~}c@{~~~~~~~}c}
\toprule
$n$ & 20 & 50 & 80 & 100 \\
\midrule
Runtime (s) & 0.007 & 0.06 & 0.06 & 0.28 \\
\bottomrule
\end{tabular}
\end{table}

Motivated by these observations, we consider the DFS sampling strategy explained above
based on stopping the traversal after a prescribed number of leaves has
been reached.
In principle, one could consider several regimes for the sample size,
including polynomial or exponential functions of the genome size $n$.
Here we choose a more economical linear regime, since our goal is to
repeat the experiment over many independent instances, including larger
values of $n$.
More precisely, during the depth-first traversal of
$\mathcal T_{X,\OO}$, we record the first $l_n=50n$ leaves at level $n-1$, corresponding to $50n$ candidate permutations
generated near the beginning of the search.
This procedure can be viewed as an early stopping rule for the
enumeration process.
The choice $l_n=50n$ keeps the sample size small compared with the
full reduced search space, while allowing the number of sampled
candidates to grow with the genome size.
At the same time, the linear sampling rule keeps the computational cost
low enough to perform $100$ independent runs for genome sizes
$n=10,20,\ldots,100$. For random input genomes, the error bound in Theorem~\ref{Algo4_error_bound} is small with high probability. Hence, with high probability, every leaf in the search space \(Y(\mathcal T_{X,\alpha})\), including the leaves obtained by early sampling, has total distance within a small additive error of the median value. This provides a theoretical explanation for why early samples already give good near-median candidates.

We begin by comparing the statistical properties of this sampled set
with those of the full search space for $n\leq 20$.
The goal of this comparison is to assess whether early sampled
candidates already provide a reliable approximation of the geometry of
$Y(\mathcal T_{X,\OO})$.
For $i\le |Y(\mathcal{T}_{X,\OO})|$, let $z_i\in Y(\mathcal T_{X,\OO})$ be the candidate genome corresponding
to the $i$-th leaf encountered during the depth-first traversal of
$\mathcal T_{X,\OO}$, and let $\hat{\mathscr{Z}}_X$ be a genome chosen uniformly at random from
$z_1,\ldots,z_{l_n}$. Define $\hat{\delta}_X=d_T(\hat{\mathscr{Z}}_X,X)$.
For each run, we compare the normalized mean total distance over the
sampled set,
\[
\frac{\mathbb{E}(\hat{\delta}_X)}{n}
=
\frac{1}{n\ell_n}\sum_{i=1}^{\ell_n} d_T(z_i,X),
\]
with the corresponding normalized mean total distance over the full
search space. Namely, if $\mathscr{Z}_X$ is chosen uniformly at random
from $Y(\mathcal T_{X,\OO})$ and
$\delta_X=d_T(\mathscr{Z}_X,X)$, then
\[
\frac{\mathbb{E}(\delta_X)}{n}
=
\frac{1}{n|Y(\mathcal T_{X,\OO})|}
\sum_{z\in Y(\mathcal T_{X,\OO})} d_T(z,X).
\]

\begin{figure}
\centering
\includegraphics[width=0.58\linewidth]{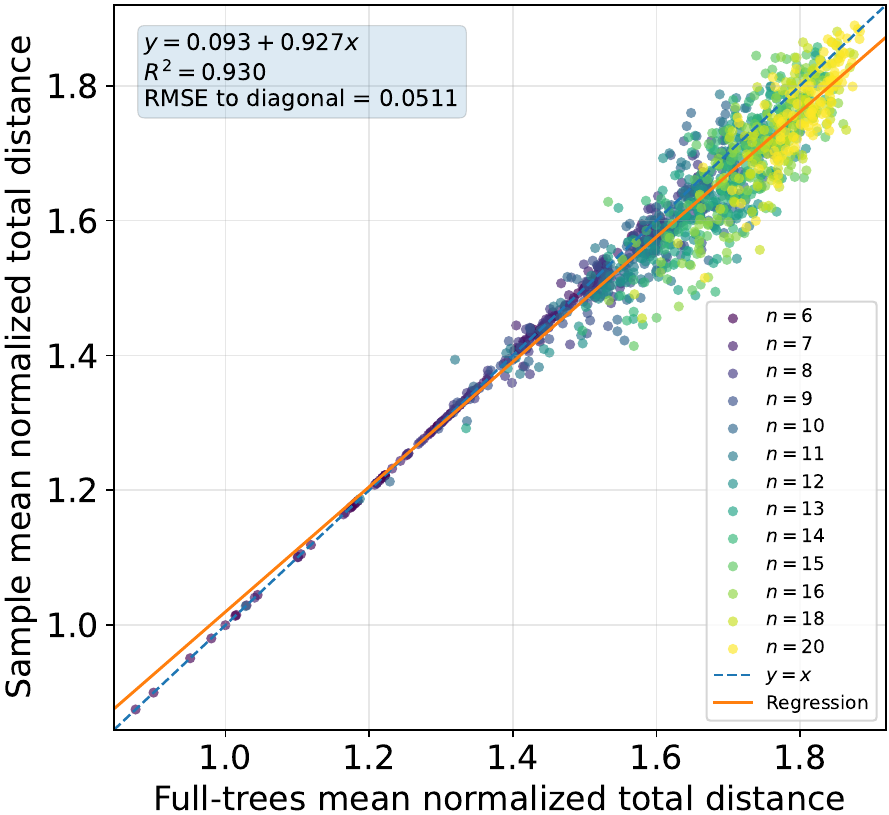}
\caption{Scatter plot comparing the normalized mean total distance
over the linear prefix sample, with $l_n=50n$, against the
corresponding mean over the full candidate set
$Y(\mathcal T_{X,\mathcal O})$. Each point corresponds to one random
instance, and colors indicate the genome size $n$. The solid line is the fitted regression line.}
\label{fig:scatter_sampling}
\end{figure}

For each instance, we write
$x=\mathbb E(\delta_X)/n$ and 
$y=\mathbb E(\hat\delta_X)/n$.
Figure~\ref{fig:scatter_sampling} shows the scatter plot of $y$ versus
$x$ across $100$ independent runs for each value of $n\leq 20$. The dashed diagonal line $y=x$ represents exact
agreement between the sampled and full-search-space means: points close
to this line indicate that the sampled mean is already close to the full
mean without any correction. We also fit the linear regression
$y=a+\beta x $.
The purpose of this regression is not only to test whether the mean over the sampled leaves is close to the full-search-space mean, but also to test whether it provides a
stable calibrated predictor of the full-search-space statistic. The regression line allows for a systematic
correction. For example, if
the early sample is consistently biased toward smaller distances, then
the regression line records how the sampled value differs from the
corresponding full-search-space value.
In this interpretation, the intercept $a$ and slope $\beta$ describe
this correction: the intercept accounts for an additive shift, while the
slope accounts for a multiplicative change in scale. The residual $e_i$ of a point $(x_i,y_i)$ is the vertical prediction error from
the fitted regression line, $e_i=y_i-(a+\beta x_i)$. The value of $R^2$ summarizes how small these residuals
are relative to the overall variation in the sampled means. Thus, a large value of $R^2$ indicates that the sampled-prefix mean
provides a stable linear predictor of the full-search-space mean across
instances.

For $l_n=50n$, the fitted regression using all
instances with $6\le n\le 20$ is
\[
y=0.0931+0.9266x,
\qquad
R^2=0.9298.
\]
Thus the statistic over the first
sampled leaves is strongly correlated with the full-search-space
statistic, but the fitted line is not exactly the diagonal. This means
that the linear prefix sample should be interpreted as a calibrated
predictor rather than as an unbiased direct estimate of the full-search-space
mean.

To examine whether this calibration remains stable as larger genome
sizes are included, we refit the regression cumulatively. For each
$r\in\{6,7,\ldots,16,18,20\}$,
we fit
$y=a_r+\beta_r x$
using all instances with $6\le n\le r$. We then measure the change in
the fitted line by the discrete increments
\[
\Delta\beta_r
=
\frac{\beta_r-\beta_{r^-}}{r-r^-},
\qquad
\Delta a_r
=
\frac{a_r-a_{r^-}}{r-r^-},
\]
where $r^-$ denotes the previous available value of $r$. Thus, for
example, the increments from $16$ to $18$ and from $18$ to $20$ are
normalized by the step size $2$.

\begin{table}[ht]
\centering
\caption{Cumulative regression diagnostics for the first
$l_n=50n$ leaves sampled. For each value of $r$, the regression is fitted using all
instances with $6\le n\le r$.}
\label{tab:linear-prefix-cumulative-regression}
\begin{tabular}{@{\!}c@{\!~~}c@{\!~~~}c@{\!~~~}c@{\!~~~}c@{\!~~}c@{\!~~}c@{\!~~}c@{\!}}
\toprule
\(r\) & \# instances & \(\beta_r\) & \(a_r\) & \(R^2\) &
mean bias & \(\Delta\beta_r\) & \(\Delta a_r\) \\
\midrule
6  & 100  & 1.0058 & -0.0066 & 0.9996 &  0.0011 & --      & -- \\
7  & 200  & 0.9872 &  0.0167 & 0.9966 & -0.0012 & -0.0186 &  0.0232 \\
8  & 300  & 0.9787 &  0.0274 & 0.9909 & -0.0033 & -0.0085 &  0.0107 \\
9  & 400  & 0.9696 &  0.0387 & 0.9822 & -0.0060 & -0.0091 &  0.0113 \\
10 & 500  & 0.9519 &  0.0618 & 0.9665 & -0.0101 & -0.0176 &  0.0231 \\
11 & 600  & 0.9469 &  0.0687 & 0.9602 & -0.0119 & -0.0050 &  0.0069 \\
12 & 700  & 0.9348 &  0.0847 & 0.9530 & -0.0156 & -0.0121 &  0.0160 \\
13 & 800  & 0.9276 &  0.0942 & 0.9449 & -0.0185 & -0.0072 &  0.0095 \\
14 & 900  & 0.9282 &  0.0926 & 0.9405 & -0.0205 &  0.0006 & -0.0016 \\
15 & 1000 & 0.9280 &  0.0925 & 0.9328 & -0.0220 & -0.0002 & -0.0001 \\
16 & 1100 & 0.9237 &  0.0981 & 0.9277 & -0.0244 & -0.0044 &  0.0056 \\
18 & 1200 & 0.9254 &  0.0951 & 0.9276 & -0.0256 &  0.0009 & -0.0015 \\
20 & 1300 & 0.9266 &  0.0931 & 0.9298 & -0.0268 &  0.0006 & -0.0010 \\
\bottomrule
\end{tabular}
\end{table}

The cumulative diagnostics in
Table~\ref{tab:linear-prefix-cumulative-regression} show that the fitted
calibration line becomes stable once moderate genome sizes are included.
From $r=14$ onward, the slope and intercept change only slightly when
larger values of $n$ are added. Quantitatively,
\[
\frac{1}{5}\sum_{r\in\{14,15,16,18,20\}} |\Delta\beta_r|
\approx 0.00133,
\qquad
\frac{1}{5}\sum_{r\in\{14,15,16,18,20\}} |\Delta a_r|
\approx 0.00198.
\]
This indicates that the correction from the mean over the
sampled leaves to the full-search-space
mean is approximately stable over this range of $n$.

The statistic of the first sampled leaves also has a consistent negative bias compared with the mean over the full search space. At $r=20$, the average bias is approximately
\[
\frac{\mathbbm E(\hat\delta_X)-\mathbbm E(\delta_X)}{n}
\approx -0.0268.
\]
Thus, the early sampled leaves tend to contain candidates with smaller total
distance than a uniformly chosen leaf of the full reduced search space.
This bias is useful for finding near-median candidates quickly, but it
also means that the sampled mean should be interpreted through the
calibration relation. Therefore, for a larger instance where only the
prefix mean $y$ is available, the fitted linear-regime calibration
suggests the estimate
\[
\widehat{x}
=
\frac{y-0.0931}{0.9266}
\]
for the corresponding full-search-space normalized mean.

\begin{figure}
    \centering
    \includegraphics[width=1\linewidth]{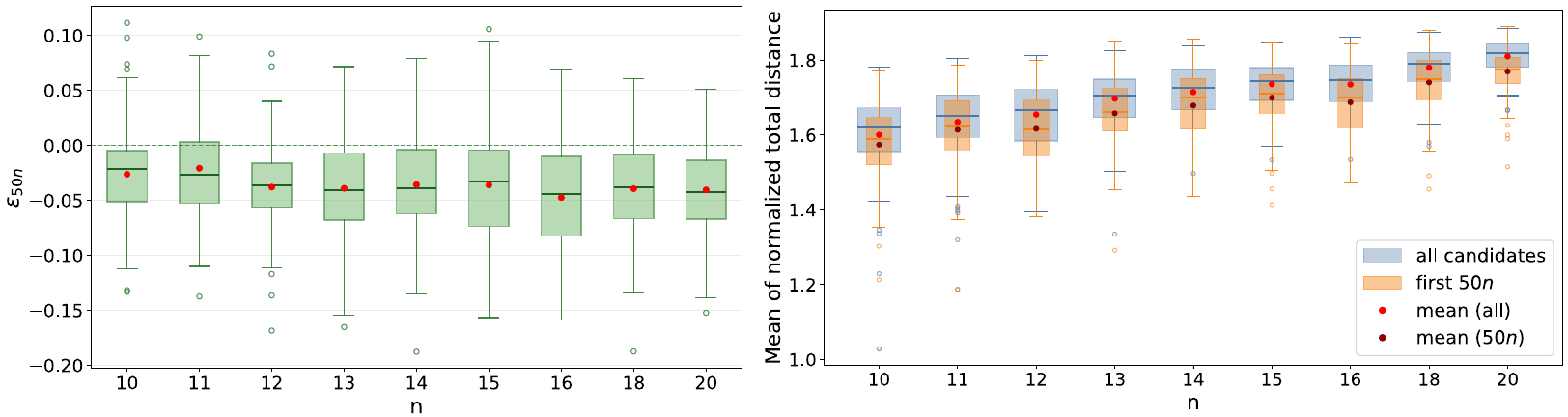}
    \caption{Comparison of normalized total distances for early candidates and the full candidate set \(Y(\mathcal T_{X,\OO})\). Left: boxplots of \(\varepsilon_{l_n}=\mathbbm{E}(\hat\delta/n)-\mathbbm{E}(\delta/n)\) for \(l_n=50n\), showing that early candidates have mean normalized total distance close to that of the full candidate set, with a slight negative bias. Right: overlay of the normalized total-distance distributions for the first \(l_n=50n\) candidates, shown in orange, and for all candidates, shown in blue. The two distributions largely overlap across the tested values of \(n\), while the early-candidate distribution is consistently shifted slightly toward lower distances.}
    \label{fig:delta_and_layered}
\end{figure}

To further quantify this agreement, we consider the difference between the
normalized mean total distances
$\varepsilon_{l_n}
=
\e(\hat\delta_X)/n
-
\e(\delta_X)/n.$
Figure~\ref{fig:delta_and_layered}~(left) presents boxplots of
$\varepsilon_{l_n}$ as a function of $n$.
The distributions are concentrated near zero but are mostly shifted to
negative values. This indicates that the first $l_n=50n$ leaves tend,
on average, to have smaller total distance than a uniformly chosen
element of the full reduced search space $Y(\mathcal T_{X,\OO})$.
Thus the linear prefix sample has a systematic bias toward near-optimal
candidates.
A more detailed comparison is provided in
Figure~\ref{fig:delta_and_layered}~(right), where we overlay the
distributions of normalized total distances obtained from the first
$l_n=50n$ candidates and from the full set $Y(\mathcal T_{X,\OO})$.
The two distributions exhibit a substantial overlap across all tested
values of $n$, with similar spreads. At the same time, the
sampled-prefix distribution is consistently shifted toward lower values,
again reflecting the tendency of the depth-first traversal to encounter
near-median candidates early.

In Table~\ref{tab:dist_stats_sampled}, we compute the same summary statistics
as in Table~\ref{tab:dist_stats}, but restricted to the first $l_n=50n$ sampled candidates. 
For each run, the minimum and maximum are
taken over the sampled prefix $z_1,\ldots,z_{l_n}$. Thus the
columns min and max report the averages, over the $100$ runs, of
\[
\min \frac{\hat\delta_X}{n}:=\min_{i=1,\ldots,l_n} \frac{d_T(z_i,X)}{n}
\qquad\text{and}\qquad
\max \frac{\hat\delta_X}{n}:= \max_{i=1,\ldots,l_n} \frac{d_T(z_i,X)}{n}.
\]
The mean and variance columns are computed with respect to the uniform
distribution on the sampled prefix, that is, for
$\hat\delta_X=d_T(\hat{\mathscr Z}_X,X)$, where
$\hat{\mathscr Z}_X$ is chosen uniformly from $z_1,\ldots,z_{l_n}$. Finally,
$\max j$ denotes the average, over runs, of the largest excess
$j=d_T(z_i,X)-\mu$
observed among the sampled candidates, where $\mu$ is the median value computed from the full reduced search space.

\begin{table}[ht!]
\centering
\caption{Summary statistics of $\hat\delta_X/n$ for the first $l_n=50n$ sampled candidates, averaged over $100$ runs.}
\label{tab:dist_stats_sampled}
\begin{tabular}{cccccc}
\toprule
$n$ &
$\min \hat\delta_X/n$ &
$\max \hat\delta_X/n$ &
$\mathbb{E}(\hat\delta_X/n)$ &
$\mathrm{Var}(\hat\delta_X/n)$ &
$\max j$ \\
\midrule
6 & 0.97 & 1.66 & 1.34 & 0.0313 & 4.13 \\
8 & 1.18 & 1.78 & 1.51 & 0.0195 & 4.96 \\
10 & 1.32 & 1.79 & 1.57 & 0.0115 & 4.99 \\
12 & 1.42 & 1.81 & 1.62 & 0.0078 & 5.24 \\
14 & 1.54 & 1.81 & 1.68 & 0.0043 & 4.59 \\
15 & 1.56 & 1.82 & 1.70 & 0.0035 & 4.71 \\
16 & 1.56 & 1.80 & 1.69 & 0.0032 & 4.88 \\
18 & 1.63 & 1.84 & 1.74 & 0.0024 & 4.65 \\
20 & 1.68 & 1.85 & 1.77 & 0.0016 & 4.46 \\
\bottomrule
\end{tabular}
\end{table}

A direct comparison with Table~\ref{tab:dist_stats} shows that the
statistics computed from the sampled candidates closely match those of
the full search space.
In particular, the mean normalized distance differs only marginally
across all values of $n$, while exhibiting a consistent slight
downward shift.
The variance and range follow the same trends, including the strong
concentration as $n$ increases.
These observations indicate that the early portion of the tree
traversal already captures the essential geometry of the reduced search
space.

Since the sampling procedure requires only a small number of leaves,
it enables us to efficiently explore much larger instances.
To assess the quality of the early sampled candidates,
Table~\ref{tab:dist_stats_large} reports summary statistics for the first
$l_n=50n$ sampled leaves. For each value of $n$, every entry in the table
is averaged over $100$ independent runs.
In addition to the minimum, maximum, mean, and variance of the normalized
total distance over the sampled prefix, we include the normalized lower
and upper bounds $\mathcal L_n/n$ and $\mathcal U_n/n$ for the median value, given in Theorem~\ref{thm:medianbounds2}. We also report the normalized gap between the
best sampled candidate and the lower bound,
$\min_{1\le i\le l_n} (d_T(z_i,X)-\mathcal L_n)/n$.

\begin{table}[ht!]
\centering
\caption{Summary statistics of $\hat\delta_X/n$ for $l_n=50n$, averaged over 100 runs.}
\label{tab:dist_stats_large}
\begin{tabular}{@{\!}c@{\!~~~}c@{\!~~~}c@{\!~~~}c@{\!~~~}c@{\!~~~}c@{\!~~~}c@{\!~~~}c@{\!}}
\toprule
$n$ &
$\frac{\min \hat\delta_X}{n}$ &
$\frac{\max \hat\delta_X}{n}$ &
$\mathbbm{E}[\hat\delta_X/n]$ &
$\mathrm{Var}(\hat\delta_X/n)$ &
$\frac{\mathcal L_n}{n}$ &
$\frac{\mathcal U_n}{n}$ &
$\frac{(\min \hat\delta_X - \mathcal L_n)}{n}$ \\
\midrule
30 & 1.80 & 1.88 & 1.84 & 0.000462 & 1.74 & 1.77 & 0.05 \\
40 & 1.85 & 1.91 & 1.88 & 0.000208 & 1.81 & 1.83 & 0.04 \\
50 & 1.88 & 1.93 & 1.91 & 0.000101 & 1.85 & 1.87 & 0.04 \\
60 & 1.90 & 1.94 & 1.92 & 0.000067 & 1.87 & 1.88 & 0.03 \\
70 & 1.92 & 1.95 & 1.93 & 0.000049 & 1.89 & 1.90 & 0.03 \\
80 & 1.92 & 1.95 & 1.94 & 0.000037 & 1.90 & 1.91 & 0.02 \\
90 & 1.93 & 1.96 & 1.94 & 0.000024 & 1.91 & 1.92 & 0.02 \\
100 & 1.93 & 1.96 & 1.95 & 0.000018 & 1.91 & 1.93 & 0.02 \\
\bottomrule
\end{tabular}
\end{table}

Several notable phenomena emerge from Table~\ref{tab:dist_stats_large}.
First, the normalized total distances of the sampled candidates remain
on a stable scale across the larger values of~$n$. The mean normalized
distance increases slowly from $1.84$ to $1.95$, while the minimum and
maximum sampled values remain in a narrow interval. This is consistent
with the behavior observed in the fully enumerated cases for smaller
$n$.

Second, the sampled distances become increasingly concentrated as $n$
grows. The range between the minimum and maximum normalized distances
decreases from about $0.08$ for $n=30$ to about $0.03$ for $n=100$, and
the variance decreases from $4.62\cdot 10^{-4}$ to $1.8\cdot 10^{-5}$.
Thus, even though only the first $l_n=50n$ leaves are sampled, their
total distances are confined to a narrow band for large $n$.

Third, the best sampled candidates lie close to the theoretical lower
bound. The interval
$[\mathcal L_n/n,\mathcal U_n/n]$ is itself narrow, and the normalized
gap \linebreak
$\min_{1\le i\le l_n} (d_T(z_i,X)-\mathcal L_n)/n$
decreases from $0.05$ to $0.02$ over the range of $n$ considered.
This indicates that the early sampled candidates are already
near-optimal, even for substantially larger instances.

\begin{figure}
    \centering
    \includegraphics[width=1\linewidth]{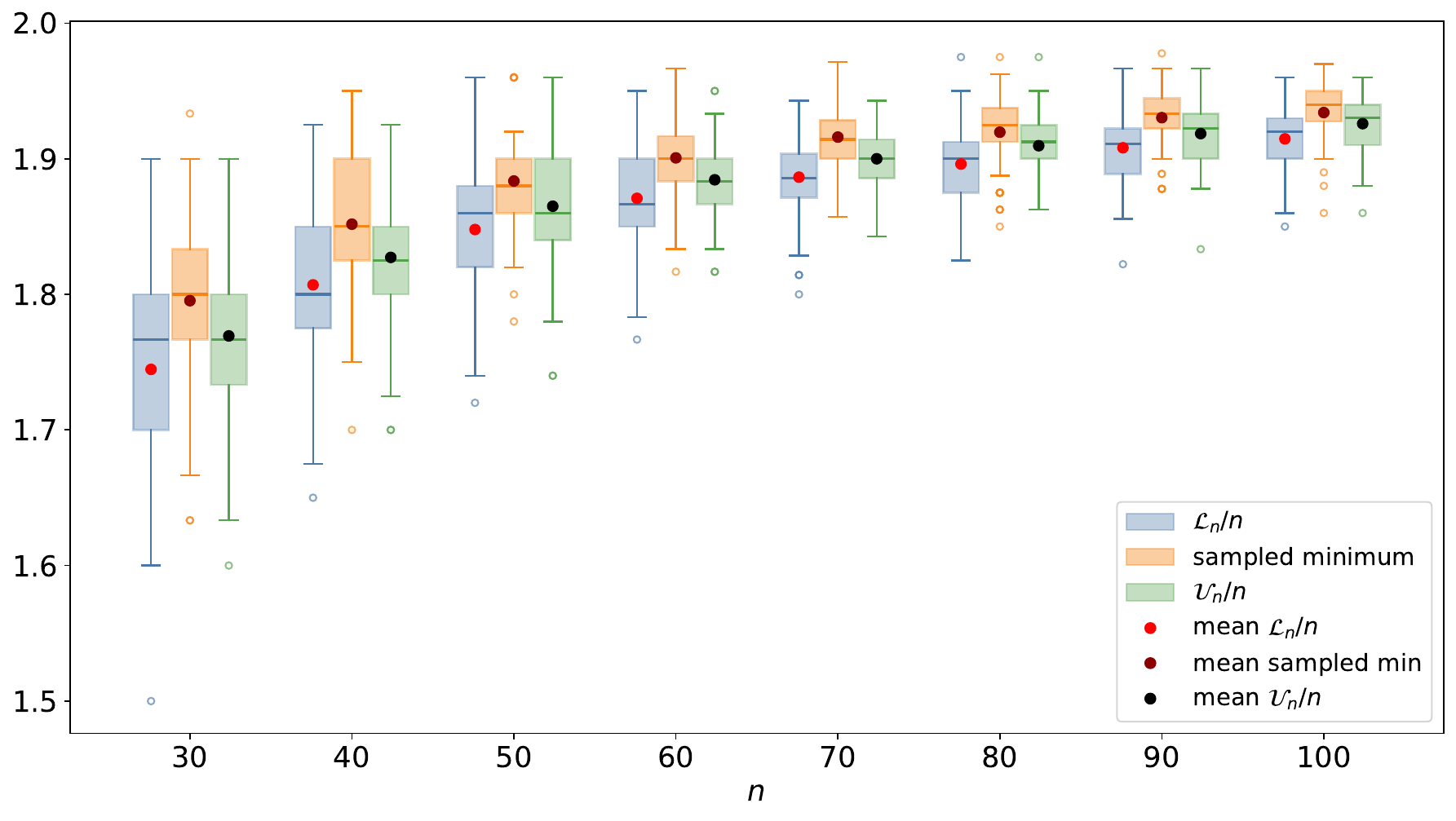}
    \caption{Run-level comparison of sampled minima with the median-value
bounds for large instances. For each $n$, the boxplots show the
distributions over $100$ runs of $\mathcal L_n/n$,
$\min_{1\le i\le l_n} d_T(z_i,X)/n$, and $\mathcal U_n/n$, with
$l_n=50n$. Dots mark the corresponding means.}
    \label{fig:large_min_bounds}
\end{figure}

Figure~\ref{fig:large_min_bounds} gives a run-level version of this
comparison. For each value of $n$, it shows the distributions over the
$100$ runs of the normalized lower bound $\mathcal L_n/n$ of the median value, the best sampled value
$\min \hat{\delta}_X=\min_{1\le i\le l_n}d_T(z_i,X)/n$,
and the normalized upper bound $\mathcal U_n/n$. The figure shows that
all three quantities lie on the same normalized scale and become more
concentrated as $n$ increases. For smaller instances, such as
$n=30$ and $n=40$, the best sampled values are typically slightly above
the theoretical upper-bound scale. However, the separation
between the sampled minima and the interval
$[\mathcal L_n/n,\mathcal U_n/n]$ decreases with $n$, and by
$n=70,80,90,$ and $100$ the three distributions are tightly clustered. Thus the
early sampled leaves not only have small average gap from the lower
bound, as reported in Table~\ref{tab:dist_stats_large}, but also show
stable run-level behavior across independent instances.

\begin{figure}
    \centering
    \includegraphics[width=1\linewidth]{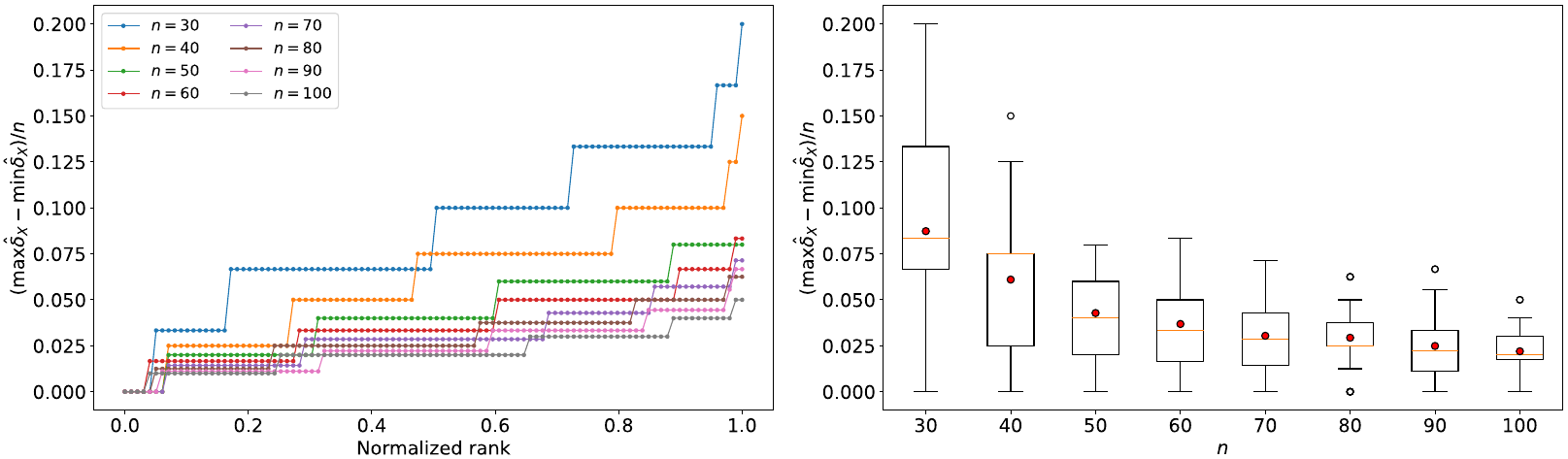}
    \caption{Concentration of total distances within the sampled prefix for
large instances. For each run, we compute $(\max \hat{\delta}_X -\min \hat{\delta}_X)/n$.
Left: ordered values of $(\max \hat{\delta}_X -\min \hat{\delta}_X)/n$ over the $100$ runs for each value of
$n$. Right: boxplots of the same quantities. The downward shift of the
ordered profiles and the contraction of the boxplots show that the
first $l_n=50n$ sampled leaves occupy an increasingly narrow normalized
distance range.}
    \label{fig:large_prefix_range}
\end{figure}

To further examine the dispersion of the sampled prefix, we consider the
normalized within-prefix range
$(\max \hat{\delta}_X -\min \hat{\delta}_X)/n=(\max_{1\le i\le l_n} d_T(z_i,X)-\min_{1\le i\le l_n} d_T(z_i,X))/n$.
For each fixed value of $n$, this quantity is computed for each of the
$100$ independent runs. In the left panel of
Figure~\ref{fig:large_prefix_range}, these $100$ values are sorted in
increasing order and plotted against their normalized rank. Thus the
left panel gives the empirical quantile profile of $(\max \hat{\delta}_X -\min \hat{\delta}_X)/n$ for each
genome size. The step-like shape of the curves reflects the fact that
the total distance is integer-valued before normalization by $n$.
Both panels show a strong decrease in the within-prefix range as $n$
increases. The quantile profiles shift downward almost uniformly from
$n=30$ to $n=100$, showing that the effect is not driven only by the
mean or by a few exceptional runs. The boxplots in the right panel
summarize the same phenomenon: the typical range drops from about
$0.09$ for $n=30$ to about $0.02$ for $n=100$, and the spread across
runs also decreases. Hence the first $l_n=50n$ sampled leaves occupy an
increasingly narrow normalized distance band. This provides further
evidence that the early depth-first traversal is concentrated around
near-optimal candidates, rather than producing candidates spread broadly
over the reduced search space.

These observations support the use of early sampling as a
practical approach for obtaining high-quality near-medians
without exhaustive enumeration, making it possible to explore
substantially larger instances.

\subsection{Sampling under additional constraints}

Beyond unconstrained sampling, the tree-based construction also allows
sampling under user-specified geometric conditions.
Rather than exploring arbitrary candidate medians, one may restrict the
search to genomes satisfying prescribed distance or adjacency
constraints.
This provides a flexible mechanism for navigating specific regions of
the reduced search space.

As an illustration, we consider the following constrained sampling
procedure.
Given a set $X=\{x_1,x_2,x_3\}\subset S_n,$ we search for the first leaf $z\in Y(\mathcal T_{X,\OO})$ satisfying $d(z,x_i)\ge B$, for all $i=1,2,3$, where
$B=\left\lfloor{2n}/{3}\right\rfloor-\gamma$.
In the experiments below we take $\gamma=\lceil 0.1n\rceil$.
This condition forces the sampled permutation to avoid neighborhoods of
the input genomes (the corners) and biases the search toward more balanced
medians lying in an intermediate region of the search space with respect to the input genomes.

As representative examples of this constrained sampling procedure, we
consider a single run for each value of $n$ with $\OO_n(X)=0$.
Table~\ref{tab:conditional_sampling} shows the distance vector
$(d(z,x_1),d(z,x_2),d(z,x_3))$, the total distance $d_T(z,X)$, and the
lower and upper bounds $\mathcal L_n(X)$ and $\mathcal U_n(X)$ for the median value
(Theorem~\ref{thm:medianbounds2}), together with the time required to
locate the first feasible leaf.

\begin{table}[ht!]
\centering
\footnotesize{
\caption{Single-run examples of constrained sampling with
$d_i=d(z,x_i)\ge B$ for all $i$,
where $B=\lfloor 2n/3\rfloor- n/10$ and $\OO_n(X)=0$.}
\label{tab:conditional_sampling}
\begin{tabular}{@{\!}c@{\!~~}c@{\!~~}c@{\!~~}c@{\!~~}c@{\!~~}c@{\!~~}c@{\!~~}c@{\!}}
\toprule
$n$ & $B/n$ & 
$(\frac{d_1}{n},\frac{d_2}{n},\frac{d_3}{n})$
& $\frac{d_T(z, X)}n$ &$\frac{\mathcal L_n(X)}n$&$\frac{\mathcal U_n(X)}n$& \footnotesize{$d_T(z, X)-\mathcal L_n(X)$} & \footnotesize{Runtime}
\\
\midrule
30  & 0.57 & $(0.57,0.63,0.73)$ & 1.93 & 1.80 & 1.80 & 4 & 8.05 s \\
40  & 0.55 & $(0.65,0.7,0.55)$ & 1.90 & 1.90 & 1.90 & 0 & 1.49 s \\
50  & 0.56 & $(0.56,0.72,0.6)$ & 1.88 & 1.82 & 1.82 & 3 & 57.17 s \\
60  & 0.57 & $(0.58,0.62,0.72)$ & 1.92 & 1.85 & 1.85 & 4 & 0.87 s \\
70  & 0.56 & $(0.56,0.66,0.69)$ & 1.90 & 1.84 & 1.84 & 4 & 3.45 min \\
80  & 0.56 & $(0.70,0.68,0.56)$ & 1.94 & 1.94 & 1.94 & 0 & 17.74 min \\
\bottomrule
\end{tabular}}
\end{table}

The sampled candidates are simultaneously far from all three input
genomes, indicating that the tree contains balanced configurations well
outside the immediate neighborhoods of the inputs.
At the same time, the total distances remain close to the theoretical
bounds $\mathcal L_n$ and $\mathcal U_n$.
In several cases the sampled candidate attains the upper bound exactly,
while in the remaining cases the excess above the lower bound is only a
small additive quantity.

The balanced constraint above targets candidates far from all three
inputs. To illustrate a different form of control, we next
impose an asymmetric distance profile. Let
$d_{(1)}(z)\le d_{(2)}(z)\le d_{(3)}(z)$ denote the ordered values of
$d(z,x_1),d(z,x_2),d(z,x_3)$. We require
$d_{(1)}(z)\ge B_1$, $d_{(2)}(z)\ge B_2$ and $
d_{(3)}(z)\ge B_3,$
where
$B_1=\left\lfloor n/3\right\rfloor-\gamma$,
$B_2=\left\lfloor 2n/3\right\rfloor-\gamma$,
$B_3=n-\gamma$.
Here again, we take $\gamma=\lceil 0.1n\rceil$. Equivalently, the sampled
candidate must be at least $B_1$ from all three inputs, at least $B_2$
from two of them, and at least $B_3$ from one of them. This imposes a
prescribed asymmetric distance profile, forcing the output to lie in a
more specific region of the search space than the balanced constraint
above.

\begin{table}[ht!]
\centering
\small{
\caption{Single-run examples of asymmetric constrained sampling with
$d_{(i)}(z)\ge B_i$,
where $B_i=\lfloor in/3\rfloor- n/10$ and $\OO_n(X)=0$.}
\label{tab:asymmetric_sampling}
\begin{tabular}{@{\!}c@{\!~~}c@{\!~~}c@{\!~~}c@{\!~~}c@{\!~~}c@{\!~~}c@{\!}}
\toprule
$n$ &
$(\frac{B_1}{n},\frac{B_2}{n},\frac{B_3}{n})$ &
$(\frac{d_{(1)}}{n},\frac{d_{(2)}}{n},\frac{d_{(3)}}{n})$ &
$\frac{d_T(z,X)}{n}$ &
$\frac{\mathcal L_n(X)}{n}$ &
$\frac{\mathcal U_n(X)}{n}$ &
Runtime \\
\midrule
30 & $(0.23,0.57,0.90)$ & $(0.23,0.70,0.93)$ & 1.87 & 1.80 & 1.80 & 0.05 s \\
40 & $(0.23,0.55,0.90)$ & $(0.23,0.78,0.90)$ & 1.90 & 1.90 & 1.90 & 0.17 s \\
50 & $(0.22,0.56,0.90)$ & $(0.22,0.78,0.90)$ & 1.90 & 1.82 & 1.82 & 3.01 s \\
60 & $(0.23,0.57,0.90)$ & $(0.23,0.73,0.90)$ & 1.87 & 1.85 & 1.85 & 7.60 s \\
70 & $(0.23,0.56,0.90)$ & $(0.23,0.80,0.90)$ & 1.93 & 1.84 & 1.84 & 2.72 min \\
80 & $(0.23,0.56,0.90)$ & $(0.23,0.84,0.90)$ & 1.96 & 1.94 & 1.94 & 1.26 min \\
\bottomrule
\end{tabular}}
\end{table}

These examples illustrate that the tree representation can be used not
only for exhaustive enumeration, but also as a structured framework for
targeted exploration of geometrically constrained regions of the search
space.
Even under strong balance constraints, feasible candidates can often be
located rapidly during traversal, making it possible to investigate
larger and more structured instances without exhaustive search. 
Thus, these examples show that the tree traversal provides a flexible
way to impose geometric constraints during the search. The same framework
can be directed toward balanced candidates far from all inputs, toward
candidates with prescribed asymmetric distance profiles, or toward other
regions of the reduced search space defined by distance or adjacency
conditions.

To our knowledge, this is the first breakpoint-median search procedure that can explicitly control the distance profile of the output genome during the search. Rather than finding only an arbitrary optimal or near-optimal solution, which may be very close to one of the input genomes, the tree-based framework allows one to impose geometric constraints and locate medians or near-medians in prescribed regions of the genome space, such as the balanced region far from all three input genomes considered here.

\section{Conclusion}\label{sec:conclusion}
In this paper, we developed tree-based algorithms for studying breakpoint medians and near-medians of linear unichromosomal genomes represented as unsigned permutations. In contrast to approaches based on reductions to optimization problems such as the Traveling Salesman Problem (TSP), which typically return only a single median, our method constructs highly restricted combinatorial search trees whose leaves encode candidate median genomes. 
By enforcing local adjacency constraints derived from the input genomes and using theoretical bounds on novel adjacencies, the resulting search spaces are guaranteed to contain all exact medians while remaining substantially smaller than the full permutation space. Furthermore, for independent random input genomes, the bounds in Theorems~\ref{thm:error-alg1}, \ref{thm:error-alg2}, and \ref{Algo4_error_bound} indicate that the majority of the permutations generated by our algorithms remain, with high probability, good approximations to medians. More generally, when the number of adjacencies shared by pairs of input genomes is small, the candidate genomes produced by Algorithm~1 provide good near-optimal approximations. Thus, even before enlarging the search space to allow adjacencies outside the input set, the basic tree construction already provides a useful and efficiently searchable space of near-medians.

This framework makes it possible not only to find optimal solutions, but also to enumerate all medians, study near-median genomes, and investigate the structure of the reduced search space, providing a more comprehensive view of evolutionary possibilities than single-solution TSP-based optimization approaches. 
This is particularly valuable in comparative genomics, where access to many median and near-median candidates allows possible ancestral genomes to be compared using additional biological or statistical criteria, such as synteny structure, balance among the input genomes, or consistency with known ancestral features.

Our experiments on three-genome instances confirm that these reduced tree-based search spaces can be extremely small, comprising, for example, less than $0.00019\%$ of $S_{16}$, while still containing all medians and a large collection of near-medians. Many of the near-median genomes have a total breakpoint distance only slightly above the minimum. We also observed that even when allowing up to three adjacencies outside the input set, external adjacencies were extremely rare: in many cases, fewer than $1\%$ of the medians contained even one such adjacency. These findings support the idea that the median structure is strongly constrained by the adjacency information already present in the input genomes.

Furthermore, by exploring the search trees through depth-first backtracking and recording leaves as soon as they are reached, we developed early sampling techniques that enable the rapid discovery of high-quality near-median approximations without requiring a complete traversal of the search trees. This is an important advantage of
the tree representation: rather than producing only one optimized genome, it
allows many candidate genomes to be sampled from different regions of the
reduced search space. More generally, the tree framework supports constrained
sampling under additional criteria. In particular, it can be used to search for
balanced near-medians whose distances to the input genomes are more evenly
distributed. Such midpoint-like candidates are especially relevant for
ancestral reconstruction and for exploring alternative evolutionary scenarios,
since they avoid solutions that are very close to one input genome and far
from the others.

Overall, our tree-based framework provides a useful unified approach for exact enumeration of all medians, efficient sampling of near-medians for larger genomes, and structural analysis of median sets under the breakpoint distance. Although this work focuses on unsigned unichromosomal genomes, the tree-based construction and median search strategy naturally extend to more general models, including signed permutations and multichromosomal genomes.

\vspace*{2mm}
\noindent\textbf{Funding:}
D.S.'s research is supported in part by NSERC Discovery Grant RGPIN-2022-05212.

\newcounter{mainequation}
\setcounter{mainequation}{\value{equation}}

\begin{appendices}
\renewcommand{\theequation}{\arabic{equation}}
\setcounter{equation}{\value{mainequation}}

\section{\!\!Proofs of Theorems\!~\ref{iff-Meidan-k=2}, \ref{thm:sandwich} and\!~\ref{thm:medianbounds2}}\label{appen:A}

\begin{proof}[Proof of Theorem~\ref{iff-Meidan-k=2}]
    First note that
$\mu(\{x,y\}) \le d_T(x,\{x,y\}) = d(x,y).$
Also, for any $z \in S_n$,
\[
d_T(z,\{x,y\}) = d(x,z) + d(z,y) \ge d(x,y).
\]
This implies
$\mu(\{x,y\}) = d(x,y).$
Therefore, $z \in M(\{x,y\})$ if and only if
\[
d_T(z,\{x,y\}) = d(x,z) + d(z,y) = d(x,y),
\]
proving “(1) $\Longleftrightarrow$ (2)”.

Now to prove “(1) $\Longleftrightarrow$ (3)”, recall the definition of $\mathcal B_{U}^X$ before Theorem~1. Omitting $X$ from the superscripts for simplicity, for $X=\{x,y,z\}$,
\begin{equation}\label{(1)}
n-1 = |\mathcal A_z|
= |\mathcal B_{xz}| + |\mathcal B_{yz}| + |\mathcal B_{xyz}|+|\mathcal B_{z}|. 
\end{equation}

Now note that $d(x,z)+d(z,y)=d(x,y)$ is equivalent to
\[
2(n-1) - |\mathcal B_{xz}| - |\mathcal B_{yz}| - 2|\mathcal B_{xyz}|
= (n-1) - |\mathcal B_{xy}| - |\mathcal B_{xyz}|,
\]
which itself is equivalent to
\begin{equation}\label{(2)}
    n-1 = |\mathcal B_{xz}| + |\mathcal B_{yz}|-|\mathcal B_{xy}|+|\mathcal B_{xyz}|. 
\end{equation}

From (\ref{(1)}), (\ref{(2)}) is equivalent to
$|\mathcal B_{z}| + |\mathcal B_{xy}| = 0,$
hence to
$|\mathcal B_{z}| = |\mathcal B_{xy}| = 0.$
The latter is equivalent to $\mathcal A_z \subseteq \mathcal A_x \cup \mathcal A_y$, and
$\mathcal A_{xy} = \mathcal B_{xy} \cup \mathcal B_{xyz} = \mathcal A_{xyz},$
which implies $\mathcal A_{xy} \subseteq \mathcal A_z$.
\end{proof}
\begin{proof}[Proof of Theorem~\ref{thm:sandwich}]
 The best possible scenario for the total distance of $m$ to $X$ is when
$m$ includes $\mathcal A_{x_i x_j}$ for all $i \ne j$ and the remainder of the adjacencies of
$m$ comes from $\bigcup_{x\in X} \mathcal A_x$. Therefore, the first part of the theorem follows immediately.

For the second part, note that from (\ref{eq:sandwich}), $\mathcal{A}_m$ can be expressed as the disjoint union
\begin{equation}\label{(3)}
\mathcal A_m
=
\left(
\bigcup_{r=2}^{k} \mathcal B^{X}_{x_{i_1},\ldots,x_{i_r}}
\right)
\cup
\left(
\bigcup_{i=1}^{k} \bigl(\mathcal A_m \cap \mathcal B^{X}_{x_i}\bigr)
\right).
\end{equation}

Thus,
\[
d(m,x_i)
=
(n-1)
-
\left|\mathcal A_m \cap \mathcal B^{X}_{x_i}\right|
-
\sum_{\substack{x_i \in U \subseteq X\\ |U|\geq 2}} \left|\mathcal B^{X}_{U}\right|.
\]

Since $|B^{X}_U|$ (on the right-hand side of the last equality) contributes $|U|$ times to $\sum_{i=1}^{k} d(m,x_i)$, we have
\[
d_T(m,X)
=
k(n-1)
-
\sum_{i=1}^{k} \left|\mathcal A_m \cap \mathcal B^{X}_{x_i}\right|
-
\sum_{1\le i_1<\cdots<i_r\le k} r \left|\mathcal B^{X}_{x_{i_1},\ldots,x_{i_r}}\right|
\]
\[
=
(k-1)(n-1)
-
\sum_{1\le i_1<\cdots<i_r\le k} (r-1)
\left|\mathcal B^{X}_{x_{i_1},\ldots,x_{i_r}}\right|,
\]
where the last equality follows from (\ref{(3)}).   
\end{proof}

\begin{proof}[Proof of Theorem~\ref{thm:medianbounds2}]
The lower bound follows from Theorem~\ref{thm:sandwich}. For the upper bound,
\[
\mu(X)
\le
\min_{x\in X} d_T(x,X)
=
\min_{x\in X}
\sum_{y\neq x}
\bigl((n-1) - |\mathcal A_{xy}|\bigr)
\]
\[
=
(k-1)(n-1)
-
\max_{x\in X}
\sum_{y\ne x} |\mathcal A_{xy}|
=
(k-1)(n-1) - v_n(X).
\]
\end{proof}

\section{Proof of Theorem~\ref{thmalpha}}\label{secA1}

Below, we include the proof of Theorem~\ref{thmalpha}, as presented in~\cite{dSJS25}.
\begin{proof}
For a permutation $\pi$ and $r\leq k$, let
$
\bar{\varepsilon}_{i_1,...,i_r}^X(\pi):=|\mathcal{A}_\pi\cap \mathcal{B}_{x_{i_1},...x_{i_r}}^X|.
$
To ease the notation, we let $\mathcal{B}_{i_1,\cdots,i_\ell}=\mathcal{B}_{x_{i_1},...,x_{i_{\ell}}}$.  Let $\eta=|\mathcal{A}_m\setminus \cup_{i=1}^k \mathcal{A}_{x_i}|$. Then
\[
\eta+\sum\limits_{r=1}^k \sum\limits_{1\leq i_1<...<i_r\leq k} \bar{\varepsilon}_{i_1,...,i_r}^X(m)=n-1.
\]
As $m$ is a median of $X$, we have
\[
\begin{array}{l}
d_T(m,X)=k(n-1)- \sum\limits_{r=1}^k [r \sum\limits_{1\leq i_1<...<i_r\leq k} \bar{\varepsilon}_{i_1,...,i_r}^X(m)]\\
=(k-1)(n-1)+\eta-\sum\limits_{r=2}^k [(r-1) \sum\limits_{1\leq i_1<...<i_r\leq k} \bar{\varepsilon}_{i_1,...,i_r}^X(m)]\\[5mm]
\leq d_T(x_k,X)=(k-1)(n-1)-(\sum\limits_{1\leq i_1< k} |\mathcal{B}_{i_1,k}^X|+2\sum\limits_{1\leq i_1<i_2< k} |\mathcal{B}_{i_1,i_2,k}^X|\\[5mm]
~~~~~~~~~~~~~~+\cdots+(k-2)\sum\limits_{1\leq i_1<...<i_{k-2}< k} |\mathcal{B}_{i_1,...,i_{k-2},k}^X|+(k-1)|\mathcal{B}_{1,...,k}^X|).
\end{array}
\]
Hence,
\begin{multline}
\eta \leq\\
(\sum\limits_{r=2}^{k} (r-1)\sum\limits_{1\leq i_1<...<i_r\leq k} \bar{\varepsilon}_{i_1,...,i_r}^X(m))-(\sum\limits_{r=2}^{k} (r-1)\sum\limits_{1\leq i_1<...<i_{r-1}< k} |\mathcal{B}_{i_1,...,i_{r-1},k}^X|)\\
\leq \sum\limits_{r=2}^{k-1} (r-1)\sum\limits_{1\leq i_1<...<i_r<k} |\mathcal{B}_{i_1,...,i_r}^X|,
\end{multline}
where the last inequality holds because $\bar{\varepsilon}_{i_1,...,i_r}^X(m) \leq |\mathcal{B}_{i_1,...,i_r}^X|$, for any $r\leq k$ and $1\leq i_1<...<i_r\leq k$.
\end{proof}
\section{Proofs of Proposition~\ref{Suff-cond-not-median}, Theorem~\ref{phi>kk} and Theorem~\ref{AdjX-Notin-M(X)}}
\begin{lemma}\label{maxflow}
Let $n,k \in \mathbb{N}$, $n \ge 2$, and $k \le n!/2$. For any simple graph $G$ on vertices $[n]$, and $X \subset S_n$ with $|X|=k \in \mathbb{N}$, and for any $a \ne b \in [n]$, we have
$w_X(a,b) \le k,$
and
$\varphi_X(a,G) \le \phi_X(a) \le 2k.$
\end{lemma}

\begin{proof}
First note that $w_X(a,b) \le k$, as $\{a,b\}$ appears in at most $k$ (all) elements of $X$.

For the other part, note that the first inequality follows from the definitions of $\varphi_X$ and $\phi_X$ directly. For the second inequality, note that $|\mathcal N_{\hat x}(a)| \le 2$, for any $x \in X$. Having $k$ genomes in $X$ concludes the lemma.
\end{proof}
\begin{lemma}\label{quadruple-max-min-1}
Let $k \in \mathbb{N}$ and $u,u',v,v' \in \mathbb{Z}_+$ such that
\[
\min\{u+u',\, v+v'\} < k,
\]
\[
\max\{u+u',\, v+v'\} \le k.
\]
Let
\[
u^* = \min\{u,u'\}, \!\!\quad u^{**} = \max\{u,u'\}, \!\!\quad
v^* = \min\{v,v'\}, \!\!\quad v^{**} = \max\{v,v'\}.
\]
Then
\[
\min\{u^* + v^{**},\, u^{**} + v^*\} < k.
\]
\end{lemma}

\begin{proof}
Without loss of generality, suppose $u+u' < k$ and $v+v' \le k$. If $u^* < v^*$, then
\[
u^* + v^{**} < v^* + v^{**} \le k.
\]
If $v^* \le u^*$, then
$v^* + u^{**} \le u^* + u^{**} < k.$
This proves the lemma.
\end{proof}
\begin{lemma}\label{quadruple-max-min-2}
Let $k \in \mathbb{N}$, $u,u',v,v' \in \mathbb{Z}_+$, such that
$u+u' = v+v' = k.$
Let
\[
u^* = \min\{u,u'\}, \!\!\quad u^{**} = \max\{u,u'\}, \!\!\quad
v^* = \min\{v,v'\}, \!\!\quad v^{**} = \max\{v,v'\}.
\]
Suppose $(u^{**},u^*) \neq (v^{**},v^*)$. Then,
$\min\{u^* + v^{**},\, u^{**} + v^*\} < k.$
\end{lemma}

\begin{proof} Without loss of generality, suppose $u^* < v^*$. Then
$u^{**} = k - u^* > v^{**} = k - v^*.$
Therefore,
\[
u^* + v^{**} < u^* + u^{**} = k.
\]
\end{proof}
\begin{proof}[Proof of Proposition~\ref{Suff-cond-not-median}]
First, note that from the assumption, we must have $n\geq 3$. For $n=1,2$, $\hat{S}_n$ is trivial, i.e. it has only one point, contradicting the assumption that for $x\in S_n$, $\{a,b\}\in \mathcal{A}_X\setminus \mathcal{A}_x$.

We first assume (C1). Suppose $|\mathcal{N}_{\hat{x}}(a)| = |\mathcal{N}_{\hat{x}}(b)| = 1$, i.e., $a$ and $b$ are telomere genes in $x$. Then there exists $\{a',b'\} \in \mathcal{A}_x$ such that $w_X(a',b') < k$. Otherwise, for all $\{a',b'\} \in \mathcal{A}_x$, we have $w_X(a',b') = k$, meaning that all $k$ genomes include a circle $a\leftrightarrow b \leftrightarrow \cdots \leftrightarrow a$, which is a contradiction.

We construct $\hat{x}' \in S_n$ from $\hat{x}$ by deleting $\{a',b'\}$ from $\hat{x}$ and adding $\{a,b\}$, hence improving the total distance of $\hat{x}$. Therefore, $x \notin M(X)$.

Now, without loss of generality, assume
$1 = |\mathcal{N}_{\hat{x}}(a)| < |\mathcal{N}_{\hat{x}}(b)| = 2.$
Suppose $\mathcal{N}_{\hat{x}}(b) = \{b',b''\}$, be such that $b'$ is the unique gene (vertex) on the path connecting $b$ to $a$ in $\hat{x}$. If $w_X(b,b') < k$, then construct a genome $x'$ whose total distance to $X$ is improved (smaller) by removing $\{b,b'\}$ from $x$ and adding $\{a,b\}$. If $w_X(b,b') = k$, then as $w_X(b,a) = k$, from Lemma~\ref{maxflow}, we have $w_X(b',b'') = 0$.

There must also be an edge $\{c,d\}$ on the path between $a$ and $b$ in $\hat{x}$ with $w_X(c,d) < k$, otherwise there is a cycle subgraph $a \leftrightarrow b \leftrightarrow b' \leftrightarrow \ldots \leftrightarrow a$ in all genomes of $X$. Construct $x'$ by removing $\{c,d\}$ and $\{b,b''\}$ and adding $\{a,b\}$ and $\{b'',c\}$. This improves the total distance of $x'$ to $X$ over $x$. Hence $x \notin M(X)$.

Now we assume (C2). If in addition (C1) holds, we proved that $x \notin M(X)$. Suppose (C2) holds but (C1) does not, that is suppose
$\mathcal{N}_{\hat{x}}(a) = \{a',a''\},$ and $ 
\mathcal{N}_{\hat{x}}(b) = \{b',b''\}$
such that $a'$ and $b'$ are located on the path connecting $a$ and $b$ on the graph $\hat{x}$.

Also, without loss of generality, suppose
\[
\varphi_X(a,\hat{x}) = w_X(a,a') + w_X(a,a'') < k.
\]
From Lemma~\ref{quadruple-max-min-1}, we have at least one of the following
\begin{align}
w_X(a,a') + w_X(b,b'') &< k, \label{temp-ineq-1} \\
w_X(a,a'') + w_X(b,b') &< k. \label{temp-ineq-2}
\end{align}

Suppose \eqref{temp-ineq-1} holds. We construct $x'$ by deleting $\{a,a'\}$ and $\{b,b''\}$ and adding $\{a,b\}$ and $\{a',b''\}$, improving the total distance of $x'$ to $X$ over that of $x$ to X by at least $1$. The case in which \eqref{temp-ineq-2} holds is similar.

This shows that if (C2) holds, we can always construct $x'$ such that
\[
\sum_{\tilde{x} \in X} d(x',\tilde{x}) \le -1 + \sum_{\tilde{x} \in X} d(x,\tilde{x}),
\]
which means $x \notin M(X)$.

The same argument used under assumption (C2), along with Lemma~\ref{quadruple-max-min-2}, implies that, under (C3), there exists $x'$ whose total distance to $X$ is strictly smaller than that of $x$ to $X$, i.e.,
\[
(C3) \;\Rightarrow\; x \notin M(X).
\]

The result under (C4) and (C5) can be proved similarly.
\end{proof}
\begin{proof}[Proof of Theorem~\ref{phi>kk}.]
Proposition~\ref{Suff-cond-not-median} implies:

\begin{enumerate}
\item[(i)] If $\{a,b\} \notin \mathcal{A}_x$ and (C2) (from the statement of Proposition~\ref{Suff-cond-not-median}), then $x \notin M(X)$.

\item[(ii)] If $\{a,b\} \notin \mathcal{A}_x$ and (C3), then $x \notin M(X)$.
\end{enumerate}

These are equivalent, respectively, with

\begin{enumerate}
\item[(i')] If $x \in M(X)$ and (C2), then $\{a,b\} \in \mathcal{A}_x$.

\item[(ii')] If $x \in M(X)$ and (C3), then $\{a,b\} \in \mathcal{A}_x$.
\end{enumerate}

From (i'), if $x \in M(X)$ and (C2), then $\{a,b\} \in \mathcal{A}_x$. On the other hand, $\{a,b\} \in \mathcal{A}_x$ means
\[
\varphi_X(a,\hat{x}), \ \varphi_X(b,\hat{x}) \geq k,
\]
which contradicts (C2). This concludes that if $x \in M(X)$, then (C2) is not true, proving the first part of the theorem.

By (ii'), whenever $x \in M(X)$ and $\varphi_X(a,\hat{x}) = \varphi_X(b,\hat{x}) = k$ and (C3), we must have $\{a,b\} \in \mathcal{A}_x$, which, together with $\varphi_X(a,\hat{x}) = \varphi_X(b,\hat{x}) = k$ implies
\[
W_X(a,\hat{x}) = W_X(b,\hat{x}) = (k,0),
\]
which contradicts (C3). This implies that if in addition we assume
$\varphi_X(a,\hat{x})\allowbreak = \varphi_X(b,\hat{x}) = k,$
then $W_X(a,\hat{x}) = W_X(b,\hat{x})$, proving the second part of the theorem.
\end{proof}
\begin{proof}[Proof of Theorem~\ref{AdjX-Notin-M(X)}]

First note that from Lemma~\ref{maxflow}, $\varphi_X(a,\hat{x}),\ \varphi_X(b,\hat{x}) \le k.$

For the first part, suppose $|\mathcal{N}_X(a)|=1$ or $|\mathcal{N}_X(b)|=1$. Then, as $\{a,b\} \notin \mathcal{A}_x$, we have
\[
\min\{\varphi_X(a,\hat{x}),\ \varphi_X(b,\hat{x})\} = 0 < k,
\]
and from Proposition~\ref{Suff-cond-not-median} we have $x \notin M(X)$, which is a contradiction.

If $|\mathcal{N}_X(a)| \ge 4$ or $|\mathcal{N}_X(b)| \ge 4$, then
\[
\varphi_X(a,\hat{x}) \le k-1 \quad \text{or} \quad \varphi_X(b,\hat{x}) \le k-1,
\]
implying that $x \notin M(X)$ by Proposition~\ref{Suff-cond-not-median}, which is again a contradiction.

Now suppose $|\mathcal{N}_X(a)| = |\mathcal{N}_X(b)| = 2$. If, in addition, \linebreak
$\min\{\varphi_X(a,\hat{x}),\ \varphi_X(b,\hat{x})\} < k,$
then from Proposition~\ref{Suff-cond-not-median}, $x \notin M(X)$, which is a contradiction. Otherwise, if
\[
\varphi_X(a,\hat{x}) = \varphi_X(b,\hat{x}) = k,
\]
then, since $\{a,b\} \notin \mathcal{A}_x$, we must have
$W_X(a,\hat{x}) = W_X(b,\hat{x}) = (k,0).$
Therefore, from Proposition~\ref{Suff-cond-not-median}, $x \notin M(X)$, which contradicts the assumptions of the theorem.

Finally, suppose
$\min\{|\mathcal{N}_x(a)|,\ |\mathcal{N}_x(b)|\} = 2, $ and $
\max\{|\mathcal{N}_x(a)|,\ |\mathcal{N}_x(b)|\} \allowbreak = 3.$
Without loss of generality, suppose $|\mathcal{N}_x(a)| = 2$ and $|\mathcal{N}_x(b)| = 3$. If
$\min\{\varphi_X(a,\hat{x}),\ \varphi_X(b,\hat{x})\} < k,$
then from Proposition~\ref{Suff-cond-not-median}, $x \notin M(X)$. If
$\varphi_X(a,\hat{x}) = \varphi_X(b,\hat{x}) = k,$
then
\[
W_X(a,\hat{x}) = (k,0), \quad W_X(b,\hat{x}) = (i, k-i),
\]
for some $k/2 \le i < k$. In this case, from Proposition~\ref{Suff-cond-not-median}, $x \notin M(X)$, which is a contradiction.

As a result, under the theorem’s assumptions, the only possible case is
\[
|\mathcal{N}_X(a)| = |\mathcal{N}_X(b)| = 3,
\]
and $\varphi_X(a,\hat{x})=\varphi_X(b,\hat{x})=k$.
Hence, we have
\[
\overline{W}_X(a) = (k,i,k-i)
\quad \text{and} \quad
\overline{W}_X(b) = (k,j,k-j)
\]
for some $k/2 \le i,j < k$. Now, as $\{a,b\} \notin \mathcal{A}_x$, either
$\min\{\varphi_X(a,\hat{x}), \varphi_X(b,\hat{x})\} \allowbreak < k,$
or
\[
W_X(a,\hat{x}) = (i,k-i)
\quad \text{and} \quad
W_X(b,\hat{x}) = (j,k-j).
\]

In both cases, from Proposition~\ref{Suff-cond-not-median}, $x \notin M(X)$, except when $i=j$, i.e.,
\[
W_X(a,\hat{x}) = W_X(b,\hat{x}) = (i,k-i).
\]
This proves the first part.

For the second part, note that for any $y \in M(X)$, $\mathcal{A}_X \setminus \mathcal{A}_y \subseteq \Gamma.$ Hence
\[
\mathcal{A}_X \setminus \mathcal{A}_x
\subseteq
\mathcal{A}_X \setminus \mathcal{A}_{M(X)}
=
\bigcup_{y \in M(X)} (\mathcal{A}_X \setminus \mathcal{A}_y)
\subseteq \Gamma,
\]
concluding the second part of the theorem.
\end{proof}

\end{appendices}

\bibliographystyle{acm}
\bibliography{ref}

\end{document}